\documentclass{aa}  
\usepackage{graphicx}
\usepackage{txfonts}
\usepackage[breaklinks, colorlinks, citecolor=blue, urlcolor = blue]{hyperref}  
\usepackage{adjustbox}
\usepackage{pifont}
\usepackage{color}

\newcommand{\cm}{{\ding{51}}}%
\newcommand{\xm}{\ding{55}}%

\newcommand{\rbul}{{\LARGE \color{red}$\bullet$}}
\newcommand{\gbul}{{\LARGE \color{green}$\bullet$}}
\newcommand{\ybul}{{\LARGE \color{yellow}$\bullet$}}

\newcommand{\upd}{{\ding{72}}}

\newcommand{\qm}{{\bf?}}

\usepackage{subfig}
\usepackage{threeparttable}
\usepackage{enumitem}

\usepackage{booktabs}
\newcommand{\tabitem}{~~~~\llap{\quad - }~~}
\begin{document}

   \title{Testing the accuracy of reflection-based supermassive black hole spin measurements in AGN}


   \author{E. S. Kammoun \inst{1}
   \and E. Nardini \inst{2}
   \and G. Risaliti \inst{3,2}
   }

   \institute{SISSA, via Bonomea 265, I-34135 Trieste, Italy; \email{\href{mailto:ekammoun@sissa.it}{ekammoun@sissa.it} }
              \and INAF - Osservatorio Astrofisico di Arcetri, Largo E. Fermi 5, I-50125 Firenze, Italy 
              \and Dipartimento di Fisica e Astronomia, Università di Firenze, via G. Sansone 1, I-50019 Sesto Fiorentino (Firenze), Italy}

   \date{Received ...; accepted ...}

\abstract{X-ray reflection is a very powerful method to assess the spin of supermassive black holes (SMBHs) in active galactic nuclei (AGN), yet this technique is not universally accepted. Indeed, complex reprocessing (absorption, scattering) of the intrinsic spectra along the line of sight can mimic the relativistic effects on which the spin measure is based.}
{In this work, we test the reliability of SMBH spin measurements that can currently be achieved through the simulations of high-quality \textit{XMM--Newton} and \textit{NuSTAR} spectra.}
{Each member of our group simulated ten spectra with multiple components that are typically seen in AGN, such as warm and (partial-covering) neutral absorbers, relativistic and distant reflection, and thermal emission. The resulting spectra were blindly analysed by the other two members.}
{Out of the 60 fits, 42 turn out to be physically accurate when compared to the input model. The SMBH spin is retrieved with success in 31 cases, some of which (9) are even found among formally inaccurate fits (although with looser constraints). We show that, at the high signal-to-noise ratio assumed in our simulations, neither the complexity of the multi-layer, partial-covering absorber nor the input value of the spin are the major drivers of our results. The height of the X-ray source (in a lamp-post geometry) instead plays a crucial role in recovering the spin. In particular, a success rate of 16 out of 16 is found among the accurate fits for a dimensionless spin parameter larger than 0.8 and a lamp-post height lower than five gravitational radii.}{} 


   \maketitle
%

\section{Introduction}
\label{sec:intro}

An isolated astrophysical black hole is characterized by two fundamental quantities: its mass, $M$, and its angular momentum, $J$; this assumes
charge neutrality, and thus the space-time is described by the Kerr metric \citep[][]{Kerr63}. The mass can be considered as a scaling factor for distances, timescales and luminosities. In other terms, this means that accreting black holes with masses ranging from stellar mass, as seen in X-ray binaries, to supermassive black holes (SMBHs), as seen in active galactic nuclei (AGN), have a similar spectro-temporal behaviour once the luminosities and timescales are scaled properly \citep{McHardy06}. However, the angular momentum (usually described in terms of the dimensionless spin parameter, $a^\ast = Jc/GM^2$), which a black hole acquires from its growth history, is arguably the most interesting parameter as it affects the Kerr metrics leading to various properties of astrophysical importance. Theoretically, the spin values range in the $[-0.998, 0.998]$ interval \citep{Thorne74}. These limits are found without considering magnetohydrodynamic (MHD) accretion. In fact, the magnetic fields of the plunging regions should give rise to torques that tend to reduce the maximum spin ($\sim 0.9-0.95$) that can be achieved by a black hole \citep[e.g.][]{Gammie04, McKinney04}.

Measurements of SMBH spins are a key ingredient for understanding the physical processes on scales ranging from the accretion disc out to the host galaxy. In fact, the spin determines the position of the innermost stable circular orbit (ISCO) of the accretion disc and of the event horizon, which are 1.24 and 1.06\,$r_{\rm g}$ for a maximally rotating black hole, and 6 and 2\,$r_{\rm g}$ for a non-spinning black hole, respectively, where $r_{\rm g} = GM/c^2$ is the gravitational radius. Hence, it has been shown that for a Schwarzschild black hole ($a^\ast = 0$) half of the energy is radiated within $\sim 30\,r_{\rm g}$, while half of the radiation emerges from within $\sim 5\,r_{\rm g}$ for a rapidly spinning black hole \citep[e.g.][]{Thorne74, Agol00}. This can be translated into an increase of the radiative accretion effeciency ($\eta$) from $\eta = 0.057$ for $a^\ast = 0$, to 0.32 for $a^\ast = 0.998$. \cite{Vasudevan16} assumed a toy model with a bimodal spin distribution and showed that a SMBH population where only 15\% of the sources are maximally rotating can produce 50\% of the cosmic X-ray background (CXB) owing to their high radiative efficiency. Moreover, these authors showed that the spin bias is even larger in flux-limited surveys, since half of the CXB can be accounted for if only 7\% of the sources have a spin of 0.998 \citep[see also][]{Brenneman11}. 

The SMBH spin distribution is also fundamental for understanding the SMBH-host galaxy co-evolution. In fact, the angular momentum of a black hole matures over cosmic time and its final value is determined by the accretion and merger history of the galaxy. For instance, mergers tend to spin down the black hole \citep{Volonteri13}, while the SMBH spins up through prograde accretion of material through the galactic disc \citep{King08}. It has also been shown that highly energetic outflows in the form of relativistic winds \citep{Gofford15} or jets \citep[e.g.][]{Blandford77,King15} are affected by the accretion flow, hence their strengths are somehow related to the SMBH spin. These forms of mechanical AGN feedback, in addition to the high radiative efficiency, seem to play a crucial role in the evolution of the host galaxy and its star formation history. Hence, understanding the growth of SMBHs and their spin distribution is a key point for our understanding of the larger scale structure of the Universe \citep[see][for a review about AGN feedback]{Fabian12}. 

In addition to the importance of SMBH spin in cosmology and galaxy evolution, the nuclear regions in AGN can be considered as unique laboratories to directly test the effects of general relativity, which manifest themselves as extreme physical phenomena such as light bending \citep[e.g.][]{Miniutti04} and reverberation lags \citep[e.g.][]{Fabian09, Emmanoulopoulos11,Kara16}. This requires high-quality X-ray observations. In fact, AGN are strong X-ray emitters and it is widely accepted that the X-rays arise from the innermost regions of the accretion disc where the primary continuum is due to the Comptonization of ultraviolet (UV) disc photons by a hot ($\sim 10^9$\,K) transrelativistic medium, usually referred to as the X-ray corona \citep[e.g.][]{Shap76,Haa93,Pet01b,Pet01a}. As the measure of the spin is strongly dependent on the irradiation and subsequent emissivity of the disc, its success is tightly connected to the study of the X-ray corona itself, whose nature and properties are still largely unknown. 

\section{Spin measurement methods and their limitations}
\label{sec:spinmethod}

The increase in number and quality of AGN spectra, at various wavelengths, allowed astronomers to attempt a determination of the SMBH spin parameters with a relatively high confidence. A variety of observed features are considered as good indicators of the black hole spin, such as continuum shape \citep[e.g.][]{Done13}, broad iron K$\alpha$ line \citep{Fab00}, and quasi-periodic oscillations \citep[QPOs; e.g.][]{Mohan14}. Recently, the detection of gravitational waves through the coalescence of black hole pairs founded a new technique to constrain the spins of non-accreting black holes \citep[][]{LIGO1, LIGO2}. There are several advantages and caveats relative to each method. Continuum fitting, for instance, can be applied to any AGN for which continuum emission is detected and has been applied to sources out to redshift $\sim 1.5$ \citep[e.g.][]{Capellupo15, Capellupo16}. However, one of the main drawbacks of this method is that it requires a broad and simultaneous wavelength coverage, which usually exceeds the capabilities of a single observatory, in order to determine properly the shape of the relevant part of the spectral energy distribution (i.e. optical to X-rays). This method requires accurate estimates for the black hole mass, disc inclination, and distance, which are typically derived from optical data. Furthermore, the continuum fitting method can be applied effectively only when the peak of the emission from the accretion disc can be reasonably probed. Since most AGN spectra peak in the extreme UV, this range is only accessible by current detectors in high-redshift objects, at the expense of a rather modest quality for the corresponding X-ray spectra \citep[e.g.][]{Collinson17}. As for the QPOs, they are common in Galactic binaries while few examples exist in AGN light curves, most of which are statistically marginal and/or controversial \citep[apart from the notable case of RE J1034+396;][]{Gierlinski08}. Their detection requires long monitoring, high signal-to-noise ratio (S/N), and a proper modelling of the continuum power spectrum \citep{Vaughan05, Vaughan06}.

The most direct and robust measurements to date are those obtained through the detection of a strong relativistic reflection feature in the X-ray spectra. This method can be applied to a wider black hole (BH) mass range. X-ray spectra of AGN can be expressed as a sum of several components, in particular a primary continuum that is well approximated by a power law with a high-energy exponential cut-off and ionized and/or neutral reflection that is detected in most of the sources, arising either from the accretion disc within a few gravitational radii from the BH or from distant Compton-thick material (the broad line region or the molecular torus), respectively \citep[e.g.][]{Light88, Geo91, Ghisellini94,Bia09}. The resulting reflection spectrum is characterized mainly by the iron K$\alpha$ emission line at $\sim 6.4-7.0$\,keV and a broad component peaked at around 20–30 keV, known as the Compton hump. Special and general relativistic effects result in blurring the ionized reflection spectrum and asymmetrically broadening the Fe K$\alpha$ emission line owing to the gravitational redshift and the motion of the emitting particles in the disc \citep[see][for reviews]{Fab00, Reynolds03}. This method consists in fitting the X-ray spectrum of a given source with a reflection model accounting for the relativistic distortions that affect these features on their way to the observer. We mention some of the models that predict the relativistic line profile for a narrow line emitted in the rest frame of the accretion disc: {\tt diskline}, {\tt laor}, {\tt kyrline}, {\tt kerrdisk}, {\tt relline} as published in \cite{Fabian89, Laor91, Dovciak04, Brenneman06, Dauser10}, respectively. The resulting shape of the reflection spectrum strongly depends on the parameters of the system. Hence, this method can be used not only to determine black hole spins but also to probe the innermost regions of the accretion discs, providing information about its inclination, ionization state, elemental abundance, and emissivity \citep[see][for a review]{Reynolds14}. However, there are known difficulties in determining the spins via X-ray reflection, which are mainly due to the complexity of (and some subjectivity in) modelling the AGN spectra, considering the various emission and absorption components that are known to be present, hence requiring high-quality data \citep[e.g.][]{Guainazzi06, Mantovani16}. 

An alternative absorption-based interpretation has been proposed to explain the apparent, broad red wing of the Fe line and the spectral curvature below 10\,keV \citep[e.g.][]{Miller08, Miller09}. According to this scenario, partial-covering absorbers in the line of sight (having column densities in the $10^{22-24}\,\rm cm^{-2}$ range) plus distant (i.e. non-relativistic) low-ionization reflection, can produce an apparent broadening of the Fe K$\alpha$ line similar to that caused by relativistic effects. Variability in the covering fraction of these absorbers would also provide a complete description of the observed spectral variability. 

Contrary to stellar-mass BHs, whose spectra are much brighter and typically less complex, both blurred reflection and partial covering are relevant to the X-ray spectra of AGN. In fact, while the former process is able, in principle, to explain the spectral and timing properties of any accreting system, the rapid Compton-thin to Compton-thick (and vice versa) transitions seen in changing-look AGN \citep[after][]{Matt03} imply that also partial covering must be taken into account. In a single-epoch AGN spectrum, the effects of disc reflection and partial covering are often hard to separate or distinguish from each other, thus leading to a long-standing debate. However, thanks to the high-quality spectra provided by {\it XMM-Newton} \citep{Jans01} and {\it NuSTAR} \citep{Har13} observations, which jointly cover a wide energy range from 0.3\,keV up to 80\,keV, it has now become possible to disentangle the two scenarios, as shown in the case of NGC\,1365 \citep{Risaliti13}. Three different variable absorbers with column densities in the range $5\times 10^{22}-6.5\times 10^{24}\rm\,cm^{-2}$ and variable covering factors would be needed to explain the spectrum of the source below 10\,keV. However, this absorption-only model fails to explain the hard X-ray spectrum in the 10--80\,keV band, where the inclusion of relativistic reflection provides a statistically better description of the data \citep{Risaliti13, Walton14}. Furthermore, the latter model is also preferred on physical grounds, as the inferred bolometric luminosity in absorption-only scenarios is significantly higher compared to other indicators such as the [\ion{O}{iii}]\,$\lambda$5007 line. The case of NGC~1365 lends weight to the idea that X-ray reflection is indeed an effective means of measuring the spin of SMBHs, even when absorption is present.

\section{Motivation and methods}
\label{sec:motivation}

Our main aim is to test the reliability of spin measurements when the spectra include additional components with respect to the simple primary continuum plus disc reflection configuration (i.e. always, in principle). In fact, the innermost emission components are generally subject to absorption by gas with column densities from $N_{\rm H} < 10^{21}\,{\rm cm^{-2}}$ to $N_{\rm H} > 10^{24}\,{\rm cm^{-2}}$ and ionization states from neutral to almost completely ionized. As demonstrated by the case of NGC\,1365, the presence of a given absorption component can be better identified thanks to its variability. NGC\,1365 is one of the unique sources showing frequent changes in its obscuration state. Recently, \cite{Risaliti16} summarized the various observational aspects of this source and proposed a multi-layer structure of the circumnuclear medium to explain all the observed absorption states and their variability. {NGC\,1365 has been observed several times in reflection-dominated states, suggesting the presence of a layer of neutral Compton-thick ($N_{\rm H}>10^{24}\rm\,cm^{-2}$) absorber located at a distance of the order of or larger than that of the broad line region \citep{Risaliti07}. The source is usually caught in a Compton-thin state, $N_{\rm H} \sim 10^{23}\,\rm cm^{-2}$, but the column can even occasionally drop down to $N_{\rm H} \sim 10^{22}\,\rm cm^{-2}$ \citep{Braito14}. Furthermore, absorption lines have been detected when the source is not heavily obscured, indicating a stratification of absorbers with ionization states ranging from highly ionized ($\log \xi > 3$)\footnote{$\xi = L/n r^2$ is the ionization parameter (units $\rm erg\,cm\,s^{-1}$), where $L$ is the luminosity of the X-ray source, and $n$ is the volume density of the absorber situated at a distance $r$ from the source.} to mildly ionized ($\log \xi \sim 1-2$) down to neutral ($\log \xi < 1$)}. {\it All} these components and absorption states can be present in {\it all} AGN. However, their detection in a single source, even if at different times, is highly dependent, on the one hand, on our line of sight, and, on the other hand, on the chance to observe any given component when variability is present.

The repetition of measurements (via X-ray reflection) can result in inconsistent values of the spin parameter for a given source. The discrepancy is mainly due to the use of different components in modelling the spectra, such as the use of dual reflectors, partial-covering, and/or warm absorbers. For example, \cite{Patrick11} analysed the {\it Suzaku} spectra of NGC\,3783, among others, and found the spin parameter in this source is $a^{\ast}<-0.04$. This result contradicts the high spin parameter ($a^\ast \geq 0.88$) found by \cite{Brenneman11} and \cite{Reynolds12}, who analysed the same observations.

The work presented in this paper is a preliminary study aiming to test, through the simulation of high-quality {\it XMM-Newton} and {\it NuSTAR} spectra (in the 0.3--79\,keV range), the reliability of reflection-based SMBH spin measurements that can currently be achieved. A similar approach has been adopted recently by \cite{Bonson16} and \cite{Choudhury17}, who simulated AGN spectra by assuming only two components: primary emission and relativistic reflection. Instead, we assume a more complex spectral configuration, closer to the real general case. This is presented below, along with a detailed description of how we simulated and fitted the data. We note that both of the aforementioned studies neglected the soft X-ray band, in which the soft excess can be a crucial driver of reflection-based spin determinations \citep[e.g.][]{Walton13}.
 
\subsection{Simulation set-up}
\label{sec:sim}

\begin{figure}
\centering
{\includegraphics[width = 0.49\textwidth]{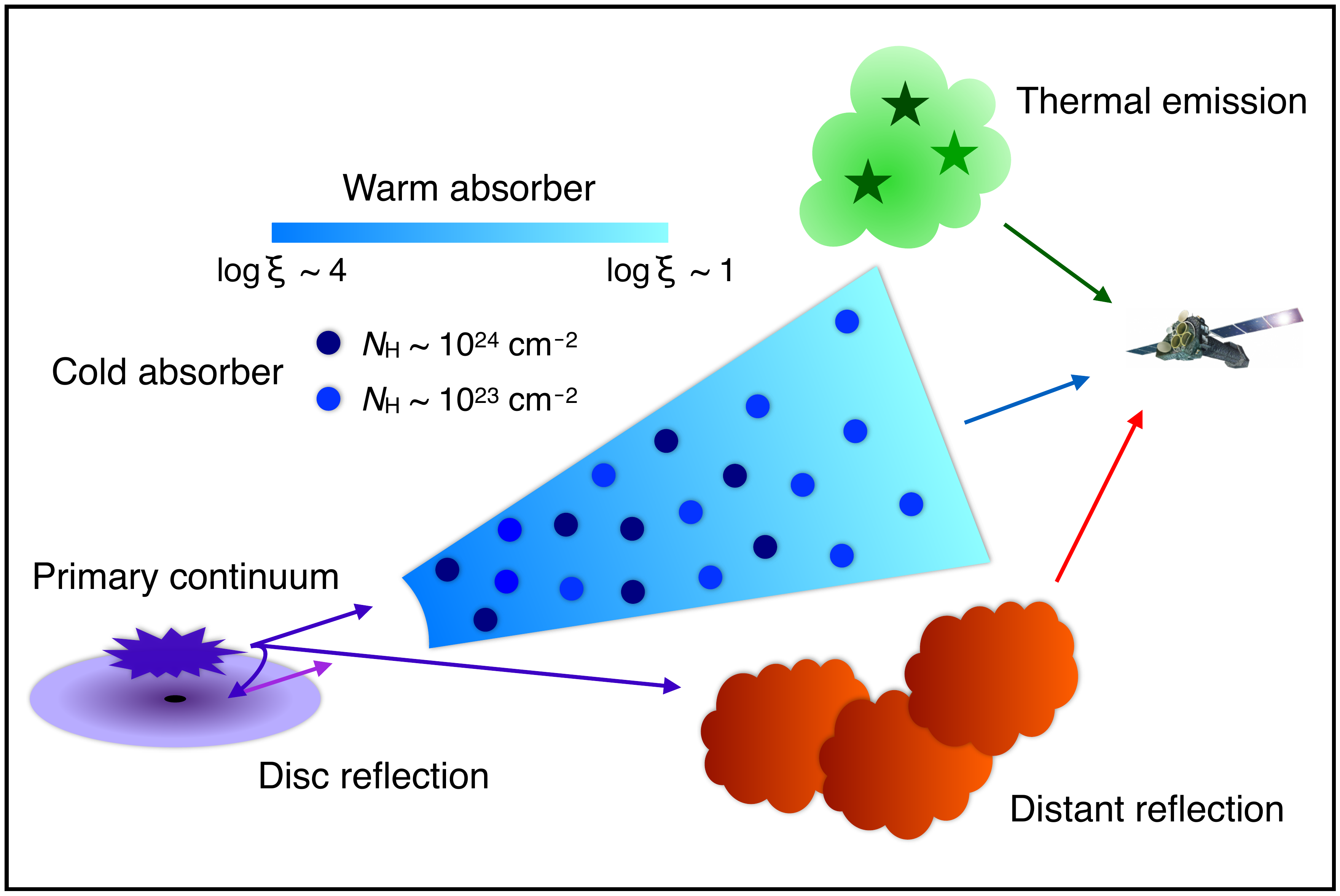}}
\caption{Schematic (not to scale) of the proposed configuration, presenting the various emission and absorption features that we consider in this work (See \S\,\ref{sec:sim} for details).}
\label{fig:configuration}
\end{figure}

As mentioned above, various emission/absorption components can be present in the X-ray spectrum of any AGN. However, depending on the state in which the source is caught, we may be able to observe all or only some of these components. Generally, the Ockham's razor argument is (or should be) applied during the spectral fitting, thereby avoiding the inclusion of unnecessary components. While this is the correct practice, we might miss a component that is actually present in the case of a single-epoch observation, either if the spectra do not cover a broad band or do not have enough S/N. We know from the literature the expected ranges of the parameters for the various components that are observed and are potentially present in any AGN spectrum. Hence, we can simulate the most general spectrum and then examine how well the model parameters are recovered using the common fitting techniques.


\begin{figure}
\centering
\includegraphics[width = 0.49\textwidth]{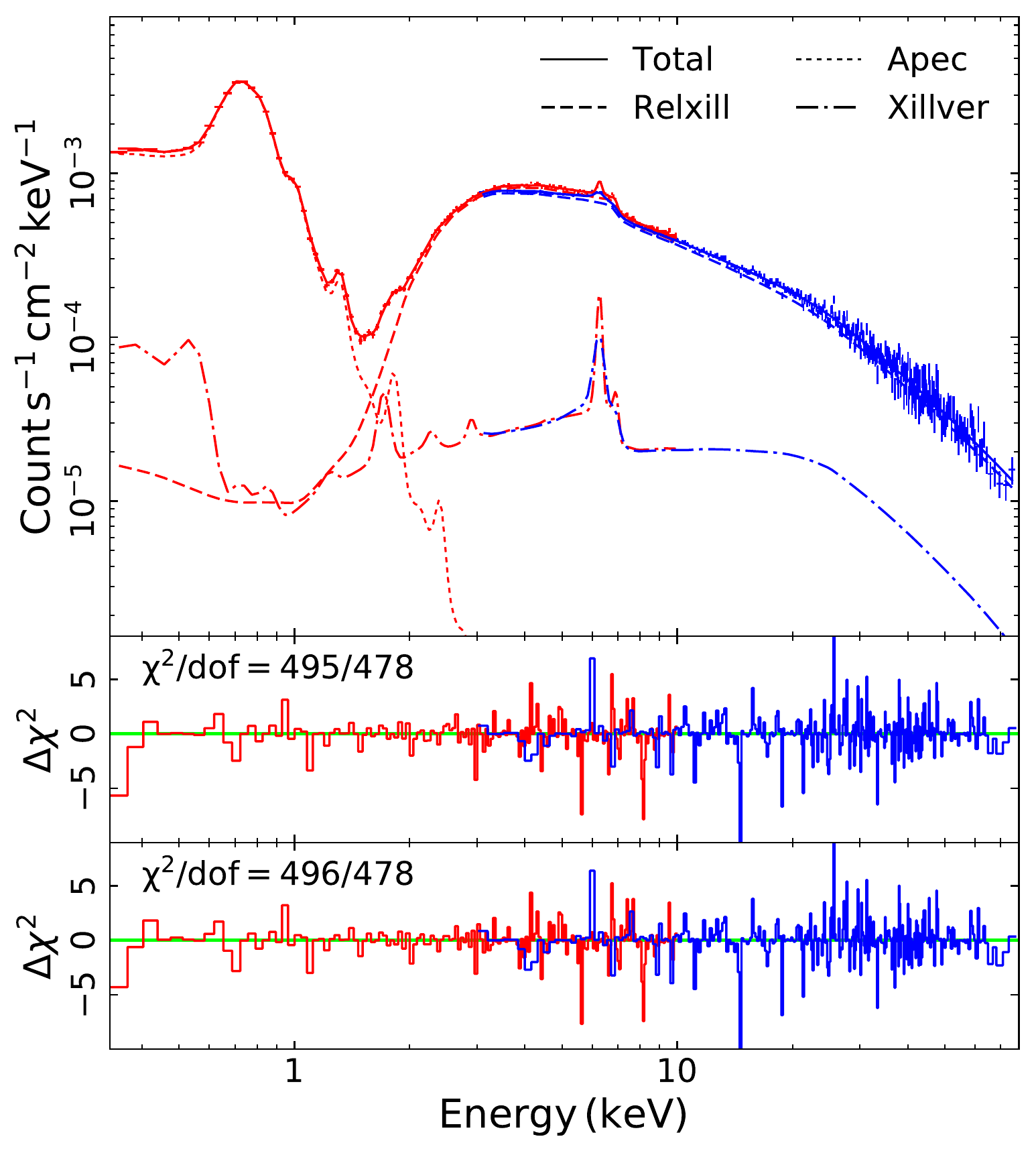} 

\caption{Top panel: Example of the simulated {\it XMM-Newton} (red) and {\it NuSTAR} (blue) spectra  (corresponding to simulation G8) together with the various components of the theoretical model assumed. The primary emission plus ionized reflection (dashed lines), neutral reflection (dash-dotted lines), and thermal emission (dotted lines) are shown. Middle and bottom panels: The $\chi^2$ residuals obtained by the two separate fits are indicated (see \S\,\ref{sec:fitting} for details).}
\label{fig:spectra}
\end{figure}

We simulated AGN spectra in the 0.3--79\,keV band via the XSPEC\,v12.9.0s\,\citep{Arnaud96} command {\tt FAKEIT} and the {\it XMM-Newton} EPIC-pn \citep{Struder01} and the {\it NuSTAR} response matrices in the  0.3--10\,keV (with an exposure time of 90\,ks)\footnote{This is approximately the maximum effective exposure per {\it XMM-Newton} orbit in Small Window mode (needed to avoid pile-up in bright sources).} and 3--79\,keV (with an exposure time of 100\,ks, i.e. 50\,ks per focal plane module) ranges, respectively. The spectra were binned not to oversample the FWHM resolution by a factor larger than 3 and 2.5 for {\it XMM-Newton} and {\it NuSTAR}, respectively. Then, we grouped the spectra, for both instruments, to ensure a minimum S/N of 5 in each energy channel. The simulations are intended to represent single-epoch observations of bright low-redshift AGN, similar to the observed sources, using {\it XMM-Newton} and {\it NuSTAR} simultaneously. Hence, we defined a generic parent model that contains the various expected emission and absorption components. The former are described below through their Xspec spectral counterparts:

\begin{itemize}
        \item {\tt APEC}: thermal diffuse emission at soft X-rays arising from the host galaxy in the cases when the star formation rate is enhanced and/or from gas photoionized by the AGN in the narrow line region \citep[see][for the reference case of NGC 1365, and references therein for other notable sources]{Nardini15}. The parameters of this model that were varied during the simulations are the temperature of the gas ($kT$) and its abundance in solar units \citep[from][]{Grevesse98}.
        
        \item {\tt RELXILLLP\,v0.4a}: primary emission plus blurred relativistic reflection from ionized material, assuming a lamp-post geometry for the  emitting source \citep{Dauser13, Garcia14}. The free parameters of this model are the photon index ($\Gamma$) of the incident continuum, the height of the lamp post ($h$, in units of $r_{\rm g}$), the spin parameter ($a^\ast$), and the inclination ($i$), ionization ($\xi_{\rm d}$), and iron abundance ($A_{\rm Fe}$ in solar units) of the accretion disc. We kept the high-energy cut-off fixed to 300\,keV. The reflection fraction is computed self-consistently within the model and fixed to the lamp-post value ({\tt fixReflFrac = 1}), as defined in \cite{Dauser16}. This also implies that the disc emissivity as a function of radius is fully determined by the height of the X-ray source.
        
        \item {\tt XILLVER}: neutral reflection arising from distant material by fixing the ionization parameter, inclination angle and iron abundance of the reflector to $\log \xi = 0$, $i =45^\circ$ and $A_{\rm Fe} = 1$, respectively. For simplicity, we tied the values of the photon index and the high-energy cut-off of this component to those of the primary continuum.
        
\end{itemize}
        
\noindent
As for the absorption components, we modelled the Galactic column using the {\tt PHABS} model and assuming $N_{\rm H} = 5\times 10^{20}\,\rm cm^{-2}$. We assumed that the intrinsic absorption can affect only the innermost emission components (primary continuum and relativistic reflection), and we considered the following configuration:

\begin{figure*}
\centering
\includegraphics[width = 0.64\textwidth]{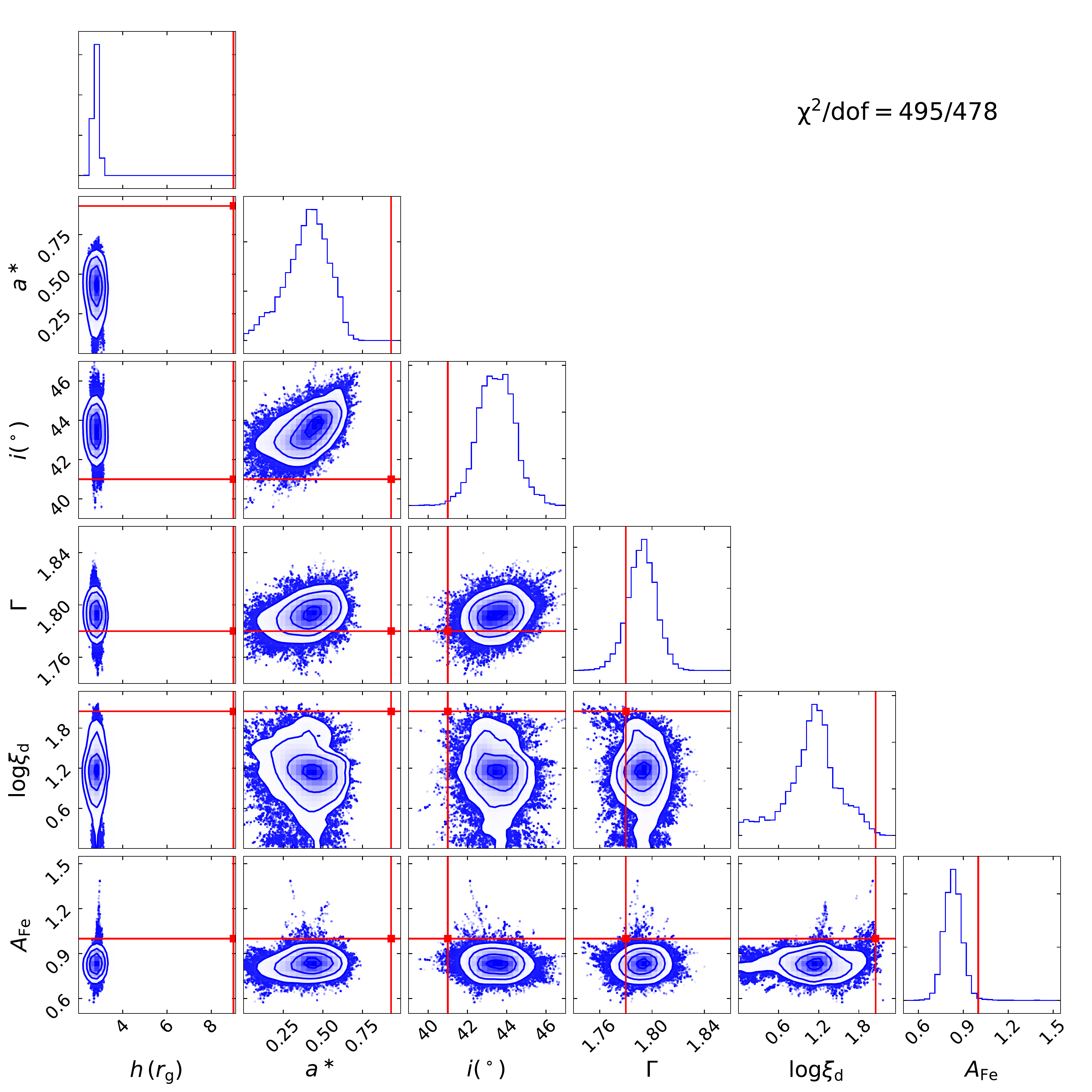} \\
\includegraphics[width = 0.64\textwidth]{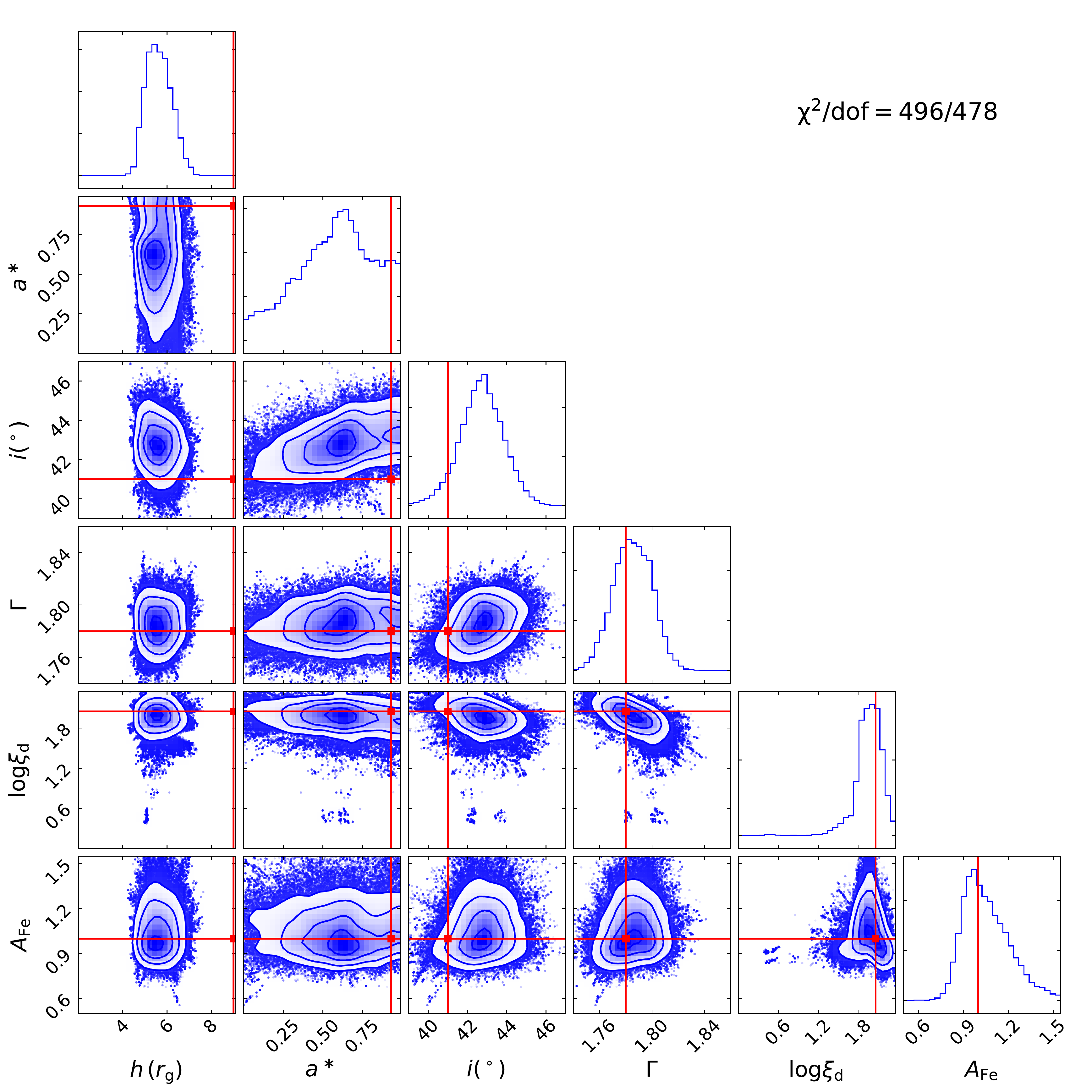} 

\caption{Results of the MCMC analysis for the relevant best-fit reflection parameters corresponding to the two different spectral fits shown in the middle and bottom panels of Fig.\,\ref{fig:spectra}. The red lines correspond to the input values assumed in order to create the simulations. We show the $\chi^2$ values obtained from the corresponding best fit, whose accuracy as a whole is excellent in both cases. The individual parameters, however, are not all correctly retrieved.}
\label{fig:contours}
\end{figure*}

\begin{itemize}

        \item {\tt WARMABS}: fully covering warm absorption modelled through an XSTAR table \citep{Kallman01} having an input continuum with a photon index of 2. Although it is usually seen in outflow \citep[e.g.][and references therein]{Braito14}, we assume for simplicity that this component is at rest in the local frame. The free parameters of this model are the column density $N_{\rm H,\, wa}$ and the ionization parameter of the absorber $\xi_{\rm wa}$. 
        
        \item {\tt ZPCFABS}: two layers of partially covering neutral absorbers that can represent the various neutral-absorption states (from Compton-thin to Compton-thick regimes). We let free to vary the column densities $N_{\rm H,\, 1/2}$ and the covering fractions $\rm CF_{1/2}$ of both absorbers.

\end{itemize}   

\noindent 
The final model can be written in XSPEC terminology, neglecting Galactic absorption, as follows:
\begin{align}
{\tt model = } & {\tt \,WARMABS \times ZPCFABS \times ZPCFABS \times RELXILLLP} \notag \\   
 & \tt + XILLVER + APEC. \notag
\end{align} 
\noindent
A scheme of the proposed configuration is shown in Fig.\,\ref{fig:configuration}. We report in Table\,\ref{table:paramInput} the input range chosen for each of the free parameters considered in the simulations. The redshift of the simulated source is fixed at 0.02. We kept the normalizations of the various emission components free, and the only limitation is that the observed flux is between 1 and 3\,mCrab in the 0.3--10\,keV range (resulting in $\sim 3\times 10^5 - 10^6$\,counts, for the {\it XMM-Newton} spectra). 

\begin{table}
\centering

\caption{Key parameters used to perform the simulations with the corresponding input range, for the various emission and absorption components. All the key  parameters were allowed to vary freely during the spectral fitting.}

\begin{tabular}{lc}
\hline\hline \\[-0.2cm]

Parameter       &       Input range     \\[0.2cm] \hline  \\[-0.2cm] 
        \multicolumn{2}{c}{Warm absorption      }       \\[0.2cm]   
$N_{\rm H,\, wa}\,({\rm cm^{-2}})$      &       $10^{18} -3\times10^{24} $       \\[0.2cm]   
$\log [\xi_{\rm wa}\,({\rm erg\,cm\,s^{-1}})]$  &       $0-5$   \\[0.2cm] \hline  \\[-0.2cm]
 
        \multicolumn{2}{c}{ Reflection}                 \\[0.2cm]    
$h\,(r_{\rm g})$        &       $2-100$ \\[0.2cm]   
$a^\ast$        &       $0-0.998$       \\[0.2cm]   
$i(^\circ)$     &       $3-89$  \\[0.2cm]   
$\Gamma$        &       $1.5-2.5$       \\[0.2cm]   
$\log [\xi_{\rm d}\,({\rm erg\,cm\,s^{-1}})]$   &       $0-4.7$ \\[0.2cm]   
$A_{\rm Fe}$\,(solar)   &       $0.5-10$        \\[0.2cm]  \hline  \\[-0.2cm]

        \multicolumn{2}{c}{PC neutral absorption}               \\[0.2cm]   
$N_{H,\,1}\, (10^{22}\,{\rm cm^{-2}})$  &       $0.01-20$       \\[0.2cm]  
CF$_1$  &       $0-1$   \\[0.2cm]  
$N_{H,\,2}\,( 10^{22}\,{\rm cm^{-2}})$  &       $0.01-500$      \\[0.2cm]  
CF$_2$  &       $0-1$   \\[0.2cm]  \hline  \\[-0.2cm] 

        \multicolumn{2}{c}{ Thermal emission}           \\[0.2cm]  
$kT$\,(keV)     &       $0.1-1.5$       \\[0.2cm]  
Abundance (solar)       &       $0-5$   \\[0.2cm]  \hline \hline

\end{tabular}
\label{table:paramInput}
\end{table}

We note that the configuration we adopted may have some caveats on physical grounds. For instance, we neglected Compton scattering out of and into the line of sight for partial coverers with column densities $N_{\rm H}> 10^{24}\,{\rm cm^{-2}}$. In fact, this would make the structure of the model much more complicated. On the one hand, accounting for the scattering into the line of sight would require arbitrary geometrical assumptions. On the other hand, the combination of partial covering and scattering out of the line of sight is not trivial in terms of model definition and handling. Even if these effects were treated properly, the actual physics of a real system would still be likely much more complex than the model we adopted. For example, the distant reflection is assumed to arise from a plane-parallel slab with an intermediate inclination of 45$^\circ$, but this is just a coarse representation of the expected geometry of the reflector \citep[see e.g.][]{Yaqoob12}. Moreover, this component might not be completely neutral; its ionization is low but not negligible, which leads to some heating near to the surface of the reflector \citep{Garcia13}. This mainly shifts the narrow iron K feature to higher energy. An additional complexity could be also due to the possible presence of multi-temperature thermal emission, a more complex structure of the warm absorption, or other forms of scattering into the line of sight. Finally, it should be kept in mind that the actual geometry of the X-ray corona is largely unknown. The point-like, lamp-post corona is a convenient approximation, but it also has some clear physical limitations, for instance a compactness problem similar to gamma-ray bursts \citep[e.g.][]{Fabian15, Dovciak16}. It is worth stressing, however, that none of our assumptions affect our results as long as the simulations and spectral fitting are performed in a self-consistent way.

\subsection{Fitting procedure}
\label{sec:fitting}

In order to reduce the observer-expectancy effect, each simulation was created by one member of the group and fitted blindly by the two other members separately. The various spectral components mentioned above were allowed to be present or absent in any simulation (and fit), except for the primary continuum plus ionized reflection component, which was always included by construction. We first simulated a general set of 15 simulations (5 simulations per person; hereafter Set\,G, we refer to these simulations as G1--G15). The simulated parameters were allowed to vary within the input range, while in the fits they were free to vary without any restriction, apart from neglecting negative spins and heights below 2\,$r_{\rm g}$. The Ockham's razor criterion was followed in the spectral analysis. A fit was considered as satisfactory at personal discretion, provided that 1) the overall statitistics was good, 2) the $\chi^2$ value represented a stable minimum, and 3) no obvious residuals were present. This does not ensure that the accepted fit is strictly the best possible. Indeed, in some cases only a local $\chi^2$ minimum is found, revealing how critical this kind of analysis can be in practice (see Section\,\ref{sec:discussion} for a more detailed discussion).

An example of the simulated data with the theoretical model and the corresponding residuals from the two different fits is shown Fig.\,\ref{fig:spectra}. Errors on the parameters are calculated from Markov chain Monte Carlo (MCMC)\footnote{We use the {\tt XSPEC\_EMCEE} implementation of the {\tt PYTHON EMCEE}  package for X-ray spectral fitting in XSPEC by Jeremy Sanders (\url{http://github.com/jeremysanders/xspec_emcee}).}, using the Goodman-Weare algorithm \citep{Goodman10} with a chain of $510,000$ elements (170 walkers and 3000 iterations), and discarding the first $51,000$ elements as part of the `burn-in' period. We show in Fig.\,\ref{fig:contours} the results of the MCMC analysis for the best-fit reflection parameters found for the simulated spectrum presented in Fig.\,\ref{fig:spectra}.

\section{Results}
\label{sec:results}

\begin{table*}
\centering

\caption{Summary of the success/failure in measuring the relevant parameters for the three simulation sets. Simulations performed assuming a spin parameter  $< 0.8$ are indicated with italics font, while the underline corresponds to a height of the lamp post that is $\leq \rm 5\,r_{\rm g}$. Symbols for the individual parameters are as follows: full success (\cm), fair success (\upd), undetermined (\qm), added component (+), missing component ($-$), failure (\xm), while blank corresponds to the cases when a given component is neither present in the simulated model nor in the fit. The qualitative classification of the fit as a whole is represented as follows: excellent fit (\gbul), good fit (\ybul), and inaccurate fit (\rbul). See text for details.}

\begin{adjustbox}{max width=0.9\textwidth}
\begin{tabular}{lccccccccccccccc}
\hline\hline\\[-0.3cm]
Parameter       &       {\it G1}        &       {\it \underline{G2}}    &       \underline{G3}  &       \underline{G4}  &       {\it G5}     &       {\it G6}        &       {\it G7}        &       G8      &       \underline{G9}  &       {\it G10}    &       {\it G11}       &       {\it\underline{G12}}    &       \underline{G13} &       {\it G14}    &       \underline{G15} \\[0.1cm]      \hline \\[-0.3cm]
$a^\ast$        &       \xm\upd &       \upd\upd        &       \cm\upd &       \upd\upd        &       \upd\qm &       \qm\qm  &       \cm\upd &       \xm\qm  &       \upd\upd        &       \qm\cm  &       \xm\xm  &       \xm\xm  &       \cm\cm  &       \qm\qm  &       \cm\cm  \\[0.1cm]   
$h$     &       \xm\xm  &       \cm\xm  &       \cm\cm  &       \cm\cm  &       \xm\xm  &       \cm\cm  &       \xm\xm  &       \xm\xm  &       \cm\xm  &       \xm\xm  &       \xm\xm  &       \xm\xm  &       \cm\upd &       \xm\xm  &       \upd\cm \\[0.1cm]   
$i$     &       \upd\upd        &       \cm\xm  &       \cm\cm  &       \cm\xm  &       \cm\upd &       \upd\upd        &       \cm\cm  &       \upd\cm &       \xm\cm  &       \xm\xm  &       \cm\cm  &       \cm\cm  &       \cm\cm  &       \xm\cm  &       \cm\cm  \\[0.1cm]   
$\Gamma$        &       \cm\upd &       \cm\cm  &       \cm\cm  &       \upd\upd        &       \upd\upd        &       \cm\cm  &       \upd\cm &       \cm\cm  &       \cm\cm  &       \upd\cm &       \cm\cm  &       \cm\upd &       \cm\cm  &       \upd\cm &       \upd\cm \\[0.1cm]   
$\xi_{\rm d}$   &       \cm\xm  &       \xm\xm  &       \cm\cm  &       \cm\cm  &       \cm\xm  &       \xm\xm  &       \cm\cm  &       \xm\cm  &       \xm\cm  &       \xm\cm  &       \xm\cm  &       \xm\cm  &       \upd\upd        &       \xm\cm  &       \cm\cm  \\[0.1cm]   
$A_{\rm Fe}$    &       \cm\xm  &       \xm\cm  &       \cm\cm  &       \cm\cm  &       \cm\xm  &       \cm\cm  &       \xm\cm  &       \xm\cm  &       \xm\cm  &       \xm\xm  &       \cm\cm  &       \xm\cm  &       \cm\xm  &       \xm\cm  &       \cm\cm  \\[0.1cm]   
$N_{\rm H,\, wa}$       &       \cm\upd &       \xm\xm  &       \cm\cm  &       \upd\xm &       \cm\cm  &       \cm\cm  &       \cm\cm  &       \upd\cm &       \cm\cm  &       \xm\cm  &       \cm\cm  &       \cm\cm  &       \cm\cm  &       \cm\cm  &       \cm\cm  \\[0.1cm]   
$\xi_{\rm wa}$  &       \xm\cm  &       \cm\cm  &       \cm\cm  &       \cm\cm  &       \cm\cm  &       \cm\cm  &       \cm\xm  &       \cm\cm  &       \cm\cm  &       \cm\cm  &       \cm\cm  &       \upd\cm &       \cm\cm  &       \cm\cm  &       \xm\xm  \\[0.1cm]   
$N_{H,\,1}$     &       $-\,$\xm        &       \xm\xm  &       \xm\xm  &       \cm\xm  &       \hspace{8pt}+   &       \cm\cm  &       \cm\upd &               &       \xm\cm  &       \cm\cm  &       \cm\cm  &       $-\,$\cm        &       \cm\cm  &       \cm\cm  &       \cm\cm  \\[0.1cm]   
CF$_1$  &       $-\,$\xm        &       \cm\cm  &       \cm\cm  &       \xm\xm  &       \hspace{8pt}+   &       \cm\cm  &       \cm\cm  &               &       \cm\cm  &       \cm\cm  &       \cm\cm  &       $-\,$\cm        &       \cm\cm  &       \cm\cm  &       \cm\cm  \\[0.1cm]   
$N_{H,\,2}$     &       \cm\upd &       \hspace{8pt}+   &       \cm\cm  &       \hspace{8pt}+   &       \hspace{8pt}+   &       \cm\cm  &       \cm\cm  &       \xm\cm  &       \cm\cm  &       $-\,$\upd       &       \cm\cm  &       \cm\cm  &       \cm\cm  &       \cm\cm  &       \cm\cm  \\[0.1cm]   
CF$_2$  &       \cm\upd &       \hspace{8pt}+   &       \cm\cm  &       \hspace{8pt}+   &       \hspace{8pt}+   &       \cm\cm  &       \cm\cm  &       \cm\cm  &       \cm\cm  &       $-\,$\cm        &       \cm\cm  &       \upd\cm &       \cm\cm  &       \cm\cm  &       \cm\cm  \\[0.1cm]   
$kT$    &       \cm\upd &       \cm\cm  &       \cm\cm  &       $+\,+$  &       \cm\cm  &       \cm\cm  &       \hspace{8pt}+   &       \cm\cm  &       \cm\cm  &       \cm\cm  &       \cm\cm  &       \cm\cm  &       \cm\cm  &       \cm\cm  &       \cm\cm  \\[0.1cm]   
$N_{\rm xillver}$       &       \cm\xm  &       \cm\cm  &       \upd\upd        &       \cm\cm  &       \cm\cm  &       \cm\cm  &       \cm\cm  &       \cm\cm  &       \cm\cm  &       \upd\cm &       \cm\cm  &       \cm\cm  &       \cm\cm  &       \cm\cm  &       \cm\cm  \\[0.1cm]   \hline 
        &               &               &               &               &               &               &               &               &               &               &               &               &               &               &               \\[-0.3cm]
Fit     &       \ybul\rbul      &       \gbul\rbul      &       \gbul\gbul      &       \rbul\rbul      &       \gbul\rbul      &       \gbul\gbul      &       \gbul\rbul      &       \gbul\gbul      &       \rbul\gbul      &       \rbul\ybul      &       \gbul\gbul      &       \ybul\gbul      &       \gbul\gbul      &       \ybul\gbul      &       \gbul\gbul
\end{tabular}
\end{adjustbox}
\begin{adjustbox}{max width = 0.9\textwidth}
\centering

\begin{tabular}{lccccccccc|cccccc}
\hline\hline
&               &               &               &               &               &               &               &               &               &               &               &               &               &               &                               \\[-0.3cm]
Parameter       &       K1      &       \underline{K2}  &       K3      &       K4      &       \underline{K5}  &       \underline{K6}  &       \underline{K7}  &       \underline{K8}  &       K9      &       B1      &       {\it \underline{B2}} &       {\it  B3}       &       \underline{B4}  &       B5      &       \underline{B6}  \\[0.1cm]  \hline
        &               &               &               &               &               &               &               &               &               &               &               &               &               &               &               \\[-0.3cm]
$a^\ast$        &       \xm\xm  &       \xm\cm  &       \xm\xm  &       \xm\xm  &       \upd\upd        &       \upd\xm &       \upd\upd        &       \cm\cm  &       \xm\xm  &       \xm\xm  &       \cm\upd &       \xm\xm  &       \upd\upd        &       \xm\xm  &       \upd\cm \\[0.1cm]   
$h$     &       \xm\xm  &       \xm\xm  &       \xm\xm  &       \xm\xm  &       \cm\xm  &       \cm\cm  &       \upd\cm &       \cm\cm  &       \cm\cm  &       \cm\cm  &       \cm\xm  &       \xm\xm  &       \xm\cm  &       \xm\xm  &       \upd\upd        \\[0.1cm]   
$i$     &       \cm\xm  &       \cm\cm  &       \cm\cm  &       \xm\xm  &       \upd\cm &       \cm\xm  &       \upd\upd        &       \cm\cm  &       \cm\cm  &       \upd\upd        &       \cm\cm  &       \cm\cm  &       \cm\cm  &       \cm\cm  &       \cm\cm  \\[0.1cm]   
$\Gamma$        &       \cm\upd &       \upd\cm &       \cm\cm  &       \upd\upd        &       \upd\upd        &       \upd\upd        &       \cm\cm  &       \upd\upd        &       \cm\cm  &       \cm\cm  &       \cm\cm  &       \cm\cm  &       \cm\cm  &       \cm\cm  &       \cm\cm  \\[0.1cm]   
$\xi_{\rm d}$   &       \cm\xm  &       \cm\xm  &       \cm\cm  &       \xm\xm  &       \xm\upd &       \cm\xm  &       \xm\xm  &       \cm\cm  &       \xm\xm  &       \upd\xm &       \cm\cm  &       \cm\cm  &       \cm\cm  &       \cm\cm  &       \cm\cm  \\[0.1cm]   
$A_{\rm Fe}$    &       \cm\xm  &       \xm\xm  &       \cm\cm  &       \cm\xm  &       \xm\cm  &       \upd\xm &       \cm\xm  &       \xm\xm  &       \xm\xm  &       \cm\cm  &       \xm\xm  &       \cm\cm  &       \cm\cm  &       \cm\cm  &       \cm\cm  \\[0.1cm]   
$N_{\rm H,\, wa}$       &               &       \xm\xm  &       \cm\cm  &       \xm\xm  &       \xm\upd &       \cm\cm  &       \cm\xm  &       \cm\cm  &       \xm\xm  &               &               &               &               &               &               \\[0.1cm]   
$\xi_{\rm wa}$  &               &       \xm\cm  &       \cm\cm  &       \upd\cm &       \xm\cm  &       \xm\cm  &       \cm\cm  &       \cm\cm  &       \cm\xm  &               &               &               &               &               &               \\[0.1cm]   
$N_{H,\,1}$     &       \cm\cm  &       \xm\xm  &       \cm\cm  &       \xm\xm  &       \xm\cm  &       \hspace{8pt}+   &       \cm\cm  &       \cm\cm  &       $- \,-$    &               &               &               &               &               &               \\[0.1cm]   
CF$_1$  &       \cm\xm  &       \xm\cm  &       \xm\xm  &       \xm\xm  &       \cm\upd &       \hspace{8pt}+   &       \cm\cm  &       \cm\cm  &       $-\, -$      &               &               &               &               &               &               \\[0.1cm]   
$N_{H,\,2}$     &       \hspace{8pt}+   &       \upd\upd        &       \upd\upd        &       \upd\upd        &       \xm\upd &       \cm\upd &       \upd\upd        &               &       \cm\cm  &               &               &               &               &               &               \\[0.1cm]   
CF$_2$  &       \hspace{8pt}+   &       \upd\upd        &       \upd\upd        &       \upd\cm &       \xm\upd &       \cm\cm  &       \upd\upd        &               &       \upd\upd        &               &               &               &               &               &               \\[0.1cm]   
$kT$    &       \cm\xm  &       \xm\cm  &       \cm\cm  &       $-\,$\cm        &       \upd\cm &       \cm\xm  &       \cm\cm  &       \cm\cm  &       \cm\cm  &       $+\, +$      &       \cm\cm  &       \cm\cm  &       \cm\cm  &       \cm\cm  &       \cm\cm  \\[0.1cm]   
$N_{\rm xillver}$       &       \cm\xm  &       \cm\cm  &       \xm\xm  &       \cm\xm  &       $+\,+$  &       \cm\cm  &       \cm\cm  &       \cm\cm  &       \cm\cm  &       \cm\cm  &       \cm\cm  &       \cm\cm  &       $-\,$\xm        &       \upd\upd        &       \cm\cm  \\[0.1cm]   \hline
        &               &               &               &               &               &               &               &               &               &               &               &               &               &               &               \\[-0.3cm]
Fit     &       \gbul\rbul      &       \rbul\gbul      &       \gbul\gbul      &       \rbul\rbul      &       \rbul\rbul      &       \ybul\rbul      &       \gbul\gbul      &       \gbul\gbul      &       \ybul\ybul      &       \rbul\rbul      &       \gbul\gbul      &       \gbul\ybul      &       \rbul\gbul      &       \gbul\gbul      &       \gbul\gbul      \\[0.1cm]   \hline \\[-0.3cm]
\end{tabular}

\end{adjustbox}
\label{table:summary}
\end{table*}

We present in Table\,\ref{table:summary} a qualitative summary of the results we obtained from the blind spectral analysis, based on the classification criteria defined below. While these criteria are to a certain extent (but unavoidably) arbitrary, none of the conclusions of the paper are substantially modified.

\noindent
$\bullet$ {\bf Individual parameters:} For all the parameters, except for the spin and, to a lesser extent, iron abundance, the measurements are generally very well constrained. Thus, we define both a \textit{full} and a \textit{fair} success criterion. The former (denoted by the \cm\,sign) is met when a measurement is consistent with the input value within a confidence level of 90\%, while the latter (denoted by the \upd\,sign) is met when the fitted and input values are formally inconsistent, but agree with each other within a 10\% uncertainty\footnote{We note that we considered the ratios $\xi_{\rm fit}/\xi_{\rm input}$ for the ionization parameters of the accretion disc ($\rm \xi_d$)
and of the warm absorber ($\rm \xi_{wa}$) rather than the ratios of the logarithms.
}. All the other cases are classified as \textit{failures} (denoted by the \xm\,sign).

\noindent
$\bullet$ {\bf Spin classification:} Since the measure of the spin is the main aim of our study and the spin is the parameter that shows the most complex behaviour in the fits, we adopted a different approach to classify the goodness of our constraints on the spin parameter. We divided the 0--0.998 spin range into three bands: low spin ($a^\ast \in [0,0.4[$),  intermediate spin ($a^\ast \in [0.4,0.8[$), and high spin  ($a^\ast \in [0.8,0.998]$). Hence, we classified the measurements based on the following criteria: (a) full success if the measured value is consistent with the input one within the 90\% confidence level and the uncertainty range is within a single spin band; (b) fair success if either the measured value is consistent with the input value within the 90\% confidence level but the uncertainty range covers two spin bands, or the measured value is not consistent with the input value but the uncertainty range is within the same single band that contains the input value; (c) {\it undetermined} (denoted by the \qm\,sign) if the measured value is consistent with the input one but the uncertainty range covers three spin bands; and (d) {\it failure} for the other cases. 

\noindent
$\bullet$ {\bf Fit accuracy:} Irrespective of the values of the individual parameters and their degree of adherence to the input values, we also defined the following quality criteria for the accuracy of the whole fit: a) {\it excellent} if the adopted model is correct and the fit statistic is within a $\Delta \chi^2$ of $2.3$ from the putative absolute minimum (see below); b) {\it good} if either the model is correct and the distance from the absolute minimum is $\Delta \chi^2 < 9.2$, or the model misses a component that turns out to be significant at less than 99\%; c) {\it inaccurate} in all the other cases, including overfitting. The absolute minimum is evaluated as $\rm min\{\chi^2_0, \chi^2_a, \chi^2_b\}$, where $\chi^2_0$ is obtained by applying a posteriori the input model and $\chi^2_{\rm a,b}$ are the results from the blind spectral analysis (provided that the correct model is used). We found that 18 out of 30 fits were excellent, 4 out of 30 were good, and 8 out of 30 were inaccurate. We note that the application of the correct model (i.e. corresponding to the input one), as becomes evident below, does not imply that the input parameters are individually recovered with success (see Fig.\,\ref{fig:contours}). We stress again, however, that even inaccurate fits are fully acceptable on statistical grounds and meet the three conditions listed in \S\,\ref{sec:fitting}. 

\begin{figure*}
\centering
{\includegraphics[width = \textwidth]{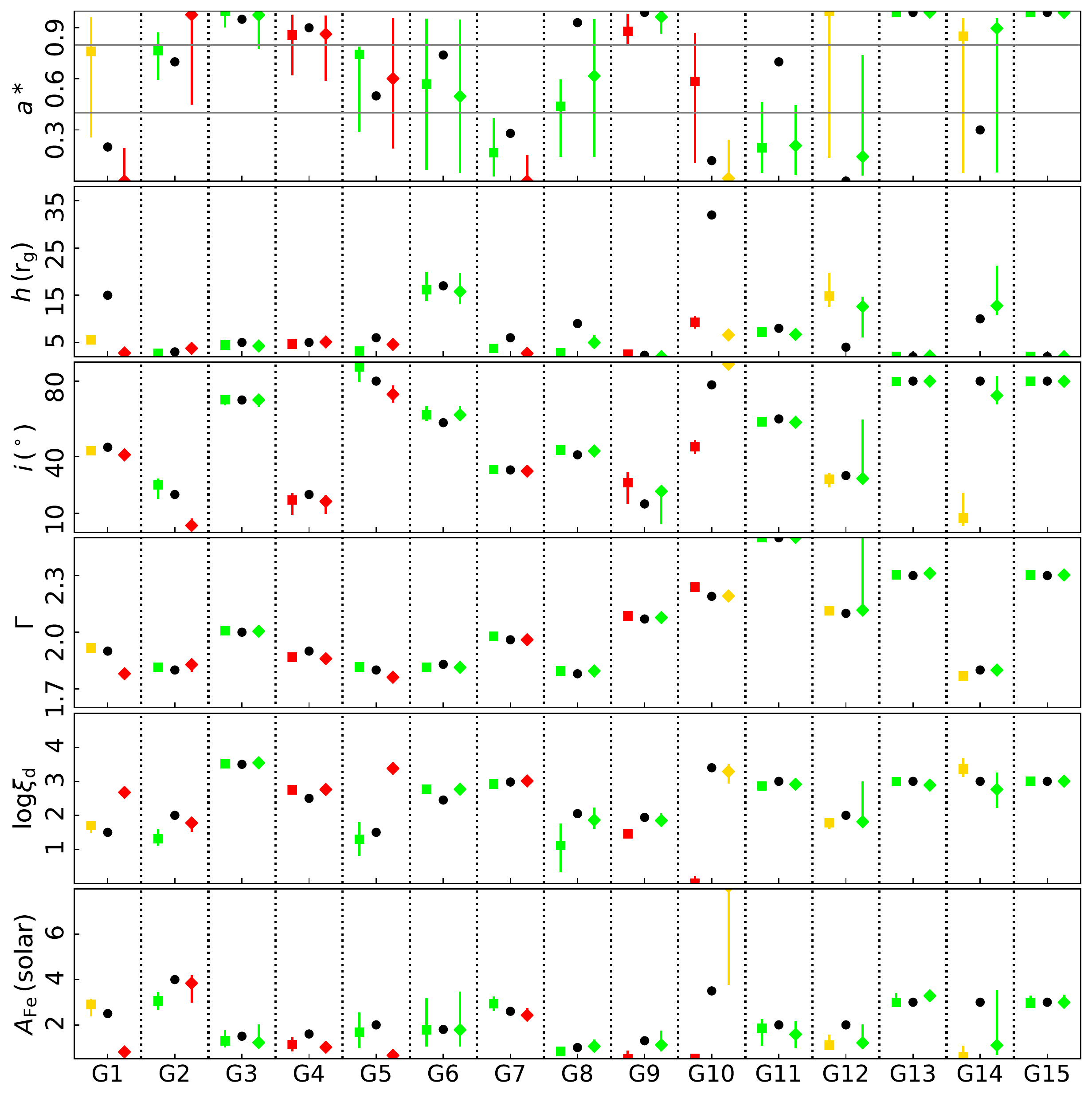}}
\caption{Input values (black dots) of the reflection parameters assumed for creating the various simulations (for Set\,G). The best-fit values obtained for the various parameters are shown as squares and diamonds for the two different realizations. The colour code refers to the quality of the fit as a whole: green for excellent, yellow for good, and red for inaccurate (see text for details). The error bars represent the  90\,\% confidence levels obtained from the MCMC analysis.}
\label{fig:refpar}
\end{figure*}

We summarize below the constraints that we obtained on the relevant parameters that are always present in the model by construction, distinguishing between accurate (either good or excellent) and inaccurate fits as follows:

\begin{itemize}
        \item The measure of the {\it spin parameter} was a full/fair success in 7+4 cases, 
        while it was undetermined/failed in 5+6 cases out of the 22 accurate fits. 
        In the 8 inaccurate fits, 6 spins were fairly retrieved and 2 were undetermined. Low and 
        intermediate spins (i.e. lower than 0.8) were present in 9 out of 15 simulations (i.e. 18 
        spectral fits, 13 of which were accurate).  A low spin value was determined correctly and 
        well constrained in only 1 out of 18 cases. The measure of the spin was fairly successful 
        in 5 cases and it was undetermined in 6. However, for the 6 high-spin simulations (9/12 
        accurate fits), the measurements were successful in 
        10 cases (5 fully, 5 fairly), undetermined once, and the only failure was in 
        a fit classified as excellent. In summary, the two different fits corresponding to 
        the same simulated spectrum might result in different best-fit values of the spin parameter if one 
        of the fits hits a secondary minimum or makes use of a {\it wrong} model, 
        thus classified as inaccurate. However, even fully successful fits are not always able to recover the correct value of the spin, with a clear preference for high values.
        
        \item The {\it height of the lamp post}, which 
        is the other key parameter that determines the strength of the reflection component
        in the observed spectra, was measured with success in 7 (full) plus 2 (fair) cases, 
        while it failed in the remaining 13 of the 22 accurate fits. However, 3 more full 
        successes were found among the 8 other fits. The role of the source height is further 
        discussed later on.
        
        \item The {\it disc inclination} was measured with success in 16 plus 4 of the accurate fits (2 failures) 
        and 2 plus 2 of the inaccurate fits (4 failures).
        
        \item The {\it photon index} was measured successfully in all cases, of which 20/30 were 
        a full success. We found a maximum difference between the measured and input photon index 
        of $\Delta \Gamma = 0.12$.
        
        \item The {disc ionization parameter} was retrieved with success in 13 plus 2 of the accurate fits 
        (with 7 failures) and 3 of the inaccurate fits (5 failures).
        
        \item The measure of {\it iron abundance} was fully successful in 15/22 accurate fits and in 4/8 inaccurate fits, 
        and unsuccessful in all the other cases.
\end{itemize} 

\noindent
As already noted, the failure in measuring the single parameters might also occur in cases in which the fit is highly accurate for both analysts (e.g. G8, G11), which is a possible indication of complex degeneracies. 

\begin{figure*}
\centering

{\includegraphics[width = 0.99\textwidth]{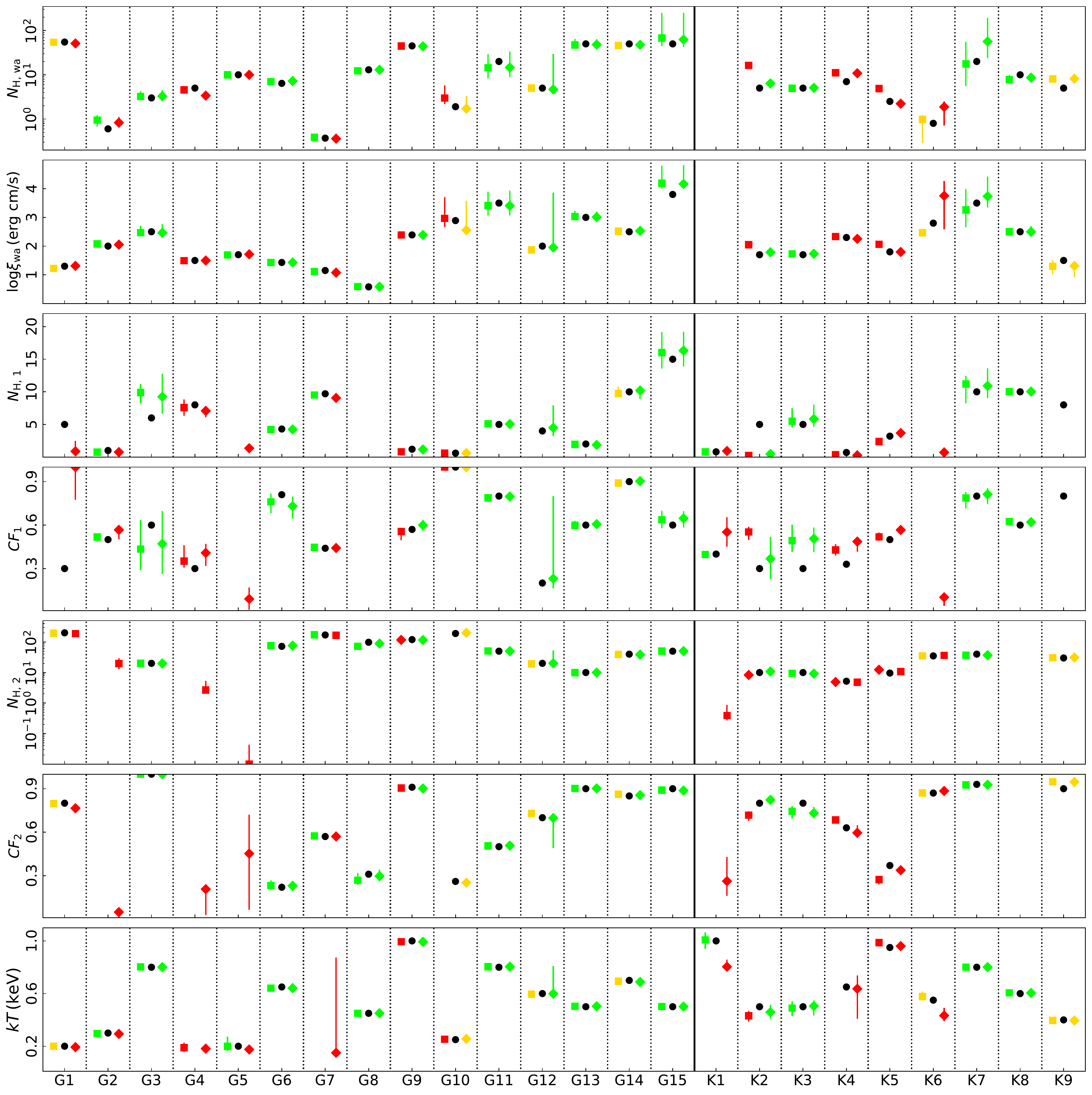}}

\caption{Similar to Fig.\,\ref{fig:refpar} but for the absorption parameters (top panels)
and the gas temperature of thermal emission component (last panel), presenting the results for 
Sets\,G and K. The column densities are in units of $10^{22}\,\rm cm^{-2}$.}
\label{fig:AbsApecpar}
\end{figure*}

The various simulated spectra for Set G together with their corresponding residuals are presented in Fig.\,\ref{fig:appendix-spectra}. The best-fit results relative to the reflection components are presented in Fig.\,\ref{fig:refpar}, while those for the absorption and thermal components are presented in Fig.\,\ref{fig:AbsApecpar}. Table\,\ref{table:App} shows the input and best-fit values of the lamp-post height and spin parameters. We report the best-fit $\chi^2/{\rm dof}$ values in the same table.

\section{Discussion}
\label{sec:discussion}

We simulated in this work high-quality single-epoch spectra of AGN at low redshift in the 0.3--79\,keV band using the responses of both {\it XMM-Newton} and {\it NuSTAR}. We assumed a general spectrum that includes, in addition to the primary emission, both ionized and neutral reflection, thermal emission, a warm absorber, and two layers of neutral partially covering absorbers. While in most cases the blind fitting procedure should be considered as successful, this fails in retrieving all of the individual input parameters. Below we examine the possible causes.

\subsection{The Kerr BH case}
\label{sec:highspin}

\begin{figure*}
\centering
{\includegraphics[width = \textwidth]{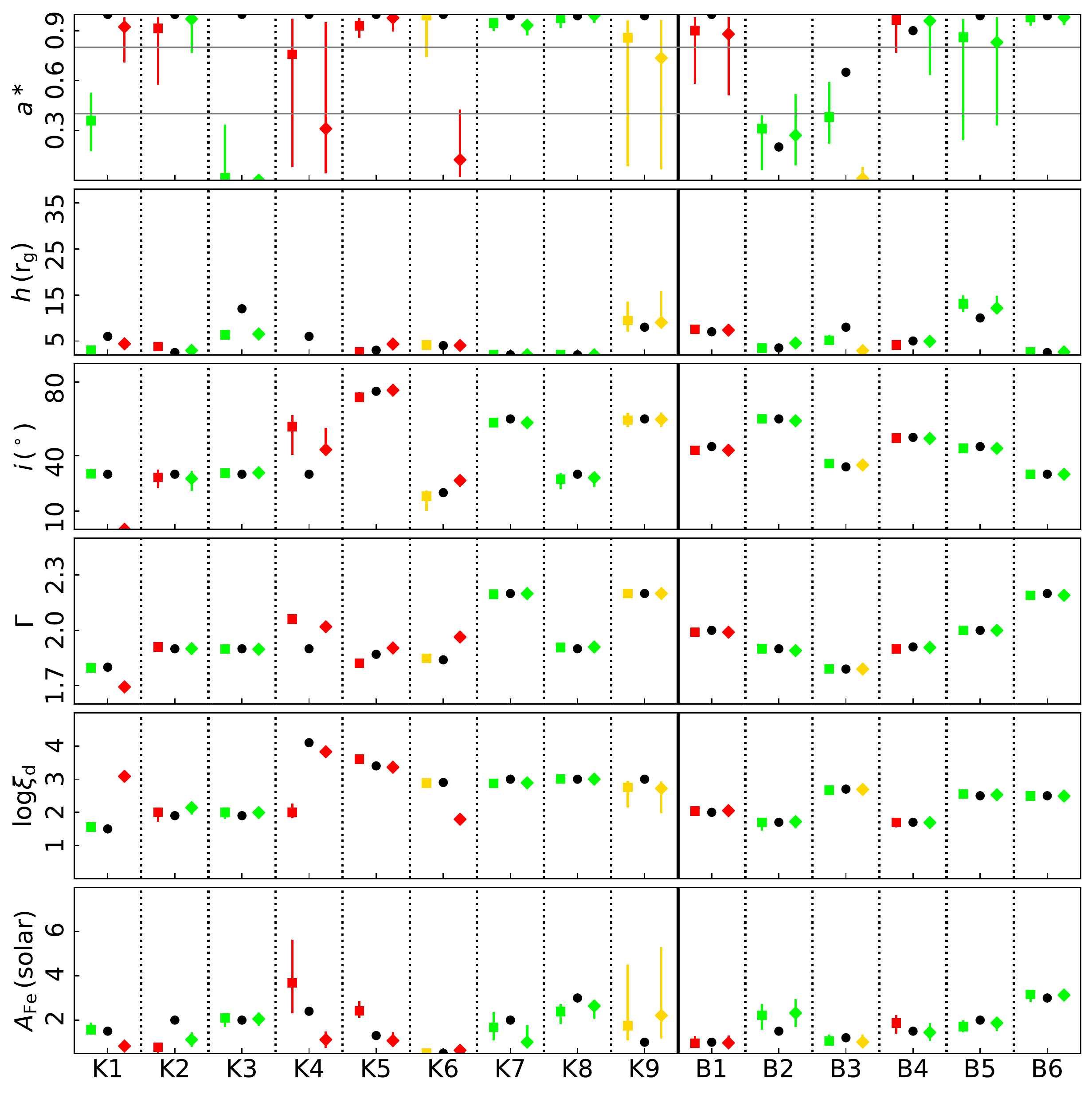}}
\caption{Similar to Fig.\,\ref{fig:refpar} but for Sets K and B.}
\label{fig:refparKB}
\end{figure*}

As noted in \S\,\ref{sec:results}, spectral analysis tends to recover high input spins better than low/intermediate spins: 10 out of 12 high-spin measurements were at least fairly successful, while only 7 out of 18 low/intermediate spins were reasonably retrieved. Furthermore, the measured spin distribution, as reported in the literature, shows a clear tendency towards high spins \citep[e.g.][]{Walton13, Reynolds14, Vasudevan16}. This evidence was the main motivation for us to simulate a set of high-spin spectra (hereafter Set\,K). Thus, we generated a set of 9 simulations (3 simulations per person) by fixing the spin parameter to its maximum allowed value; we refer to these simulations as K1--K9. However, the spin parameter was free to vary within the 0--0.998 range during the spectral fitting. The constraints on the best-fit parameters of the reflection and absorption components are presented in Table\,\ref{table:summary}, and plotted together with the corresponding input values in Figs.\,\ref{fig:refparKB} and \ref{fig:AbsApecpar}, respectively. The spin was retrieved successfully in 6 (3 fully and 3 fairly) of the 11 accurate fits, with 5 failures, while the 7 inaccurate fits returned 2 fairly constrained spins (with 5 failures). In total, only 18 out of 30 high-spin cases (in sets G and K) were a success: 13 (8 plus 5) are obtained in the 20 accurate fits, while other 5 fair successes still emerge from the 10 inaccurate fits. This suggests that, even though it plays an important role, the spin  is not the only factor that may lead to a positive result in recovering its input value.

\subsection{Effects of absorption}
\label{sec:absorption}
We summarize below the constraints we obtained on the absorption components in Sets G and K. We note that these components can be added/removed arbitrarily. 1) The fully covering warm absorber is included in 23 simulations (equivalent to 46 fits). Its column density and ionization parameter were positively recovered in 34 (30 full plus 4 fair successes) and 38 (36 plus 2) cases, respectively. Both rates are higher than the incidence of accurate fits (32/46). 2) The partial-covering low-column absorber is present in 21 simulations. Its column density and covering fraction were measured successfully in 25 plus 1 and in 28 plus 1 cases, respectively, against 25 out of 38 accurate fits (in 4 cases this component is missed in the spectral analysis). 3) The partial-covering high-column absorber is present in 19 simulations. Its column density and covering fraction were measured successfully in 23 plus 12 and 24 plus 12 cases, respectively, when the accurate fits are 24 out of 37 (this component is missed only once).
        
Summarizing, the properties of the absorbers are correctly estimated in the majority of the blind fits. Even though the number of simulations that we performed is statistically small, we can still derive a general idea of the degeneracy that may be present between the reflection-based models and the complex absorption model. It may happen that the inclusion of an absorber that is not present intrinsically mimics some of the relativistic effects on the spectrum, thus resulting in a wrong measurement of the spin parameter. However, it seems that absorption plays only a marginal role in the ability of measuring spins, as the overall absorption configuration in the fits was correct in 37 out of 48 cases. These issues are further investigated in sections\,\ref{sec:modeldependence} and \ref{sec:PCabs}.

\subsection{The bare sources case}
\label{sec:baresource}

In order to completely remove the uncertainties associated with absorption effects, we performed an additional set of 6 simulations (two per person) of bare sources, without including any intrinsic absorption (hereafter Set B, we refer to these simulations as B1-B6). The input model of this set can be written in XSPEC terminology, neglecting Galactic absorption, as follows: {\tt model =  RELXILLLP + XILLVER + APEC}. The simulated spectra and the $\chi^2$ residuals are presented in Fig.\,\ref{fig:appendix-spectra}. The input parameters together with the best-fit values are plotted in Figure\,\ref{fig:refparKB}. We summarize in Table\,\ref{table:summary} the qualitative constraints on the parameters that we obtained for this set. Four of these simulations involved a high spin value, and two cases each correspond to a lamp-post height $\leq$ or $>$ $5\,r_{\rm g}$. All the successful spin measures (1 full and 3 fair, with 3 out of 4 accurate fits) occurred for a low height of the source. Interestingly, at low height even a small spin is correctly retrieved (1 full and 1 fair success in 2 out of 2 accurate fits). Conversely, when the height of the lamp post is larger than 5\,$r_{\rm g}$, the measure of the spin always fails, irrespective of its value and despite the good rate (4/6) of accurate fits. The other disc reflection parameters were all well constrained in all cases, except for the Fe abundance in a single instance. As for the thermal and distant reflection components, they were both correctly assessed in 10 out of 12 fits.

In total, 19 out of the 30 simulations (i.e. 38 out of 60 fits) were performed asssuming maximally rotating black holes. The spectral analysis resulted in 22 successes in the measure of the spin (9 full and 13 fair), with 1 undetermined and 15 failed cases, for 25/38 accurate fits. Remarkably, when we consider only the simulations performed with a low lamp-post height, the total count features all the 22 successes and only 2 failures, for the same fraction (16/24) of accurate fits. This strongly suggests that the height of the primary X-ray source is the most critical ingredient for an accurate measure of black hole spin.

\subsection{Effects of the lamp-post height}

In the light of these findings, we further explored the dependence of our results on the input lamp-post height. Half of the simulations were perfomed assuming an input lamp-post height lower or equal to 5\,$r_{\rm g}$. In general, these heights were measured successfully in 21 (16 plus 5) fits, with 5 full successes coming from the 9/30 inaccurate fits. Three of these 15 simulations were performed assuming a low input spin parameter. The fit was accurate in 5 out of 6 cases, returning 2 successes in the measure of the height and 4 (1 plus 3) in that of the spin.

By considering the second half of the simulations, with a lamp-post height larger than 5\,$r_{\rm g}$, we find that the height was correctly estimated in 6/30 fits only despite the 21 out of 30 accurate fits. In 5 out of 30 spectral fits we were able to recover the spin (2 full and 3 fair successes), while in 6 out of 30 cases the spin was undetermined. Eight of these 15 simulations were performed assuming a low spin. As already noted, the high spin and large height case gave 1 undetermined and 13 failures, despite the 9 out of 14 accurate fits. This apparently suggests that, at large heights, the value of the spin has little weight, and the chance of success in its measure is not only small but also random (i.e. depending on several other factors such as iron abundance). In this sense, the preference for low spins is most likely a bias, so any measure at large heights should be taken with caution \citep[see also][]{Fabian14}.

The full summary is given in Table\,\ref{table:spinvsheight}. 

\begin{table}
\caption{Summary of the constraints on determining the spin showing its dependence on the lamp-post height. The values/fractions between parentheses refer to the accurate fits only.}
\centering
\begin{adjustbox}{max width=0.98\linewidth}
\begin{tabular}{l|l}
\hline \hline
                                        &                               \\[-0.3cm]
        $a^\ast < 0.8$: 11 models                                       &       $a^\ast \geq 0.8$: 19 models                                            \\
                                                &                                                       \\
        $h \leq 5 \,r_{\rm g}$: 6 (5) fits                                      &       $h \leq 5 \,r_{\rm g}$: 24 (16) fits                                               \\
\tabitem        Full success:   1/6     (       1/5     )       &       \tabitem        Full success:        9/24    (       9/16    )       \\
\tabitem        Fair success:   3/6     (       2/5     )       &       \tabitem        Fair success:        13/24   (       7/16    )       \\
\tabitem        Undetermined:   0/6     (       0/5     )       &       \tabitem        Undetermined:   0/24    (       0/16    )       \\
\tabitem        Failure:        2/6     (       2/5     )       &       \tabitem        Failure:        2/24    (       0/16    )       \\
                                                &                                                       \\
        $h > 5 \,r_{\rm g}$: 16 (12) fits                                       &               $h > 5 \,r_{\rm g}$: 14 (9) fits                                   \\
\tabitem        Full success:   2/16    (       2/12    )       &       \tabitem        Full success:        0/14    (       0/9     )       \\
\tabitem        Fair success:   3/16    (       1/12    )       &       \tabitem        Fair success:        0/14    (       0/9     )       \\
\tabitem        Undetermined:   6/16    (       4/12    )       &       \tabitem        Undetermined:   1/14    (       1/9     )       \\
\tabitem        Failure:        5/16    (       5/12    )       &       \tabitem        Failure:        13/14   (       8/9     )       \\ \hline

\end{tabular}
\end{adjustbox}
\label{table:spinvsheight}
\end{table}

\subsection{Model dependence}
\label{sec:modeldependence}
By construction, the disc reflection component is always included in both the simulated models and the fitted models, while the presence of all the other components, either additive (distant reflection, thermal emission) or multiplicative (warm, cold absorption) is arbitrary. This allows us also to investigate the impact of model dependence on the ability to accurately recover the reflection parameters. Considering all the simulations from sets G, K, and B, soft X-ray thermal emission was present in 27 out of 30 cases, and was missed in only one out of the 54 relevant fits (K4a). Of the 3 out of 30 cases where it was not required, it was included in 5 out of 6 fits (G4a,b, G7b, B1a,b). Correctly accounting for this component or not seems to have little effect on the spin determination. Interestingly, its inclusion does not necessarily undermine the measure of the spin, but the accuracy is lower (compare the results of G7b versus G7a; Table\,\ref{table:summary}. We note that model G7 is a low-spin, moderate-height case). In total, this component was measured successfully in 50 out of 53 spectral fits.

While the soft thermal component is easier to distinguish from the smooth, blurred reflection, the contribution from the distant reflector can significantly modify the shape of the Compton hump above 10 keV. This is present in 29 out of 30 models. It is missed once, and added instead in both fits of the single model (K5) where it was not originally included. This is a maximum-spin, low-height case, which, as we have seen, should have a higher chance of success. Indeed, both fits meet the fair success criterion for the spin. However, we could argue that the inclusion of distant reflection prevents us from obtaining more stringent constraints. In total, the normalization of this component was measured succesfully in 51 out of 57 spectral fits.

Absorption is allowed in 24 models only (G and K sets). The fully covering, warm layer is present in 23 out of 24 simulations. Remarkably, it is never missed and never added. This suggests that the features imprinted on the spectrum from mildly (or even highly) ionized gas in the line of sight are relatively easy to identify, at least at the X-ray brightness level of the simulated spectra. Hence, we expect this component to have no significant effect on the measure of the spin. In reality, however, the different treatment of warm absorption might lead to incompatible spin measures for the same data set of the same source, as in NGC\,3783 (\cite{Brenneman11} versus \cite{Patrick11}). The micro-calorimeters on board {\it ATHENA} and, possibly, earlier X-ray missions such as Arcus \citep{Smith16} and XARM will conclusively remove this source of ambiguity.

The lower column partial-covering absorber is included in 21 out of 24 models, while the higher column partial-covering absorber in 19 out of 24. The former is missed 4 times and added in 2 out of 6 fits, the latter is missed once and added in 4 out of 10 fits. Without distinguishing between the relative column densities, the configuration of the partial-covering, cold absorber consists of a single layer in 6 out of 24 cases and of a double layer in 17 out of 24 cases. No cold layers are included in the remaining case (G5), but they are both used in G5b, leading to an undetermined spin measure (as opposed to a fair success for G5a, where the correct model is applied). Two layers instead of the single one required are adopted in 5 out of 12 fits, while one of the two layers is missed in 5 out of 34 fits. Surprisingly, the addition of a layer does not always preclude a decent (or even good) measure of the spin (e.g. G2b), although the statistical significance of the second layer in these cases is most likely marginal. Conversely, all the 5 cases in which a layer is missed correspond to failures or indetermination in the spin measurement. We conclude that the effects of partial covering and of relativistic reflection, when high-quality broadband spectra are available, can be generally well distinguished from each other. We further discuss this point in the next section. 

\subsection{Reflection versus partial covering absorption}
\label{sec:PCabs}
According to some interpretations \citep[e.g.][]{Miller08}, no relativistic signature is needed to explain the spectra (and variability) of most AGN. This is a natural consequence of the substantial statistical equivalence between absorption- and reflection-based models, especially when the spectra are complex and require some combination of both ingredients. The relative dispute was initially concentrated on the nature of the Fe K line broadening since the partial covering absorption could reproduce both the smooth, extended red wing of the putative relativistic line as a gentle continuum curvature and its blue horn as an absorption edge. The appearance of tentative hard X-ray excesses in the \textit{Suzaku} era added a further controversial element, which could be explained either as a Compton reflection hump \citep[e.g.][]{Walton13} or as a signature of Compton-thick absorption \citep[e.g.][]{Tatum13}, thus reinforcing the polarity between the two mainstream scenarios. The advent of \textit{NuSTAR}, providing high-quality spectra also above 10 keV, accurate background subtraction, and substantial overlap with the band covered by the other X-ray observatories, can greatly reduce this persistent ambiguity. 

While, in principle, the ambiguity works in both ways, in our simulations we did not assume any pure partial-covering configuration (i.e. with no disc reflection), so we cannot verify whether an absorption layer could be missed by overestimating the amount of reflection. This is not the scope of our work. In fact, in this context it is more interesting to explore the possibility for the simulated models to be adequately reproduced without considering any relativistic component. We therefore checked the consequences of replacing the \texttt{relxilllp} component in our fits with a simple power law, in which the cut-off  is fixed at 300 keV for consistency with the primary continuum in the parent model. This is equivalent to fixing the reflection fraction in \texttt{relxilllp} to zero. In order to compensate for the lack of disc reflection, we allow for a larger complexity in the absorption configuration. It turns out that only 1 of the 30 simulated spectra could be perfectly described also by a pure partial-covering model, as the relative fits would meet all of our acceptance criteria, namely good statistics and lack of residuals. This is G10 ($\chi^2/\rm dof = 397/394$), where three cold layers are required: $N_{\rm H,1} = 6 \times 10^{21}$ cm$^{-2}$ (CF$_1 = 1$), $N_{\rm H,2} = 5.5 \times 10^{23}$ cm$^{-2}$ (CF$_2 \sim 0.1$), and $N_{\rm H,3} = 3.6 \times 10^{24}$ cm$^{-2}$ (CF$_3 \sim 0.25$). The fully covering, thinner layer is perfectly matched to a component of the input model, whose second layer has $N_{\rm H} = 1.9 \times 10^{24}$ cm$^{-2}$ and CF = 0.26. Although G10 is a low-spin ($a^\ast = 0.12$), large-height ($h = 32\,r_{\rm g}$) case, a much more complex (and rather extreme) absorption pattern is required to compensate for the lack of disc reflection in the fit. 

In three other cases (G14, K4, and K9) the reduced $\chi^2$ is fair (1.02, 1.10, and 1.04, respectively), and two layers have similar (even if not strictly consistent) properties to the input components. There are, however, some clear residual structures that make the absorption-only models not satisfactory. A third partial-covering layer is not statistically required. At low S/N, these spectra (with maximal spin but $h > 5 \,r_{\rm g}$) could be easily misinterpreted. A peculiar case is that of G5, where no cold absorbers are included in the simulation. This spectrum can be well fitted by a power-law continuum ($\chi^2/\rm dof = 479/477$) that is subject to warm absorption only with the exact input parameters. Even if the disc reflection component in this case is very smooth and featureless, a model where this is correctly accounted for (G5a; $\chi^2/\rm dof = 449/473$) is still statistically preferred \citep[at the $>4\sigma$ level based on the corrected Akaike Information Criterion usually adopted for non-nested models;][]{Akaike74}. It is likely that the contribution from disc reflection would be missed in lower quality data. A handful ($\leq 5$) of other cases, in principle, could become acceptable after the inclusion of a third partial-covering layer if the spectra were fainter by at least an order of magnitude with respect to the simulated spectra, which could mask the residuals within the photon noise. We note, however, that such a complex absorption configuration (even if real) should be rejected as a form of data overfitting at low S/N. We conclude that in most cases (26 out of 30), including all the set~B of bare spectra, disc reflection cannot be missed or mimicked by absorption effects in high-quality broadband X-ray spectra. The problems with its identification arise when the reflected spectrum is extremely smooth, so this could become a non-negligible issue if the X-ray corona is radially extended and thus responsible for the Comptonization of the relativistic signatures from the inner disc \citep[see][]{Steiner17}.

\begin{figure}
\includegraphics[width = 0.48\textwidth]{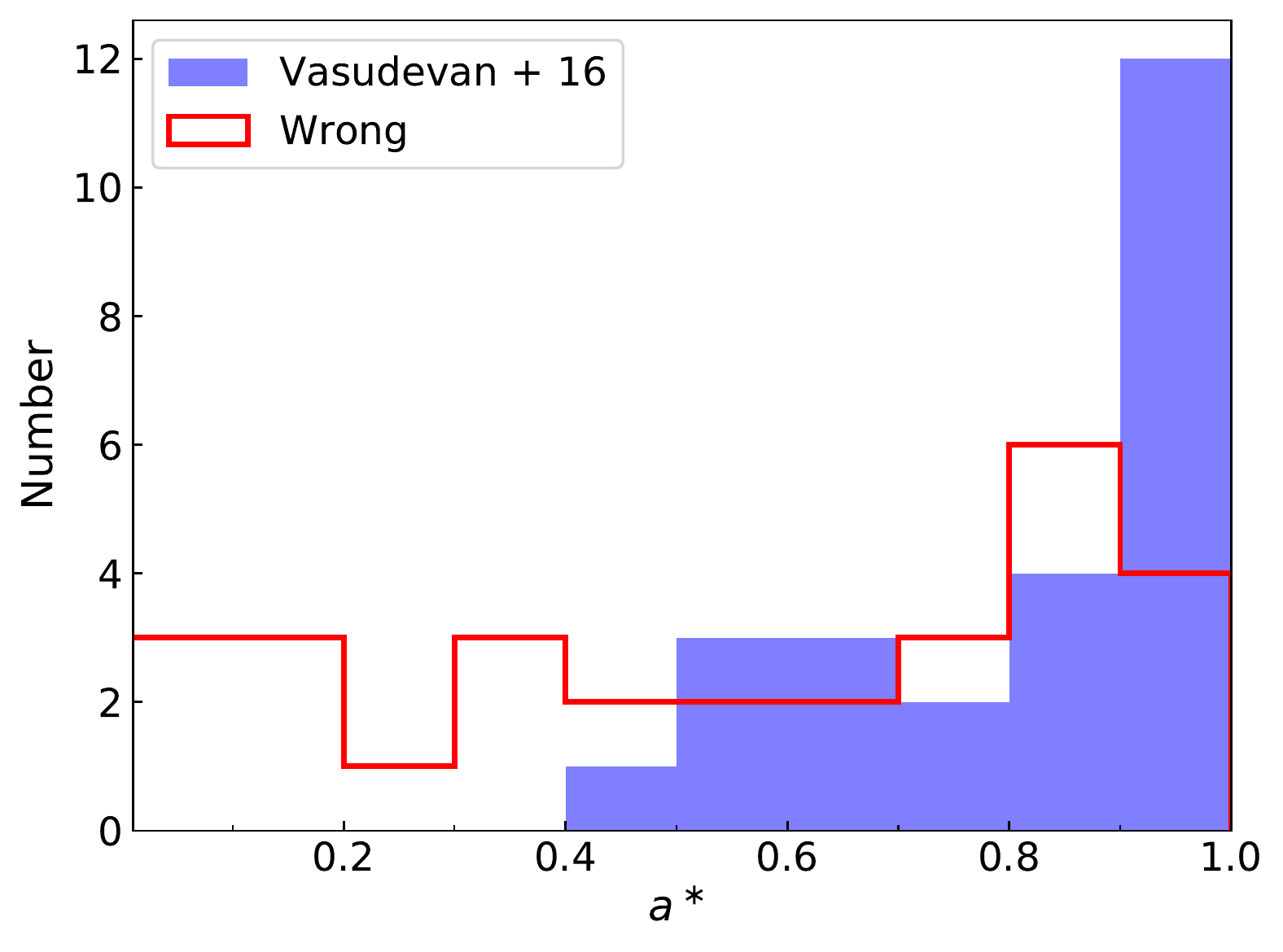}
\caption{Distribution of spin measurements from \cite{Vasudevan16} (blue histograms) and the distribution of wrong and undetermined measurements (red histograms) obtained from the simulations performed in the current work.}
\label{fig:histogram}
\end{figure}

An interesting outcome of our analysis is that the failed spin measurements have a nearly flat distribution, which is clearly different from those reported in the literature. We plot in Fig.\,\ref{fig:histogram} the distribution of spin measurements listed in \cite{Vasudevan16} together with that of wrong and undetermined measurements from our blind spectral analysis. As the uncertainties on the individual entries are rather large for both samples, we do not attempt any statistical test to compare quantitatively the two distributions. We note, however, that, while the selection effects leading to an observed spin distribution peaked towards higher values are well known \citep[e.g.][]{Brenneman11}, our results seem to discard any systematics or biases associated with possible reflection versus absorption spectral degeneracies.

\subsection{Simulations with ATHENA}
\label{sec:ATHENA}

\begin{figure}[!]
\centering

\includegraphics[width = 0.33\textwidth]{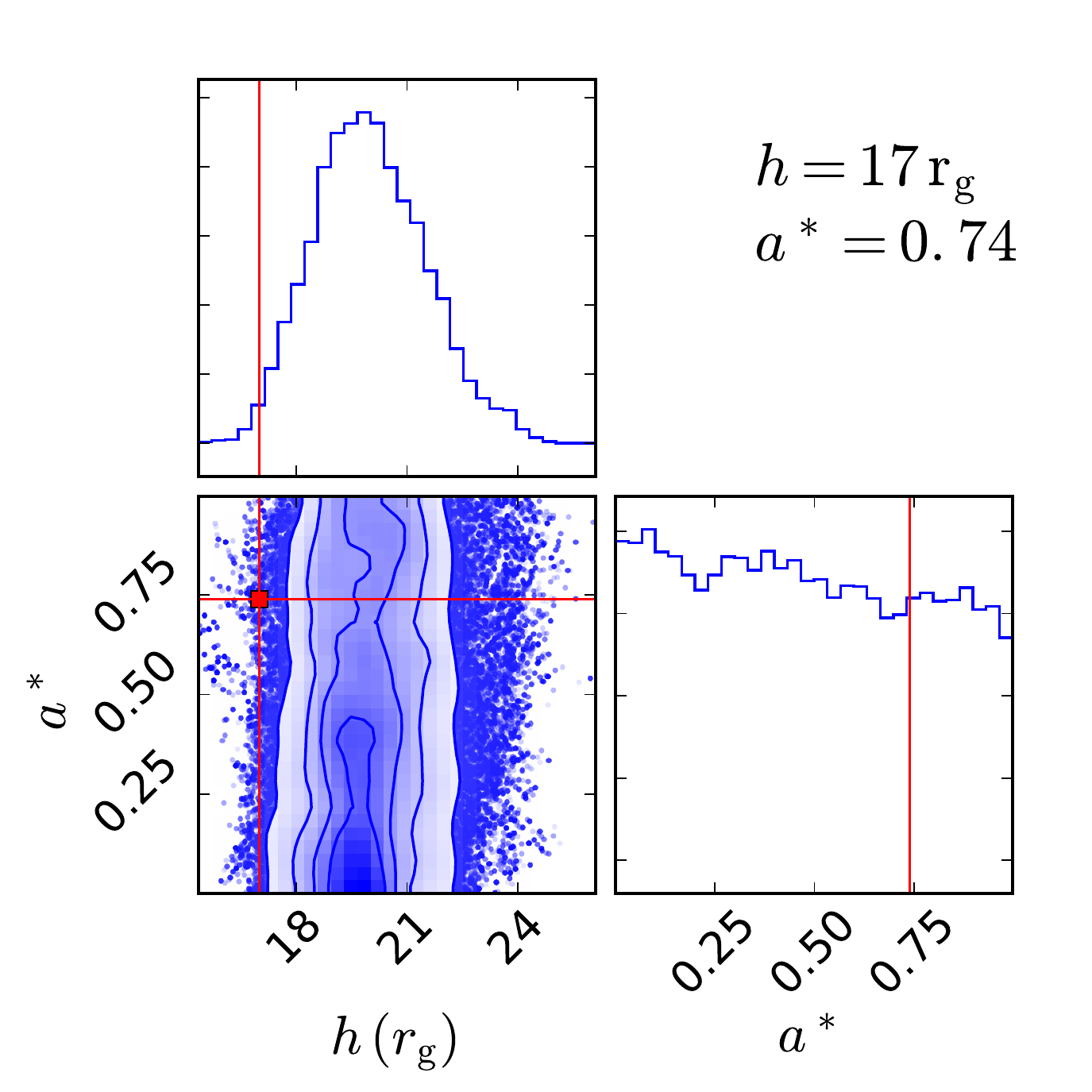} \\
\includegraphics[width = 0.33\textwidth]{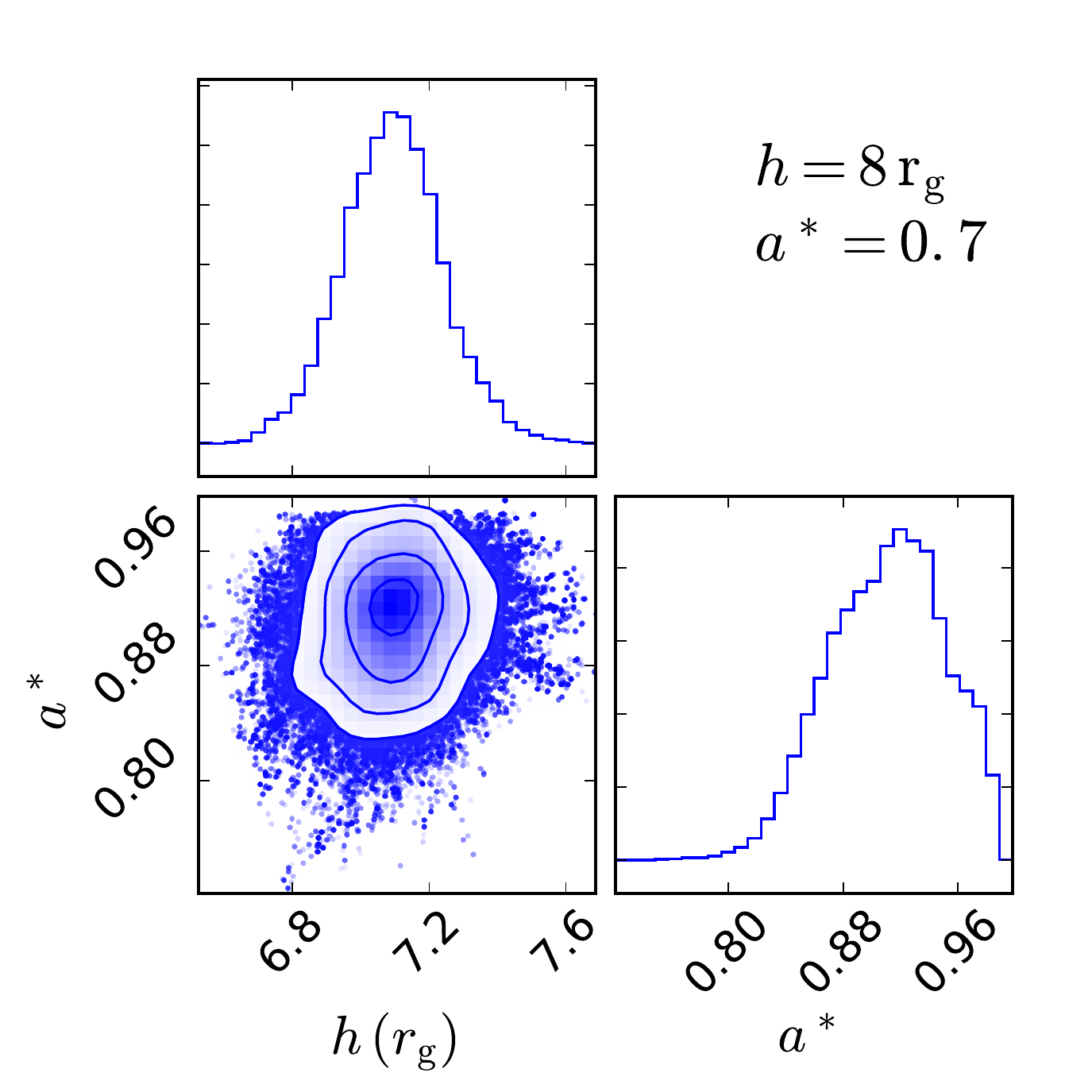} \\

\caption{Black hole spin vs. lamp-post height contour plots obtained by the MCMC analysis obtained from the best-spectral fits of G6 (top panel) and G11 (bottom panel), performed using the {\it ATHENA}-WFI response files. The input spin and height values (red lines) are listed on the top right corner of each panel for the corresponding simulation.}
\label{fig:ATHENA-corner}
\end{figure}

To verify whether the failures require an even larger data quality (hence inaccessible to the current X-ray observatories), we chose two models for which the measured values of the spin were either undetermined (G6) or wrong (G11) despite the excellent accuracy of both fits. Based on our results, both cases are expected to be rather challenging, as they involve intermediate black-hole spin and large lamp-post height. We then simulated the same input model using the response files (with an exposure time of 100\,ks) of the Wide Field Imager \citep[WFI;][]{WFI}, one of the two scientific instruments proposed for the {\it ATHENA} X-ray observatory \citep{ATHENA}. We performed an MCMC analysis as described in Section\,\ref{sec:fitting}. For model G6, which has $h = 17 \,r_{\rm g}$, the results are similar to those obtained by the spectral analysis of the joint {\it XMM-Newton} and {\it NuSTAR} spectra. Even with WFI, the measured value of the spin remains undetermined, as shown by the $a^\ast-{\rm vs}-h$ contour plots in Fig.\,\ref{fig:ATHENA-corner}. For model G11, which has almost the same input spin but lower height ($\rm 8\,r_{\rm g}$), the measured spin remains inconsistent with the actual value of 0.7, but it is now much closer to this value ($\sim 0.9$ against 0.2). This test, although unsuccessful, confirms the main indication of our study, i.e. the importance of a small lamp-post height (hence an effective illumination of the innermost disc) for an accurate measure of the BH spin. However, one of the limitations of fitting the spectra with {\it ATHENA} is the incapability of probing the Compton hump at high energies. We will further investigate this issue in a future work, where we will also consider the potential of high-resolution spectroscopy.

\section{Conclusions and future work}
\label{sec:conclusion}

The measure of black-hole spin in AGN has many important implications and the modelling of X-ray reflection features from the inner accretion disc provides a powerful method in this sense. The reliability of the available reflection-based SMBH spin measurements, however, is not fully established yet. In this work, we have investigated this issue through the simulation of high-quality broadband spectra, representive of the best possible data that can be achieved with a single simultaneous \textit{XMM--Newton} and \textit{NuSTAR} observation of a local, bright AGN. A similar attempt has been carried out recently by \cite{Bonson16} and by \cite{Choudhury17}. Both studies, however, only considered the spectra in the \textit{NuSTAR} energy range (2.5--79 keV), thereby neglecting the statistically dominant soft X-ray excess component. Moreover, the ideal scenario of pure reflection was assumed in both cases. We allowed, instead, for the general spectral complexity observed in real AGN spectra, including absorption, thermal, and distant reflection components in our parent model. The spectra were simulated by one member of the team and blindly fitted by the other two, where the only constraint was to use the same number of components employed in the parent model or less.

We have shown that the analysis of single-epoch AGN spectra can be really challenging. In fact, our simulations suggest that a correct determination of the BH spin parameter is not straightforward, and that the height of the X-ray source (in a lamp-post geometry) plays a major role in the spin measurement. This is not surprising and agrees with the conclusions reached by \cite{Fabian14}, \cite{Bonson16}, and by \cite{Choudhury17}: the closer the source to the black hole, the stronger the relativistic distortions that allow an accurate spin measurement. However, we also demonstrated that complex (i.e. partial-covering, multi-layer) absorption does not seem to have a critical impact on the ability to measure the spin correctly, at least at very high S/N. In summary, 42 out of 60 blind fits (from the 30 simulations) turned out to be accurate in that the model employed in the analysis corresponds to the input model and the overall $\chi^2$ is equal or very close to the expected absolute minimum (Section \,\ref{sec:results}). The spin was retrieved perfectly or reasonably well in 12 out of 42 and 10 out of 42 cases, respectively. It was unconstrained or wrong in 5 out of 42 and 15 out of 42 fits. The remaining 18 fits, formally inaccurate, actually return 9 additional acceptable spin measures (with 2 undetermined and 7 failed). By dividing the simulations over the four quadrants of the spin/height plane identified by the $a^\ast = 0.8$ and $h = 5\,r_{\rm g}$ values (Table \ref{table:spinvsheight}), we obtain a remarkable 16 out of 16 success rate for the accurate fits in recovering the correct spin in the high-spin/low-height quadrant. We note that the fraction of accurate fits is virtually constant over the whole parameter space.

Several lines of evidence suggest a compact primary source that is located at a few gravitational radii from the BH. Spectral-timing and reverberation studies, for example, are suggestive of a physically small corona that lies within 3--10 $\rm r_g$ above the central BH \citep[e.g.][]{Fabian09, Demarco13, Emma14, Gallo15}. Moreover, X-ray microlensing analyses of some bright lensed quasars suggest that the hard X-rays are emitted from compact regions with half-light radii less than $6\,\rm r_g$ \citep{Char09, Mosquera13, Reis13}. Our findings imply that X-ray reflection is indeed an effective method to measure the BH spin provided that reflection dominates the broadband intrinsic (i.e. before foreground absorption) spectrum, which might not always be the case \citep[e.g.][]{Parker17Ton}. Furthermore, our analysis implies that the actual nature of the X-ray source (as yet unknown) should heavily affect any reflection-based spin measure. Several other factors may lead to a wrong determination of the spectral parameters in single-epoch observations, such as the choice of a \textit{wrong} model. In fact, at low spectral quality, some absorption configurations can indeed mimic the relativistic effects. Moreover, for large lamp-post heights, the chances of reliably assessing the spin are small, and apparently independent on the value of the spin itself. Simply increasing the total number of counts or effective area does not bring any substantial improvement. For single-epoch, low-resolution spectra, indirect or complementary arguments, such as energy conservation, fractional variability, or model-independent techniques \citep[e.g.][]{Kammoun17}, are still recommended to support the conclusions of the spectral analysis. Spectral variability, however, could greatly help in constraining the constant parameters in the reflection models, such as the spin, inclination, and iron abundance, as already proved by the NGC~1365 campaign, while high resolution can remove the ambiguities associated with the introduction of ad hoc absorption components. The importance of variability in measuring the spins and the impact of future X-ray missions, carrying calorimeters and polarimeters, will be investigated in a future work.

\begin{acknowledgements}
We would like to thank the anonymous referee for his/her constructive comments that helped clarifying our manuscript. We would like to thank Prof. Andrew Fabian for valuable suggestions and comments. EN received funding from the European Union's Horizon 2020 research and innovation programme under the Marie Sk\l odowska-Curie grant agreement no. 664931. The figures were generated using {\tt matplotlib} \citep{Hunter07}, a {\tt PYTHON} library for publication of quality graphics. The MCMC results were presented using the open source code {\tt corner.py} \citep{corner}.
\end{acknowledgements}

%
   \bibliographystyle{aa} 
   \bibliography{ek-sim} 
%

\appendix
\section{Tables and figures}
We present in this appendix the best-fit results obtained from the blind fitting procedure. We list in Table\,\ref{table:App} the best-fit heights and spin values obtained for each fit, compared to the input values. We also report the minimum $\chi^2$/dof found for each fit and the reference value that we use to evaluate the accuracy of the fit (see \S\,\ref{sec:results} for details). Fig.\,\ref{fig:appendix-spectra} shows all the simulated spectra in addition to the input models and the residuals of the best-fit model for the two blind spectral fits.

\begin{table*}
\centering

\caption{Input and best-fit values of the height ($h$) and the spin parameter ($a^\ast$) found for the two fits performed to the spectra of Sets\,G, K, and B. We also report the best-fit $\chi^2/{\rm dof}$ we found and the reference value against which the accuracy of a fit is evaluated.}
\begin{adjustbox}{max width=0.98\textwidth}
\begin{threeparttable}
\begin{tabular}{lccc|lccc|lccc}
\hline \hline\\[-0.3cm]
        \multicolumn{12}{c}{Set\,G} \\[0.1cm]   \hline
        &       $       h\,({r_{\rm g}})                                                        $       &       $       a^\ast                                                  $       &       $       \chi^2/{\rm dof}                    $       &               &       $       h\,({r_{\rm g}})                                                    $       &       $       a^\ast                                                  $       &       $       \chi^2/{\rm dof}    $       &               &       $       h\,({r_{\rm g}})                                                        $       &       $       a^\ast                                                  $       &       $       \chi^2/{\rm dof}    $       \\[0.1cm]   \hline
        &       $       15                                                      $       &       $       0.2                                                     $       &       $       538/581                 $       &               &       $       17                                                      $       &       $       0.74                                                    $       &       $       429/441 $       &               &       $       8                                                       $       &       $       0.7                                                     $       &       $       338/357 $       \\[0.1cm]   
G1      &       $       5.55    _{      -0.48   }       ^{+     0.96    }       $       &       $       0.761   _{      -0.506  }       ^{+     0.202   }       $       &       $       546/583                 $       &       G6      &       $       16.21   _{      -2.50   }       ^{+     3.73    }       $       &       $       0.569   _{      -0.507  }       ^{+     0.384   }       $       &       $       429/441 $       &       G11     &       $       7.16    _{      -0.66   }       ^{+     0.73    }       $       &       $       0.196   _{      -0.149  }       ^{+     0.269   }       $       &       $       338/357 $       \\[0.1cm]   
        &       $       2.77    _{      -0.09   }       ^{+     0.16    }       $       &       $       0       _{      p       }       ^{+     0.193   }       $       &       $       597/581                 $       &               &       $       15.79   _{      -2.65   }       ^{+     3.87    }       $       &       $       0.498   _{      -0.449  }       ^{+     0.450   }       $       &       $       429/441 $       &               &       $       6.71    _{      -0.80   }       ^{+     1.04    }       $       &       $       0.208   _{      -0.173  }       ^{+     0.239   }       $       &       $       338/358 $       \\[0.1cm]   \hline
        &       $       3                                                       $       &       $       0.7                                                     $       &       $       370/388                 $       &               &       $       6                                                       $       &       $       0.28                                                    $       &       $       407/455 $       &               &       $       4                                                       $       &       $       0                                                       $       &       $       410/384 $       \\[0.1cm]   
G2      &       $       2.74    _{      -0.12   }       ^{+     0.28    }       $       &       $       0.765   _{      -0.171  }       ^{+     0.109   }       $       &       $       370/389                 $       &       G7      &       $       3.74    _{      -0.28   }       ^{+     0.40    }       $       &       $       0.166   _{      -0.139  }       ^{+     0.205   }       $       &       $       407/455 $       &       G12     &       $       14.85   _{      -2.28   }       ^{+     4.94    }       $       &       $       0.998   _{      -0.861  }       ^{         p       }       $       &       $       418/385 $       \\[0.1cm]   
        &       $       3.76    _{      -0.32   }       ^{+     0.40    }       $       &       $       0.976   _{      -0.527  }       ^{+     0.009   }       $       &       $       383/386                 $       &               &       $       2.70    _{      -0.04   }       ^{+     0.12    }       $       &       $       0       _{      p       }       ^{+     0.155   }       $       &       $       418/452 $       &               &       $       12.61   _{      -6.53   }       ^{+     2.08    }       $       &       $       0.143   _{      -0.113  }       ^{+     0.597   }       $       &       $       410/384 $       \\[0.1cm]   \hline 
        &       $       5                                                       $       &       $       0.950                                                   $       &       $       430/419                 $       &               &       $       9                                                       $       &       $       0.93                                                    $       &       $       495/478 $       &               &       $       2                                                       $       &       $       0.99                                                    $       &       $       431/400 $       \\[0.1cm]   
G3      &       $       4.44    _{      -0.58   }       ^{+     1.15    }       $       &       $       0.998   _{      -0.096  }       ^{         p       }       $       &       $       430/419                 $       &       G8      &       $       2.77    _{      -0.15   }       ^{+     0.21    }       $       &       $       0.439   _{      -0.299  }       ^{+     0.158   }       $       &       $       495/478 $       &       G13     &       $       2.00    _{      p       }       ^{+     0.20    }       $       &       $       0.989   _{      -0.001  }       ^{+     0.002   }       $       &       $       431/400 $       \\[0.1cm]   
        &       $       4.25    _{      -0.39   }       ^{+     1.10    }       $       &       $       0.975   _{      -0.200  }       ^{+     0.015   }       $       &       $       430/419                 $       &               &       $       4.97    _{      -0.18   }       ^{+     1.63    }       $       &       $       0.617   _{      -0.476  }       ^{+     0.334   }       $       &       $       496/478 $       &               &       $       2.13    _{      -0.07   }       ^{+     0.14    }       $       &       $       0.991   _{      -0.003  }       ^{         p       }       $       &       $       431/400 $       \\[0.1cm]   \hline
        &       $       5                                                       $       &       $       0.9                                                     $       &       $       410/404                 $       &               &       $       2.3                                                     $       &       $       0.99                                                    $       &       $       392/445 $       &               &       $       10                                                      $       &       $       0.3                                                     $       &       $       480/481 $       \\[0.1cm]   
G4      &       $       4.67    _{      -0.35   }       ^{+     0.77    }       $       &       $       0.86    _{      -0.24   }       ^{+     0.12    }       $       &       $       401/402                 $       &       G9      &       $       2.51    _{      -0.34   }       ^{+     0.03    }       $       &       $       0.880   _{      -0.085  }       ^{+     0.102   }       $       &       $       415/446 $       &       G14     &       $       360     _{      -143    }       ^{+     132     }       $       &       $       0.851   _{      -0.803  }       ^{+     0.106   }       $       &       $       483/482 $       \\[0.1cm]   
        &       $       5.09    _{      -0.22   }       ^{+     0.81    }       $       &       $       0.86    _{      -0.27   }       ^{+     0.12    }       $       &       $       401/399                 $       &               &       $       2.02    _{      -0.01   }       ^{+     0.16    }       $       &       $       0.964   _{      -0.097  }       ^{+     0.016   }       $       &       $       393/445 $       &               &       $       12.78   _{      -2.06   }       ^{+     8.50    }       $       &       $       0.896   _{      -0.844  }       ^{+     0.061   }       $       &       $       479/481 $       \\[0.1cm]   \hline 
        &       $       6                                                       $       &       $       0.5                                                     $       &       $       449/473                 $       &               &       $       32                                                      $       &       $       0.12                                                    $       &       $       381/391 $       &               &       $       2                                                       $       &       $       0.99                                                    $       &       $       412/412 $       \\[0.1cm]   
G5      &       $       3.20    _{      -0.23   }       ^{+     0.47    }       $       &       $       0.744   _{      -0.455  }       ^{+     0.047   }       $       &       $       449/473                 $       &       G10     &       $       9.25    _{      -1.31   }       ^{+     1.39    }       $       &       $       0.586   _{      -0.480  }       ^{+     0.284   }       $       &       $       429/394 $       &       G15     &       $       2.04    _{      -0.03   }       ^{+     0.14    }       $       &       $       0.990   _{      -0.001  }       ^{+     0.002   }       $       &       $       412/412 $       \\[0.1cm]   
        &       $       4.58    _{      -0.73   }       ^{+     0.90    }       $       &       $       0.602   _{      -0.411  }       ^{+     0.358   }       $       &       $       453/468                 $       &               &       $       6.60    _{      -0.77   }       ^{+     1.01    }       $       &       $       0.016   _{      p       }       ^{+     0.229   }       $       &       $       385/391 $       &               &       $       2.00    _{      p       }       ^{+     0.20    }       $               &       $       0.990   _{      -0.001  }       ^{+     0.003   }       $       &       $       412/412 $       \\[0.1cm]   \hline \hline

\\[-0.3cm]
\multicolumn{12}{c}{Set\,K} \\[0.1cm]   \hline
        &       $       6                                                       $       &       $       0.998                                                   $       &       $       438/433                 $       &               &       $       6                                                       $       &       $       0.998                                                   $       &       $       371/376 $       &               &       $       2                                                       $       &       $       0.99                                                    $       &       $       400/401 $       \\[0.1cm]   
K1      &       $       3.01    _{      -0.10   }       ^{+     0.19    }       $       &       $       0.359   _{      -0.183  }       ^{+     0.167   }       $       &       $       438/434                 $       &       K4      &       $       134     _{      -57     }       ^{+     204     }       $       &       $       0.758   _{      -0.680  }       ^{+     0.215   }       $       &       $       412/379 $       &       K7      &       $       2.04    _{      -0.03   }       ^{+     0.15    }       $       &       $       0.945   _{      -0.049  }       ^{+     0.022   }       $       &       $       401/401 $       \\[0.1cm]   
        &       $       4.39    _{      -0.17   }       ^{+     0.54    }       $       &       $       0.923   _{      -0.216  }       ^{+     0.058   }       $       &       $       474/433                 $       &               &       $       184     _{      -132    }       ^{+     56      }       $       &       $       0.31    _{      -0.268  }       ^{+     0.640   }       $       &       $       389/376 $       &               &       $       2.01    _{      -0.01   }       ^{+     0.20    }       $       &       $       0.933   _{      -0.061  }       ^{+     0.029   }       $       &       $       400/401 $       \\[0.1cm]   \hline
        &       $       2.5                                                     $       &       $       0.998                                                   $       &       $       397/443                 $       &               &       $       3                                                       $       &       $       0.998                                                   $       &       $       321/396 $       &               &       $       2                                                       $       &       $       0.99                                                    $       &       $       384/441 $       \\[0.1cm]   
K2      &       $       3.80    _{      -0.15   }       ^{+     0.82    }       $       &       $       0.913   _{      -0.339  }       ^{+     0.069   }       $       &       $       431/444                 $       &       K5      &       $       2.57    _{      -0.26   }       ^{+     0.45    }       $       &       $       0.930   _{      -0.075  }       ^{+     0.046   }       $       &       $       367/395 $       &       K8      &       $       2.00    _{      p       }       ^{+     0.17    }       $       &       $       0.975   _{      -0.059  }       ^{+     0.019   }       $       &       $       385/441 $       \\[0.1cm]   
        &       $       2.95    _{      -0.22   }       ^{+     0.35    }       $       &       $       0.971   _{      -0.204  }       ^{+     0.022   }       $       &       $       398/443                 $       &               &       $       4.33    _{      -0.91   }       ^{+     0.46    }       $       &       $       0.977   _{      -0.081  }       ^{+     0.019   }       $       &       $       325/396 $       &               &       $       2.00    _{      p       }       ^{+     0.17    }       $       &       $       0.993   _{      -0.046  }       ^{+     0.004   }       $       &       $       384/441 $       \\[0.1cm]   \hline 
        &       $       12                                                      $       &       $       0.998                                                   $       &       $       472/461                 $       &               &       $       4                                                       $       &       $       0.998                                                   $       &       $       466/449 $       &               &       $       8                                                       $       &       $       0.99                                                    $       &       $       377/424 $       \\[0.1cm]   
K3      &       $       6.32    _{      -0.80   }       ^{+     0.75    }       $       &       $       0.015   _{      -0.008  }       ^{+     0.319   }       $       &       $       473/462                 $       &       K6      &       $       4.11    _{      -0.21   }       ^{+     0.49    }       $       &       $       0.991   _{      -0.252  }       ^{+     0.003   }       $       &       $       471/449 $       &       K9      &       $       9.45    _{      -2.39   }       ^{+     4.12    }       $       &       $       0.857   _{      -0.774  }       ^{+     0.105   }       $       &       $       382/424 $       \\[0.1cm]   
        &       $       6.53    _{      -1.14   }       ^{+     0.08    }       $       &       $       0       _{      p       }       ^{+     0.003   }       $       &       $       472/461                 $       &               &       $       4.03    _{      -0.58   }       ^{+     0.10    }       $       &       $       0.123   _{      -0.105  }       ^{+     0.303   }       $       &       $       481/446 $       &               &       $       9.03    _{      -1.27   }       ^{+     6.83    }       $       &       $       0.736   _{      -0.670  }       ^{+     0.228   }       $       &       $       385/424 $       \\[0.1cm] \hline \hline
\\[-0.3cm]
        \multicolumn{12}{c}{Set\,B} \\[0.1cm]   \hline
        &       $       7                                                       $       &       $       0.998                                                   $       &       $       350/341                 $       &               &       $       8                                                       $       &       $       0.65                                                    $       &       $       373/383 $       &               &       $       10                                                      $       &       $       0.99                                                    $       &       $       392/366 $       \\[0.1cm]   
B1      &       $       7.56    _{      -0.97   }       ^{+     0.95    }       $       &       $       0.9     _{      -0.32   }       ^{+     0.08    }       $       &       $       350/339                 $       &       B3      &       $       5.12    _{      -0.50   }       ^{+     1.25    }       $       &       $       0.38    _{      -0.16   }       ^{+     0.21    }       $       &       $       373/383 $       &       B5      &       $       13.09   _{      -1.81   }       ^{+     1.85    }       $       &       $       0.86    _{      -0.62   }       ^{+     0.11    }       $       &       $       392/367 $       \\[0.1cm]   
        &       $       7.35    _{      -0.90   }       ^{+     1.08    }       $       &       $       0.88    _{      -0.37   }       ^{+     0.10    }       $       &       $       350/339                 $       &               &       $       2.91    _{      -0.12   }       ^{+     0.08    }       $       &       $       0.01    _{      p       }       ^{+     0.07    }       $       &       $       378/382 $       &               &       $       12.14   _{      -0.98   }       ^{+     2.67    }       $       &       $       0.83    _{      -0.50   }       ^{+     0.15    }       $       &       $       392/367 $       \\[0.1cm]   \hline
        &       $       3.5                                                     $       &       $       0.2                                                     $       &       $       367/372                 $       &               &       $       5                                                       $       &       $       0.9                                                     $       &       $       328/355 $       &               &       $       2.5                                                     $       &       $       0.99                                                    $       &       $       324/370 $       \\[0.1cm]   
B2      &       $       3.47    _{      -0.44   }       ^{+     0.40    }       $       &       $       0.31    _{      -0.25   }       ^{+     0.08    }       $       &       $       368/373                 $       &       B4      &       $       4.12    _{      -0.24   }       ^{+     0.40    }       $       &       $       0.964   _{      -0.197  }       ^{+     0.025   }       $       &       $       350/356 $       &       B6      &       $       2.60    _{      -0.01   }       ^{+     0.15    }       $       &       $       0.981   _{      -0.050  }       ^{+     0.006   }       $       &       $       324/371 $       \\[0.1cm]   
        &       $       4.55    _{      -0.54   }       ^{+     0.45    }       $       &       $       0.27    _{      -0.18   }       ^{+     0.25    }       $       &       $       367/372                 $       &               &       $       4.93    _{      -0.50   }       ^{+     0.64    }       $       &       $       0.96    _{      -0.33   }       ^{+     0.03    }       $       &       $       330/355 $       &               &       $       2.62    _{      -0.06   }       ^{+     0.10    }       $       &       $       0.981   _{      -0.048  }       ^{+     0.012   }       $       &       $       324/371 $       \\[0.1cm]   \hline \hline
\end{tabular}

\begin{tablenotes}
        \item[$p$] pegged to its maximum/minimum allowed value.
\end{tablenotes}
\end{threeparttable}
\end{adjustbox}
\label{table:App}
\end{table*}

\clearpage

\begin{figure*}
\centering
\includegraphics[width = 0.29\textwidth]{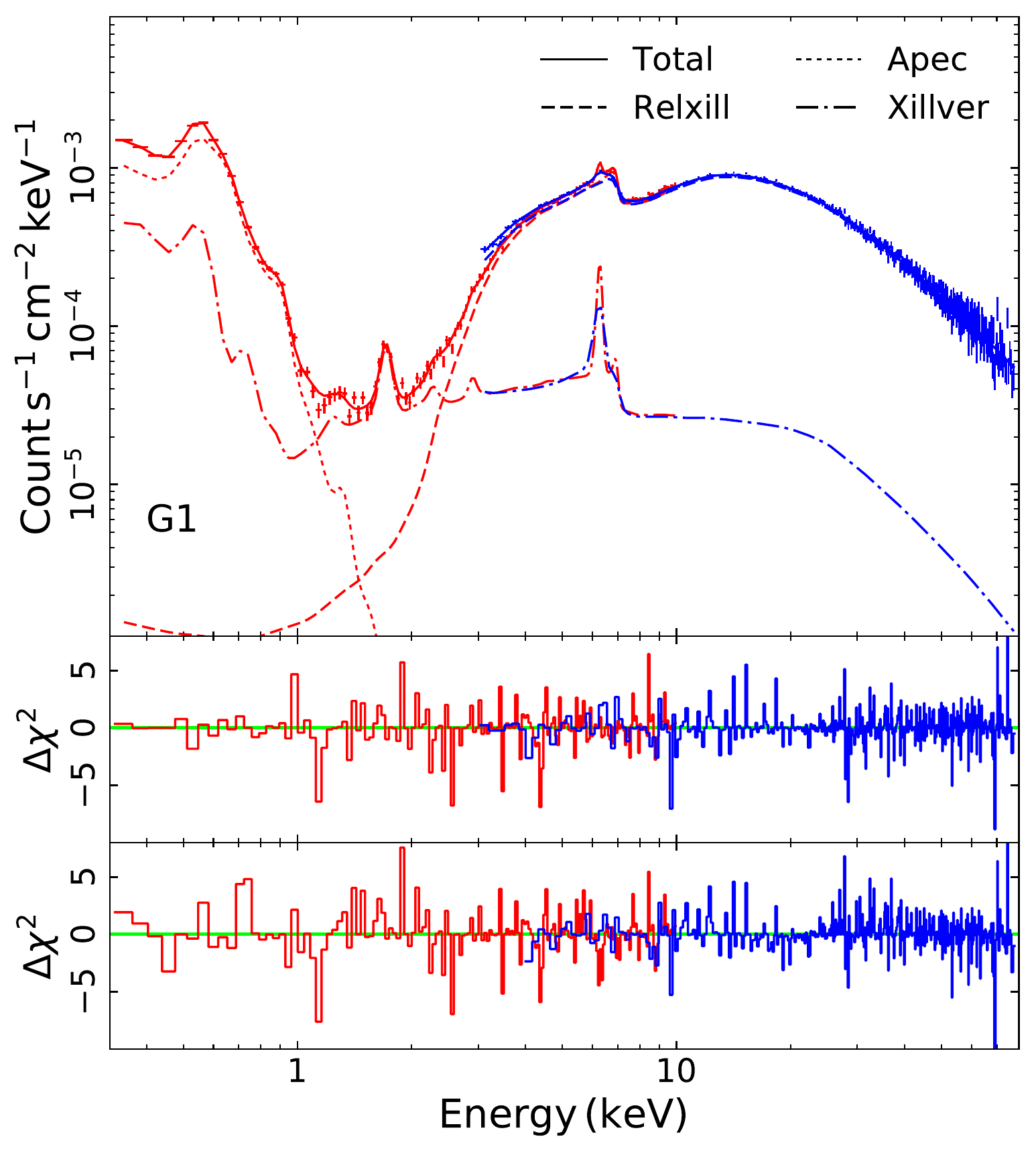} 
\includegraphics[width = 0.29\textwidth]{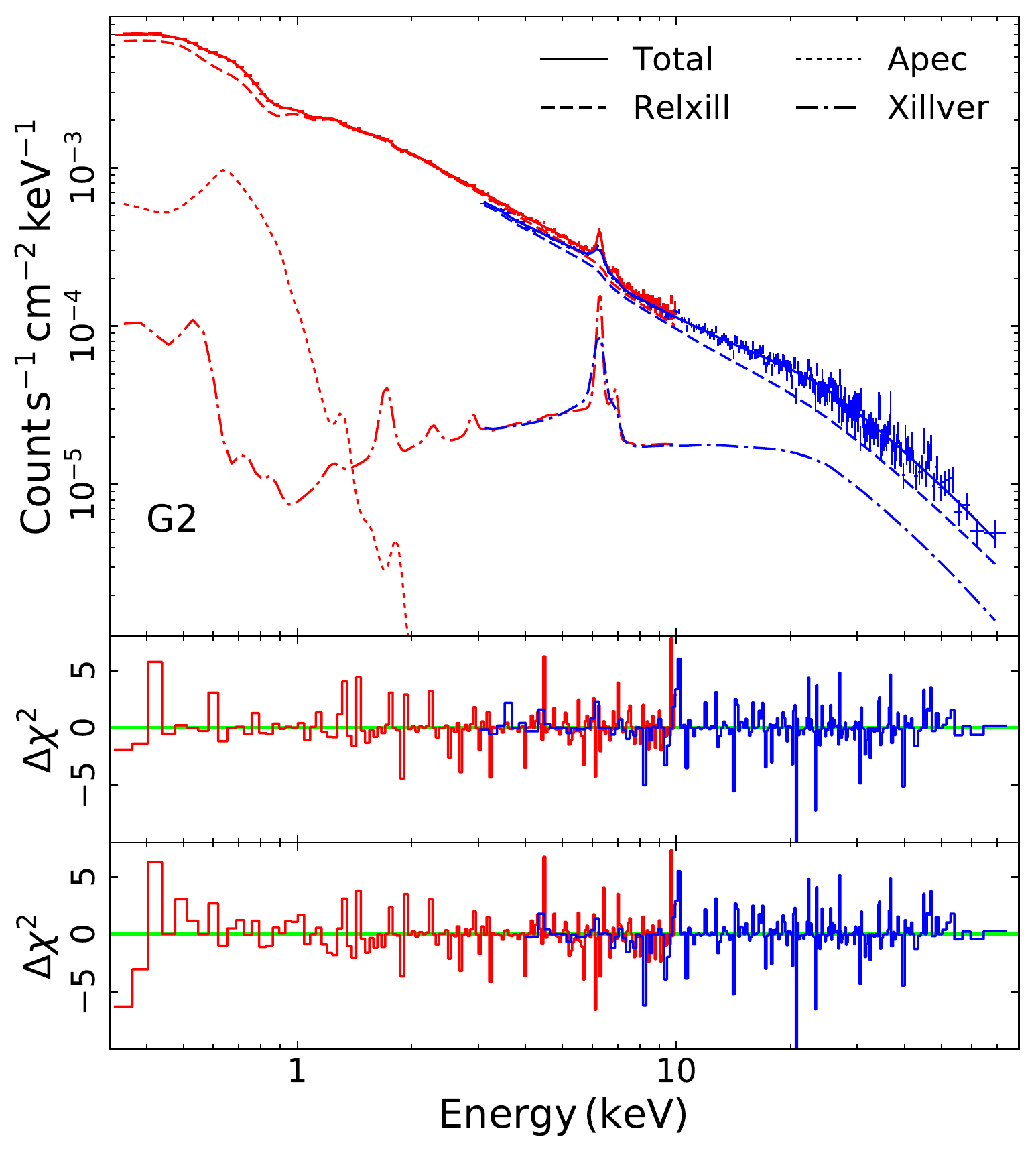} 
\includegraphics[width = 0.29\textwidth]{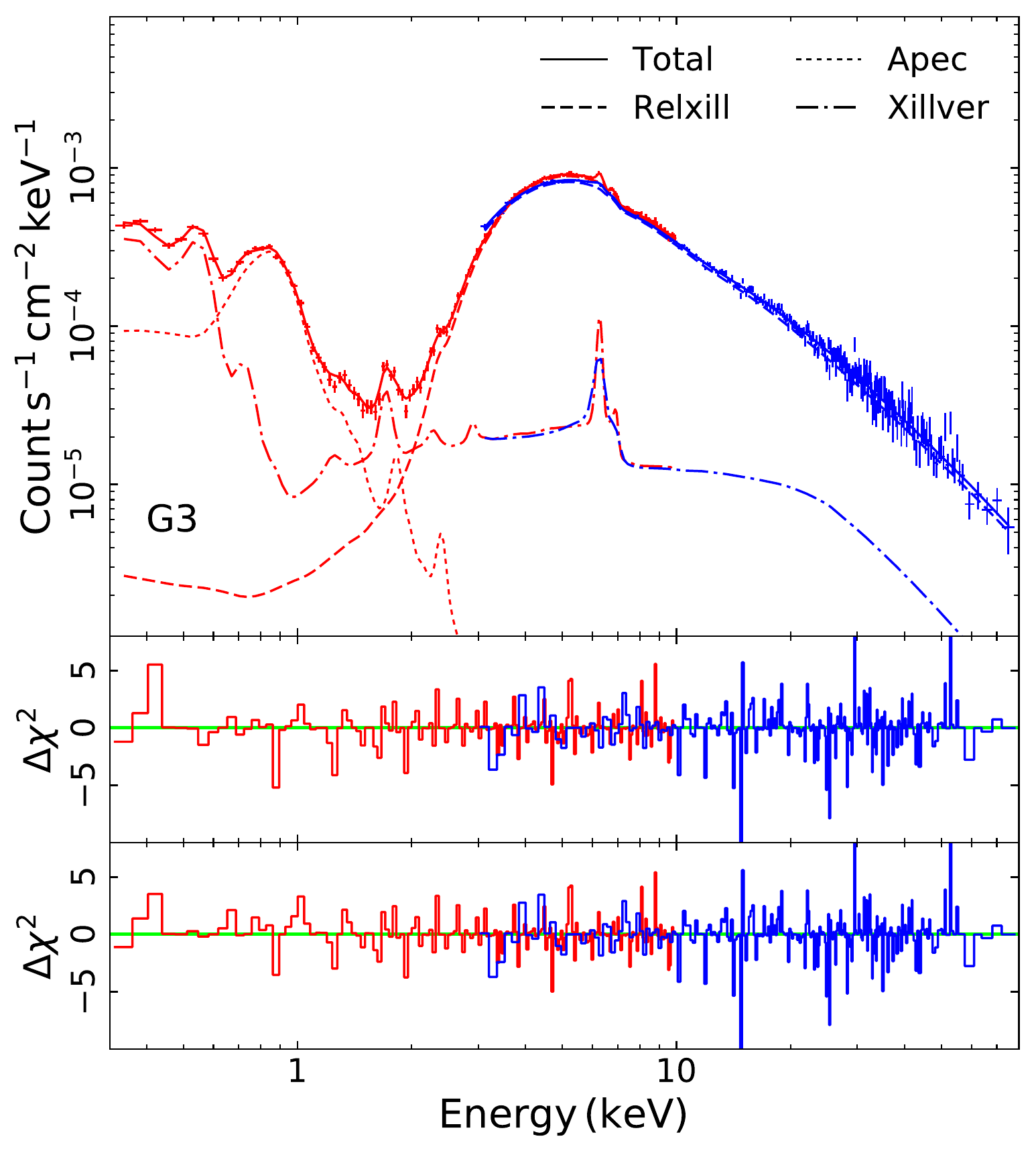} 
\includegraphics[width = 0.29\textwidth]{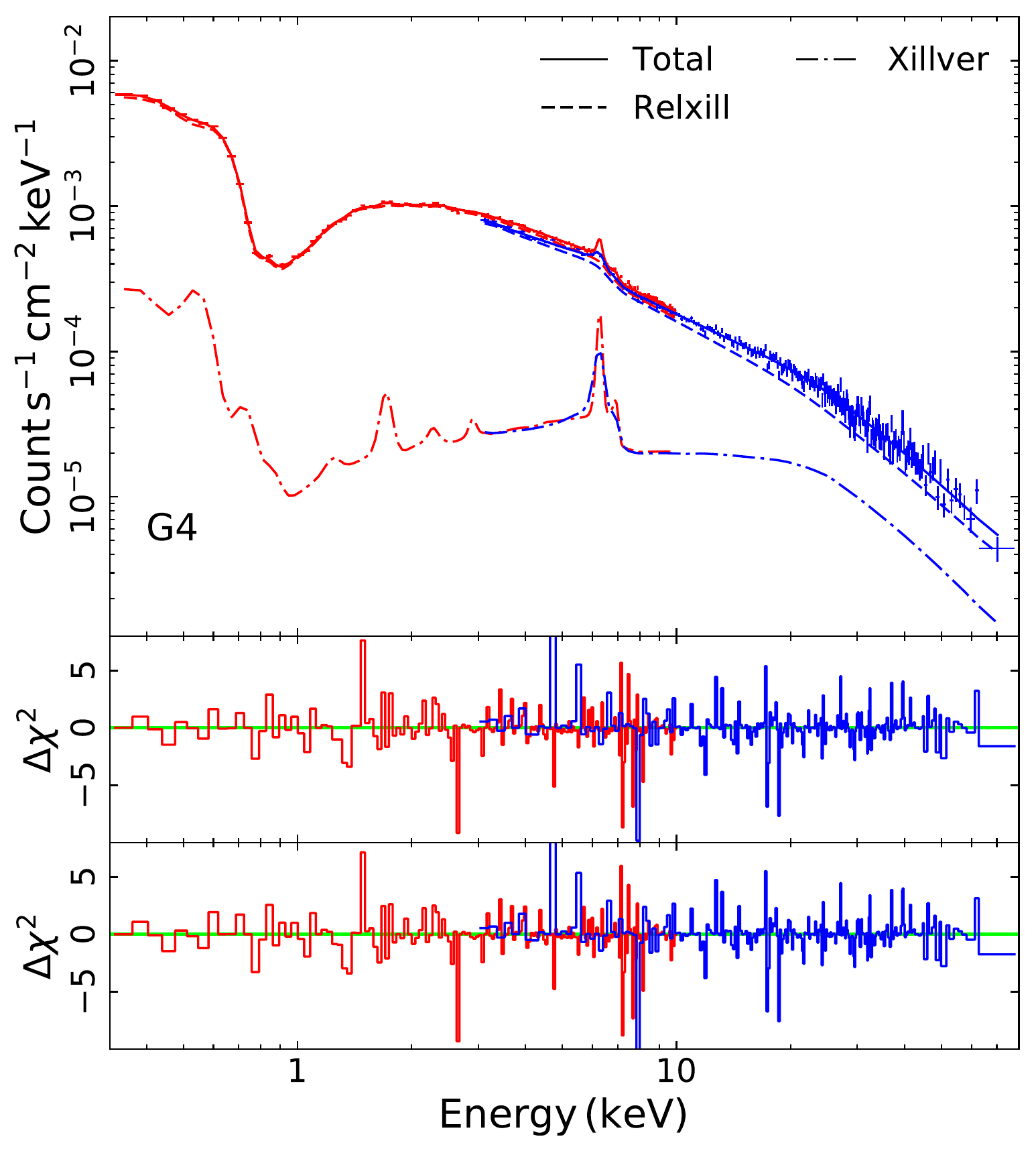} 
\includegraphics[width = 0.29\textwidth]{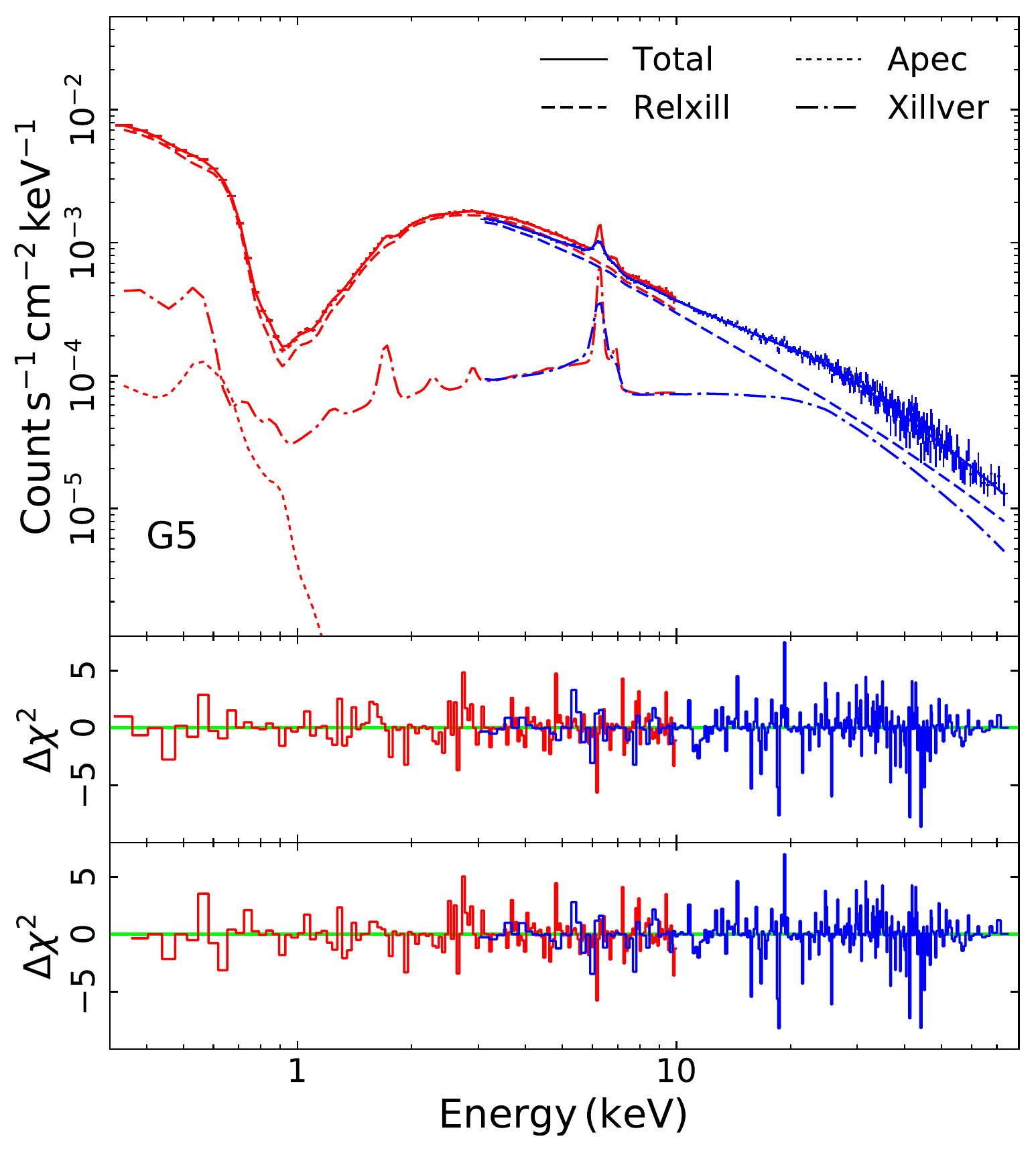} 
\includegraphics[width = 0.29\textwidth]{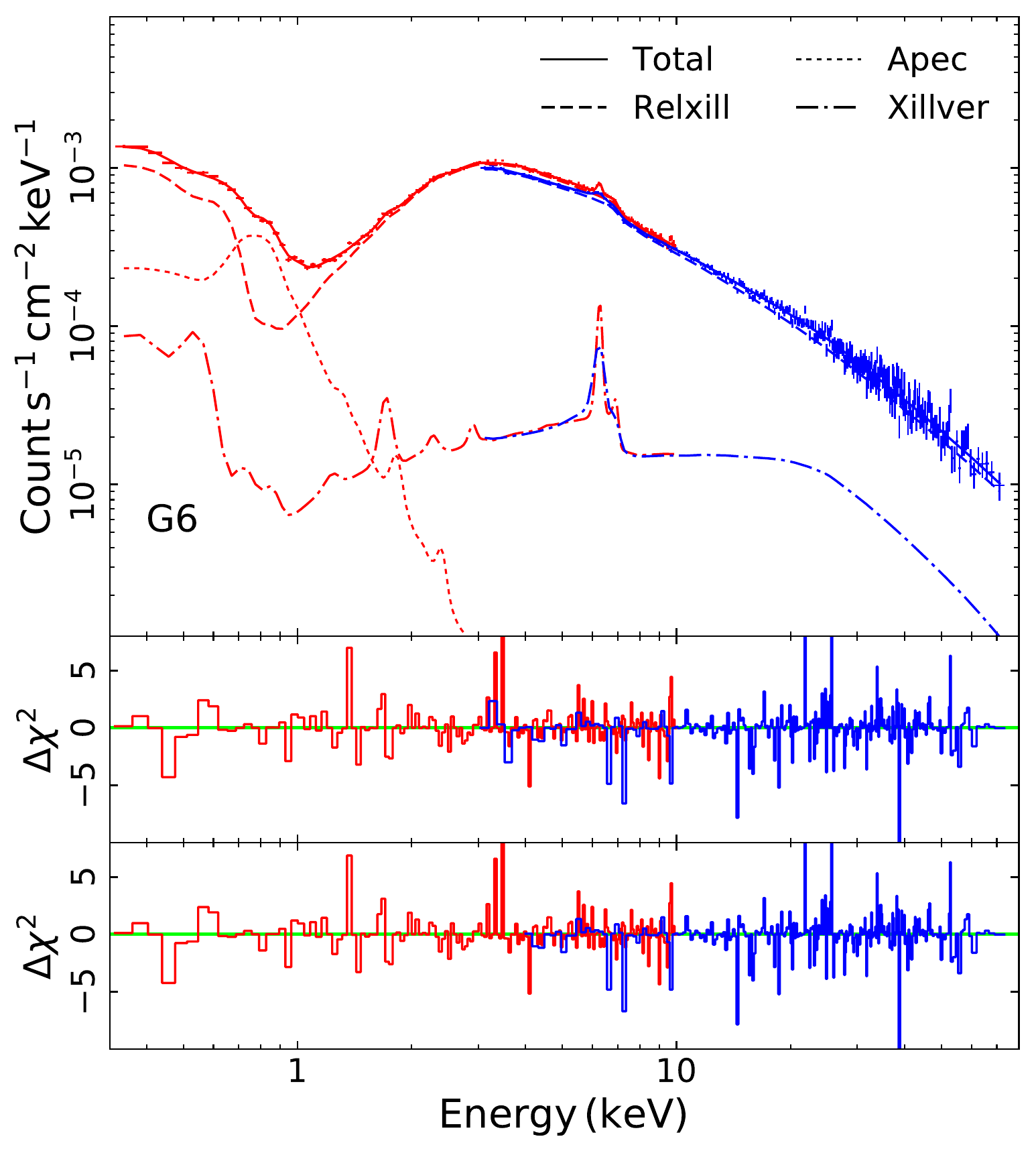}
\includegraphics[width = 0.29\textwidth]{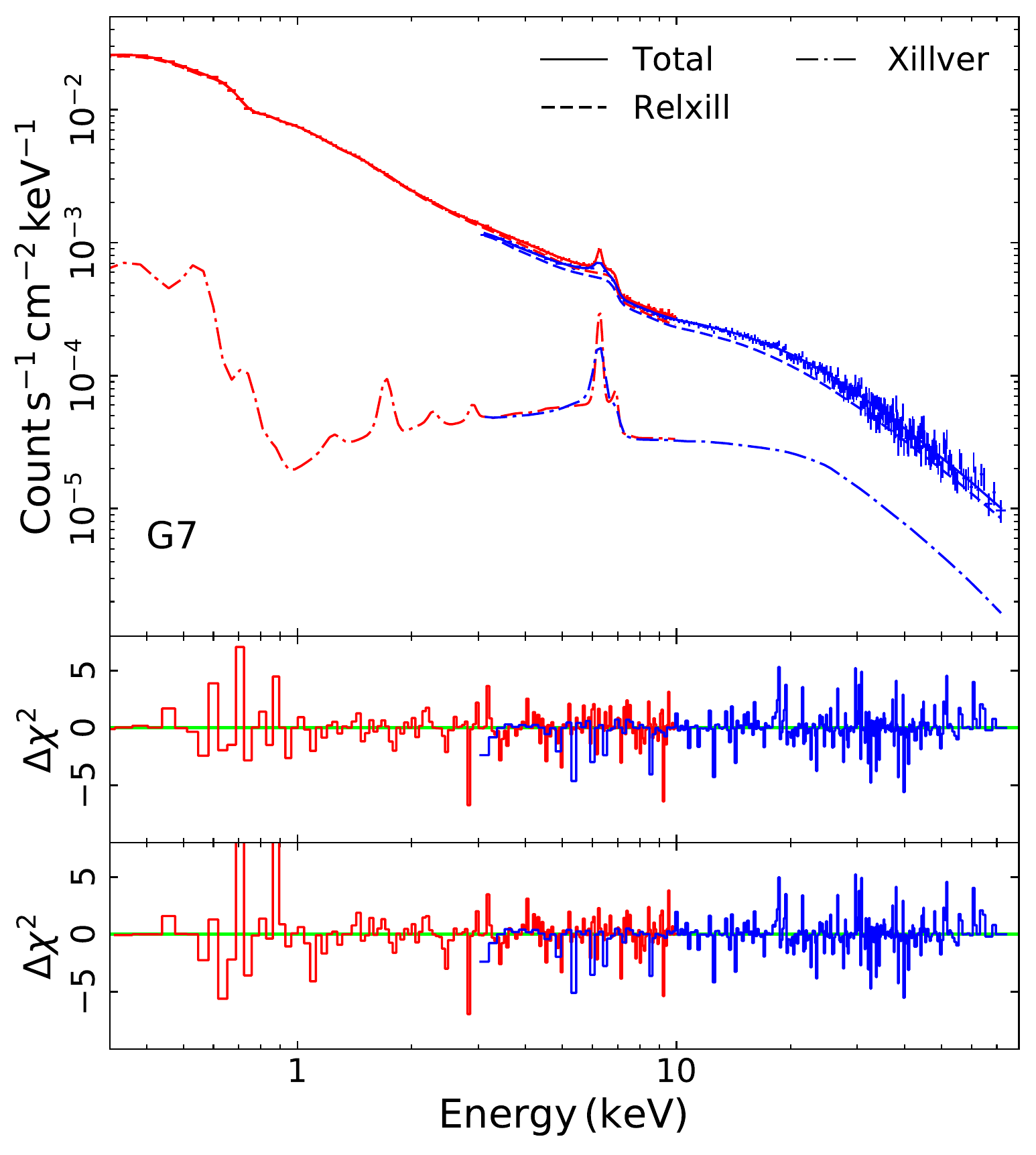}
\includegraphics[width = 0.29\textwidth]{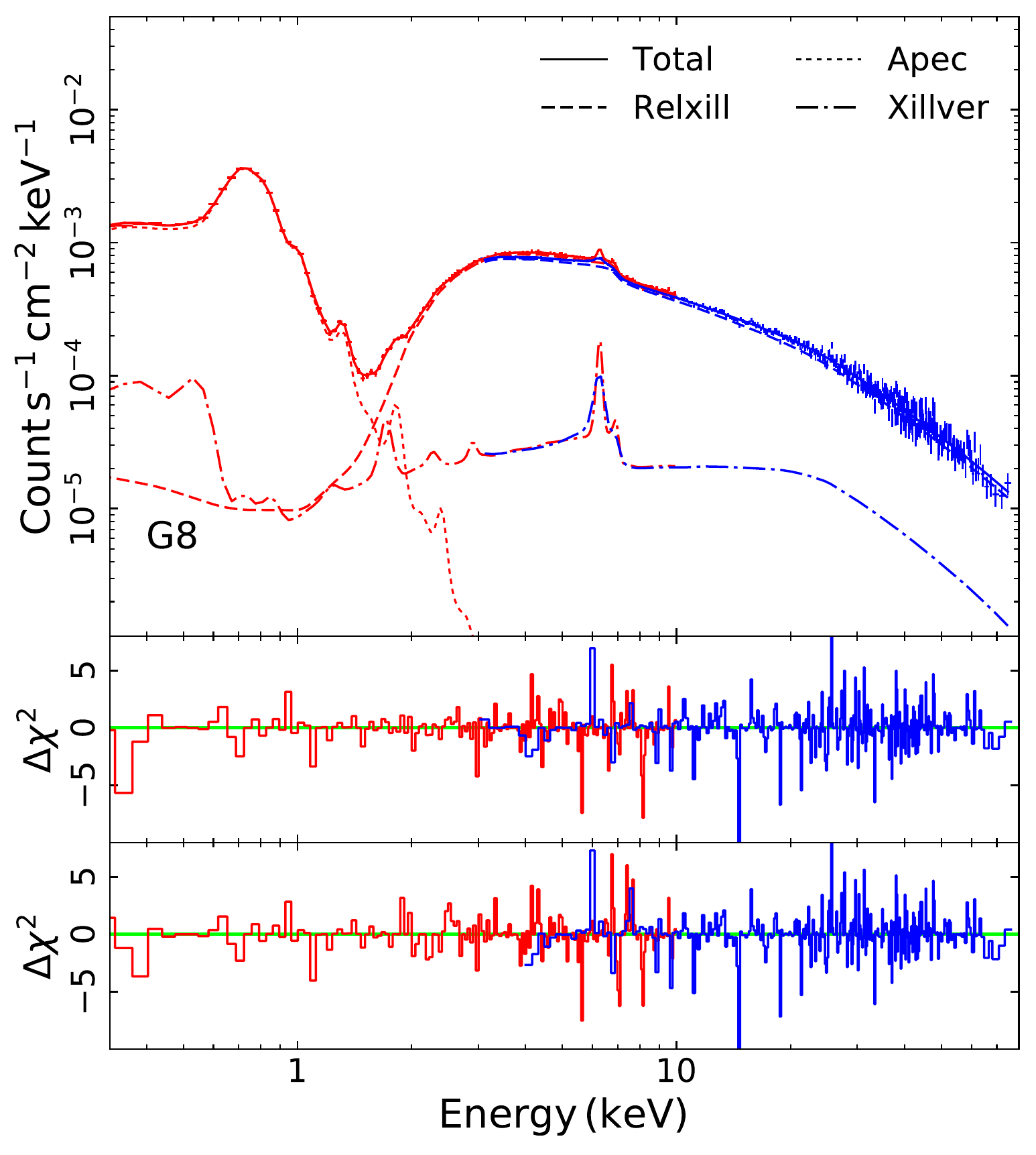}
\includegraphics[width = 0.29\textwidth]{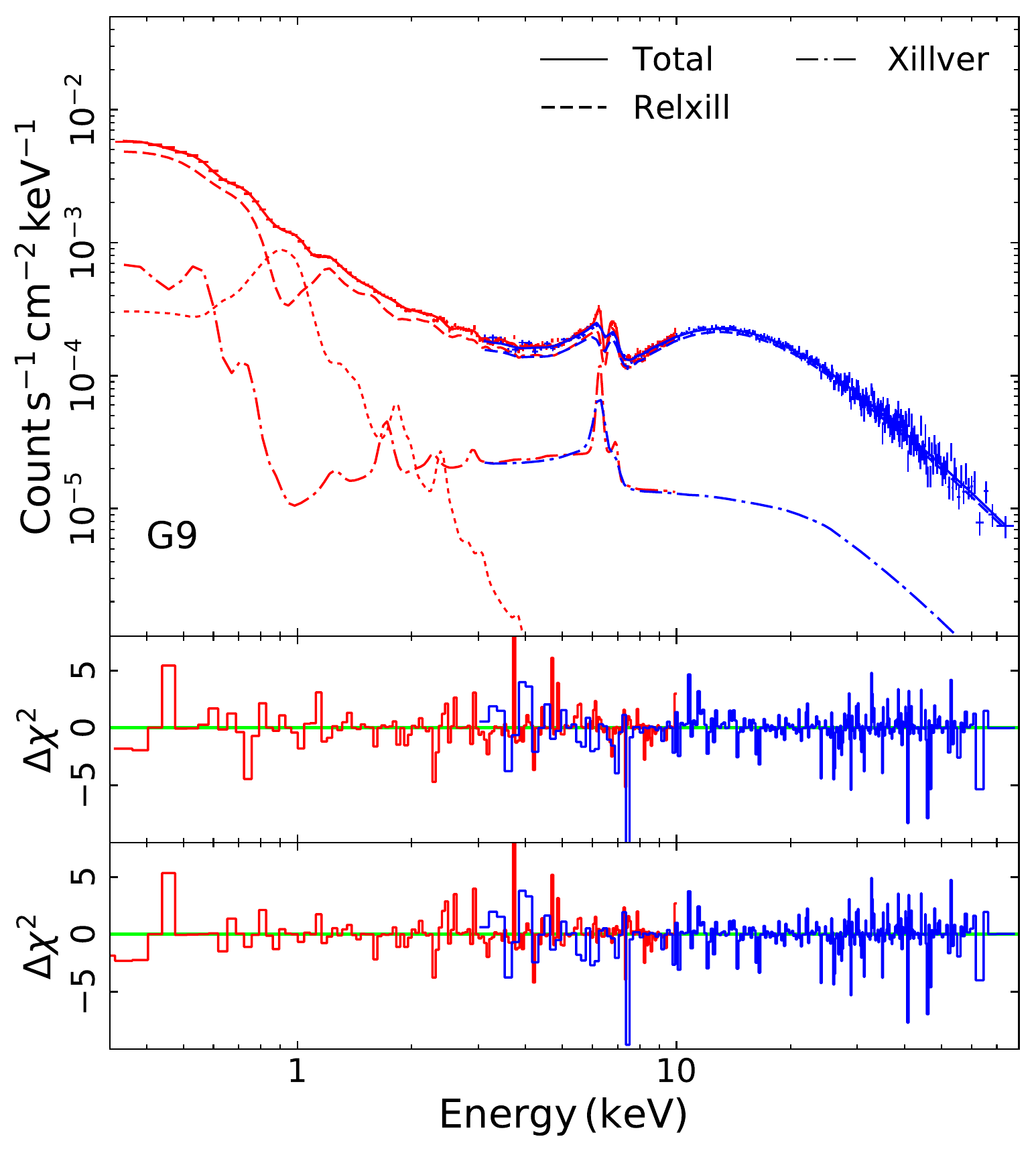}
\includegraphics[width = 0.29\textwidth]{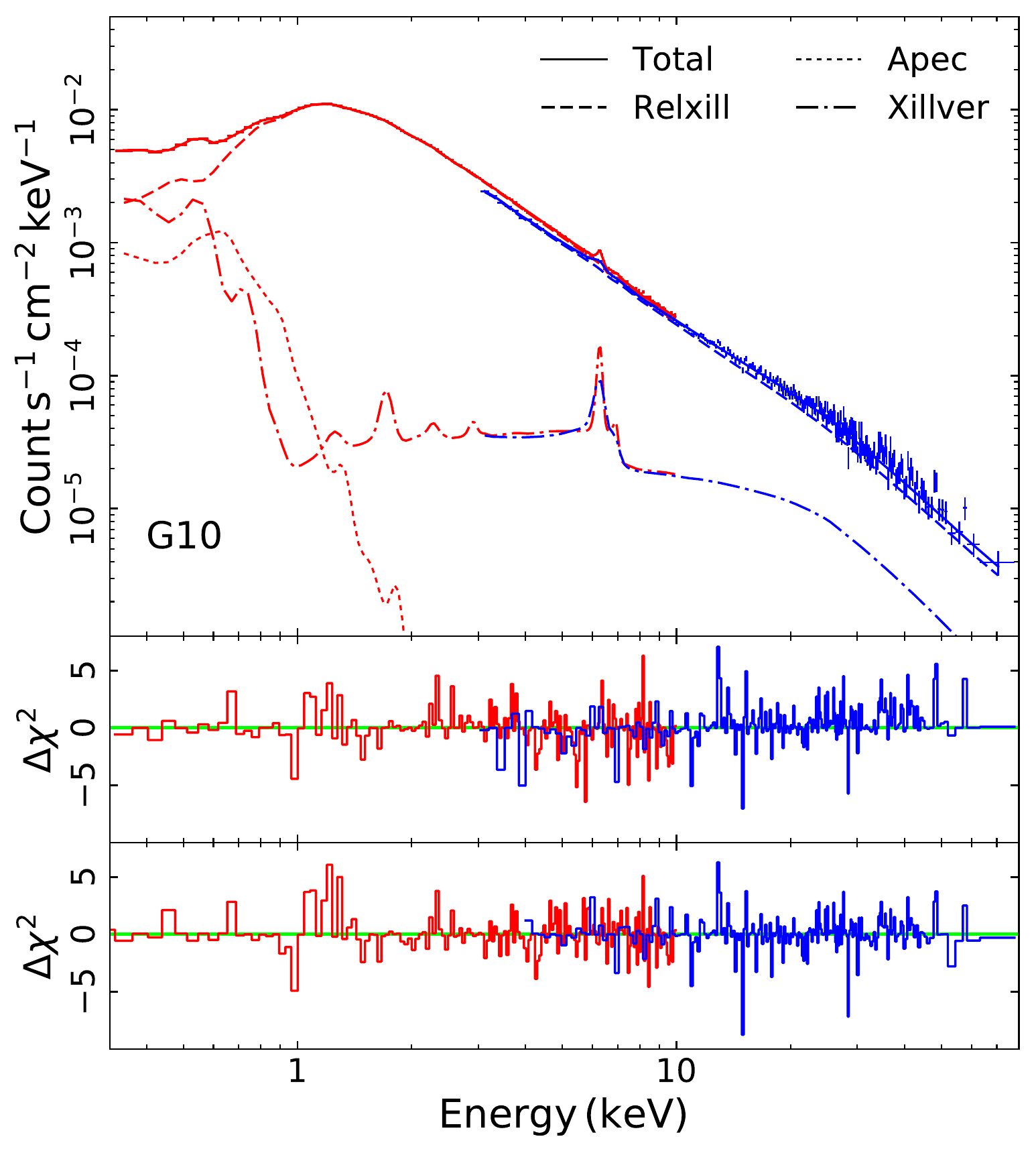}
\includegraphics[width = 0.29\textwidth]{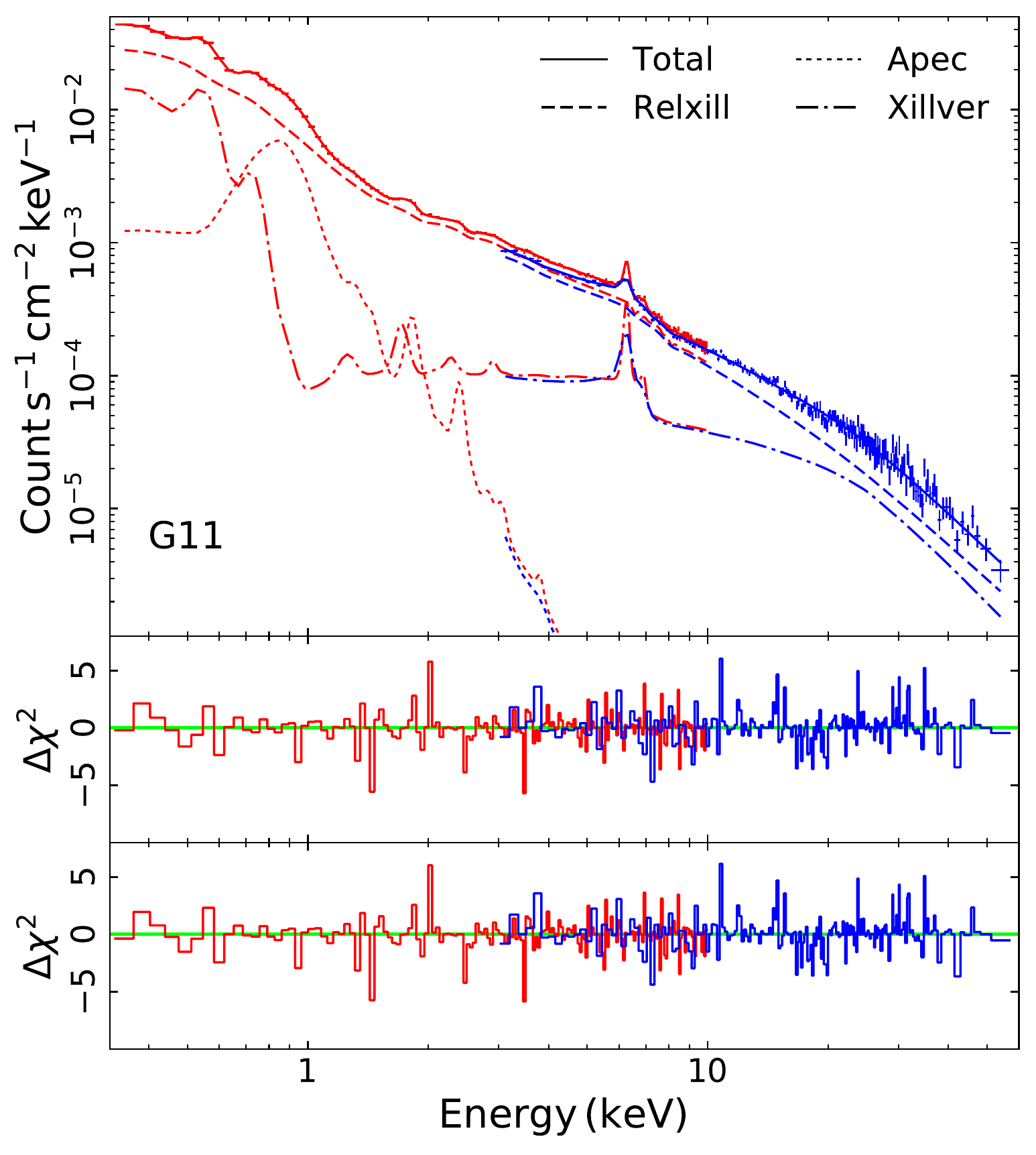}
\includegraphics[width = 0.29\textwidth]{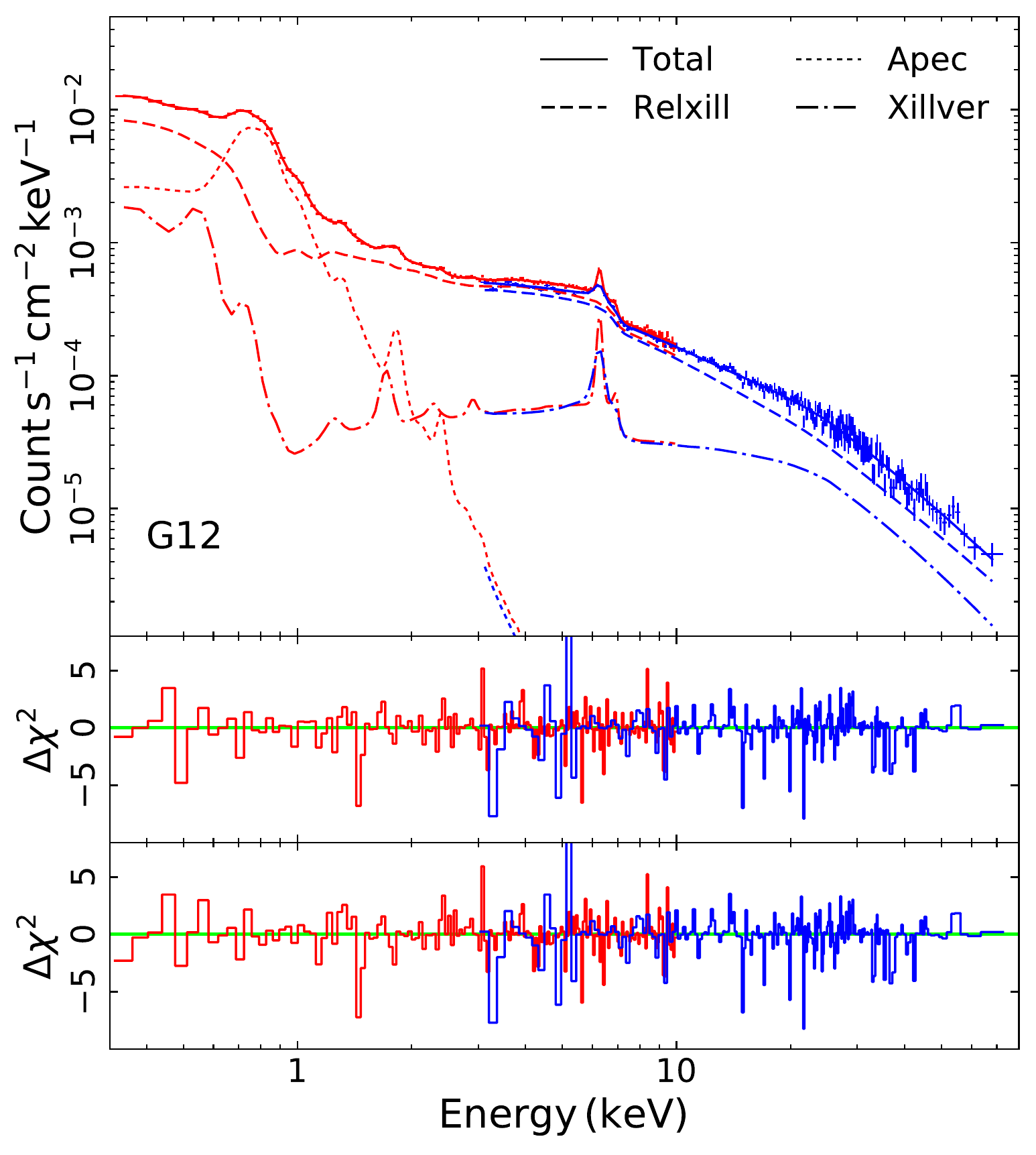}
\caption{Top panel: Simulated {\it XMM-Newton} (red) and {\it NuSTAR} (blue) spectra together with the various components of the theoretical model assumed. Primary emission plus ionized reflection (dashed lines), neutral reflection (dash dotted lines), and thermal emission (dotted lines) are indicated. Middle and bottom panels: The $\chi^2$ residuals obtained by the two blind fits (See \S\,\ref{sec:fitting} for details) are shown.}
\label{fig:appendix-spectra}
\end{figure*}

\begin{figure*}
\ContinuedFloat 
\centering
\includegraphics[width = 0.29\textwidth]{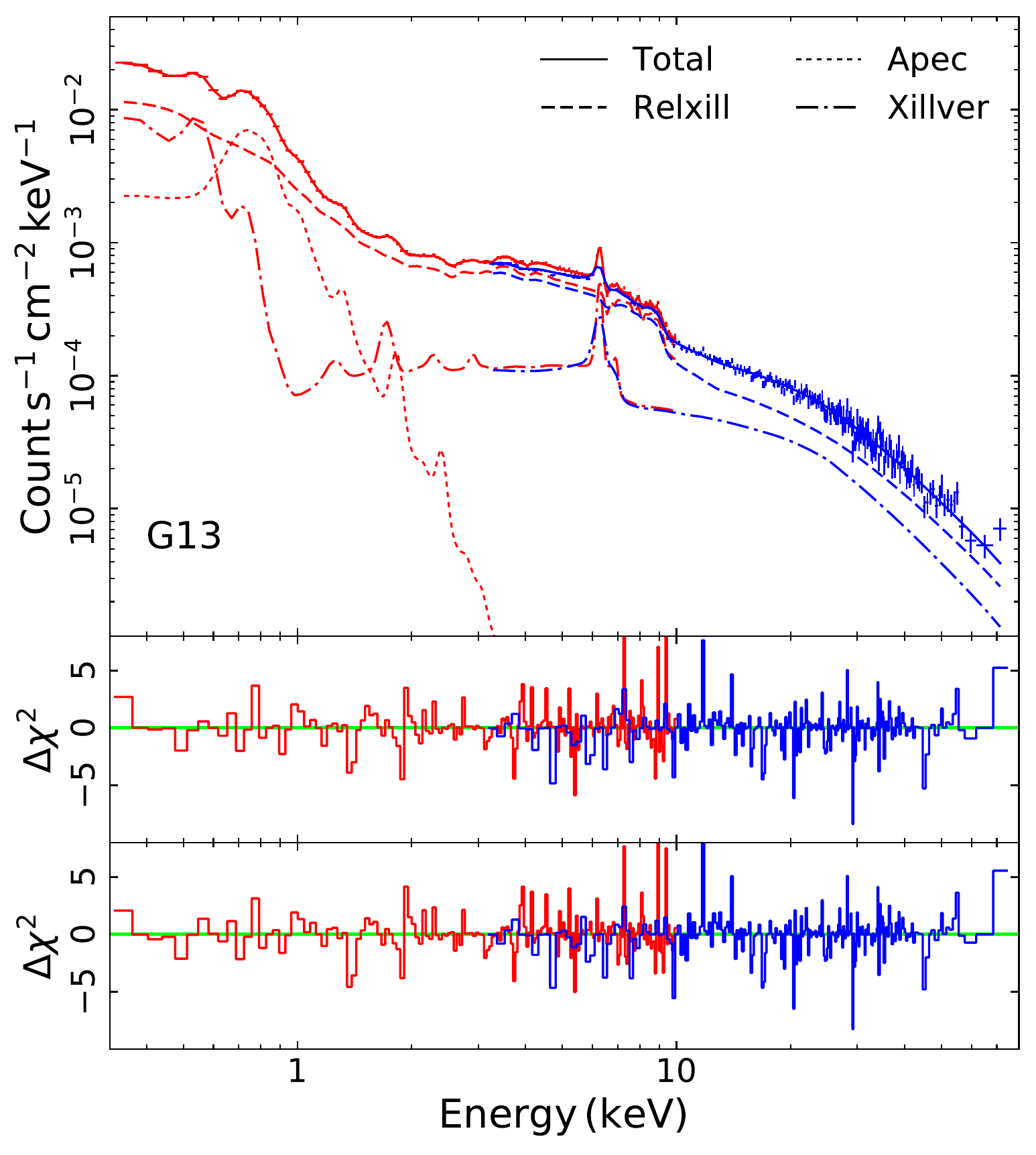}
\includegraphics[width = 0.29\textwidth]{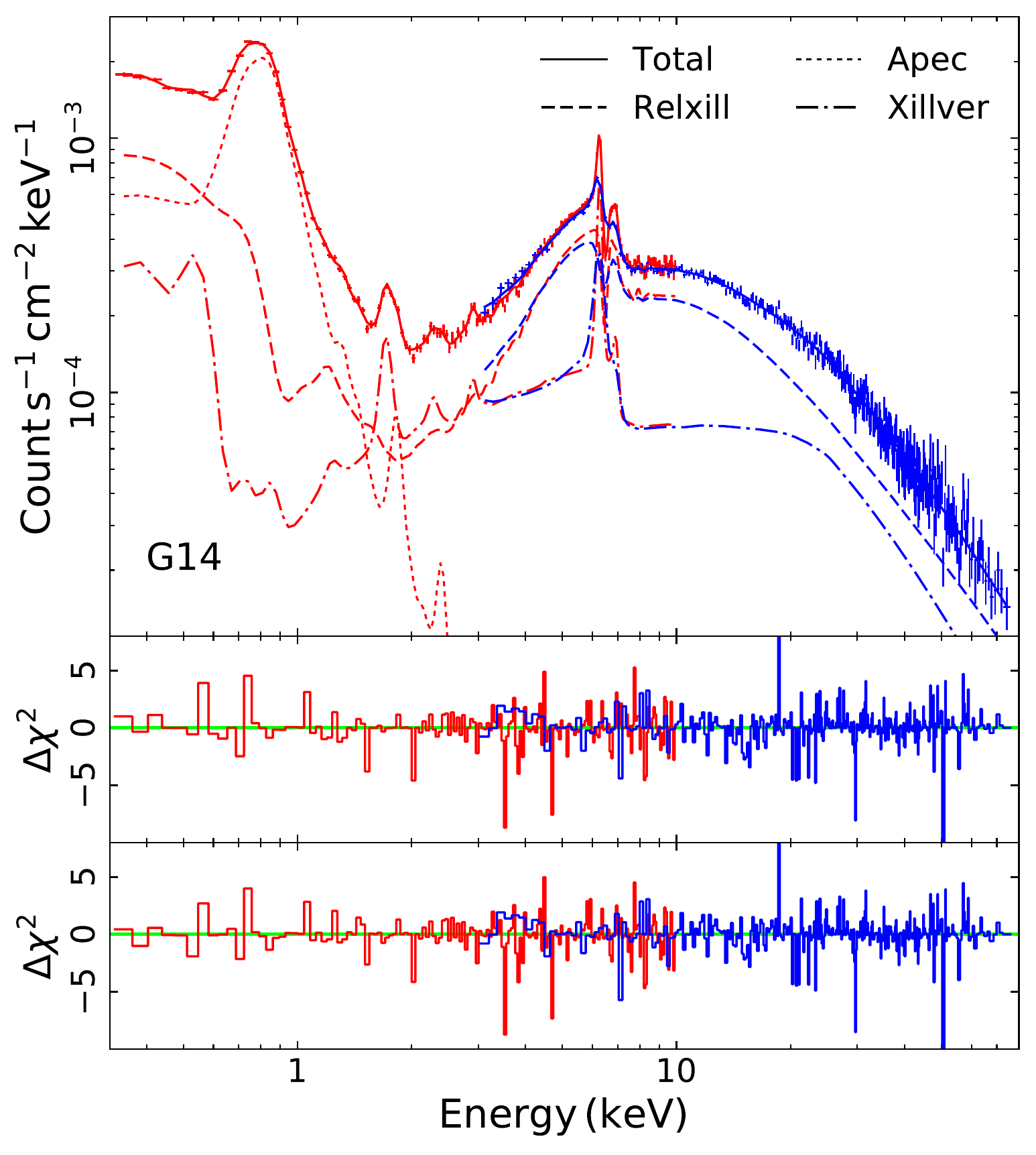}
\includegraphics[width = 0.29\textwidth]{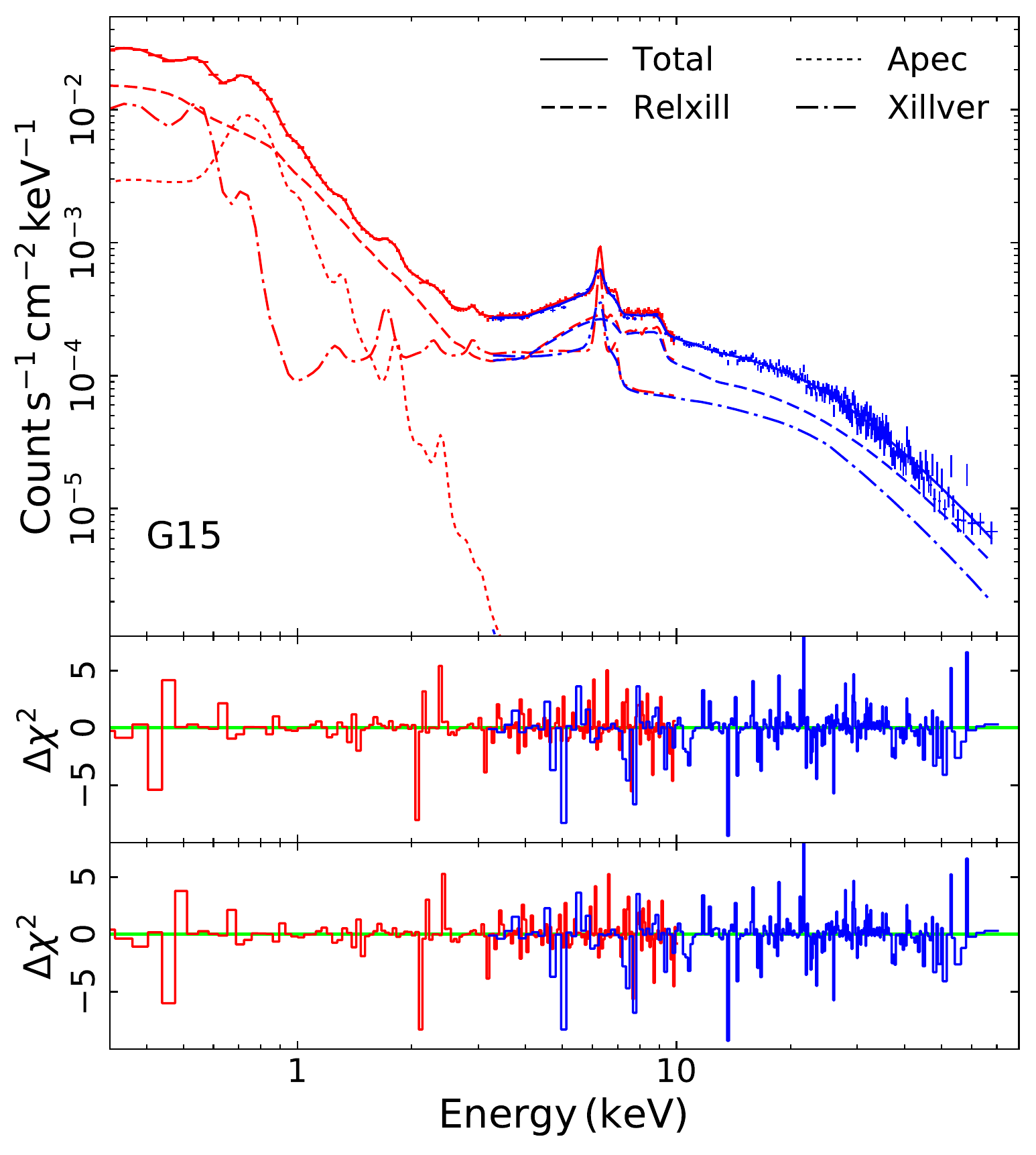}
\includegraphics[width = 0.29\textwidth]{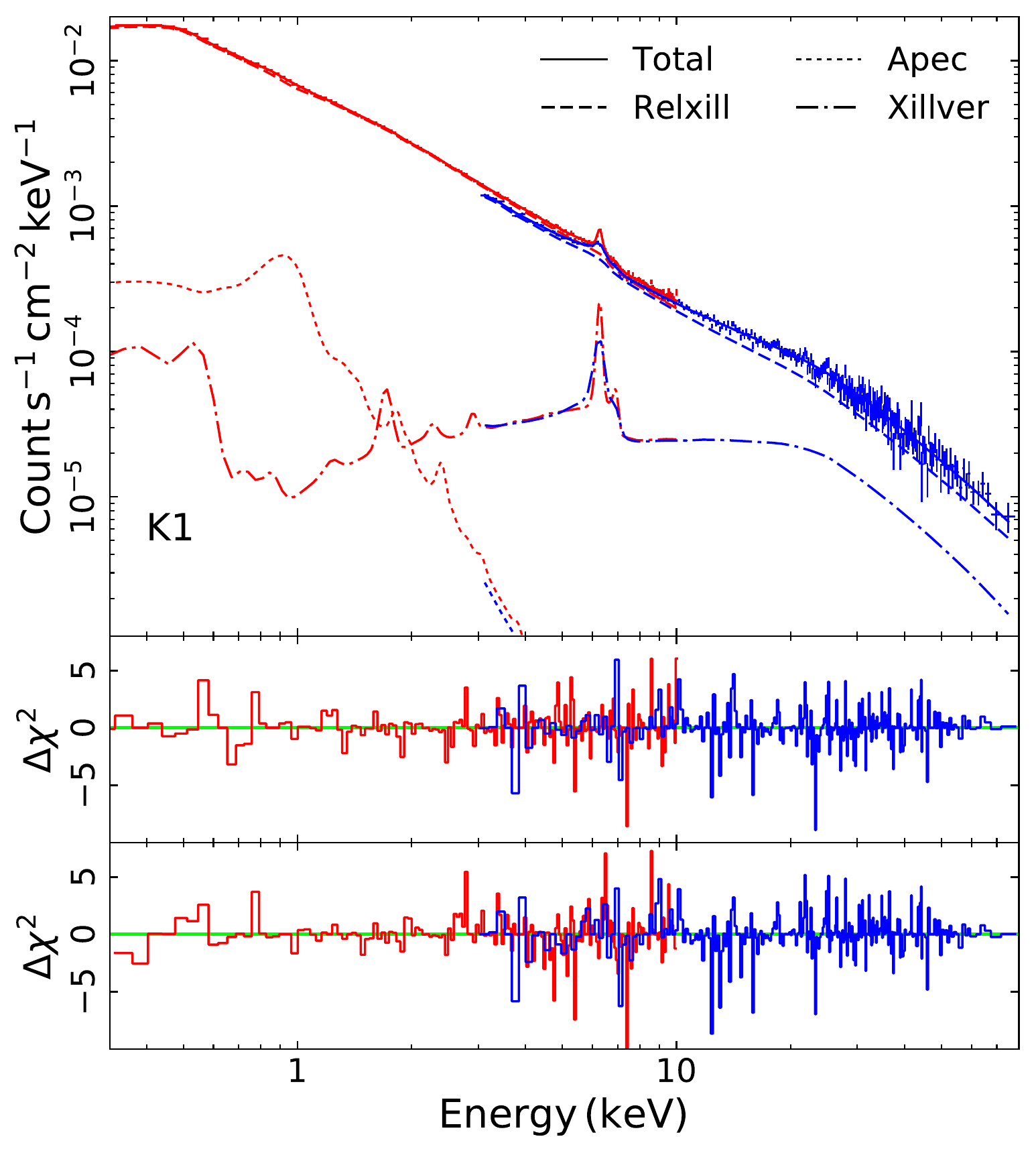}
\includegraphics[width = 0.29\textwidth]{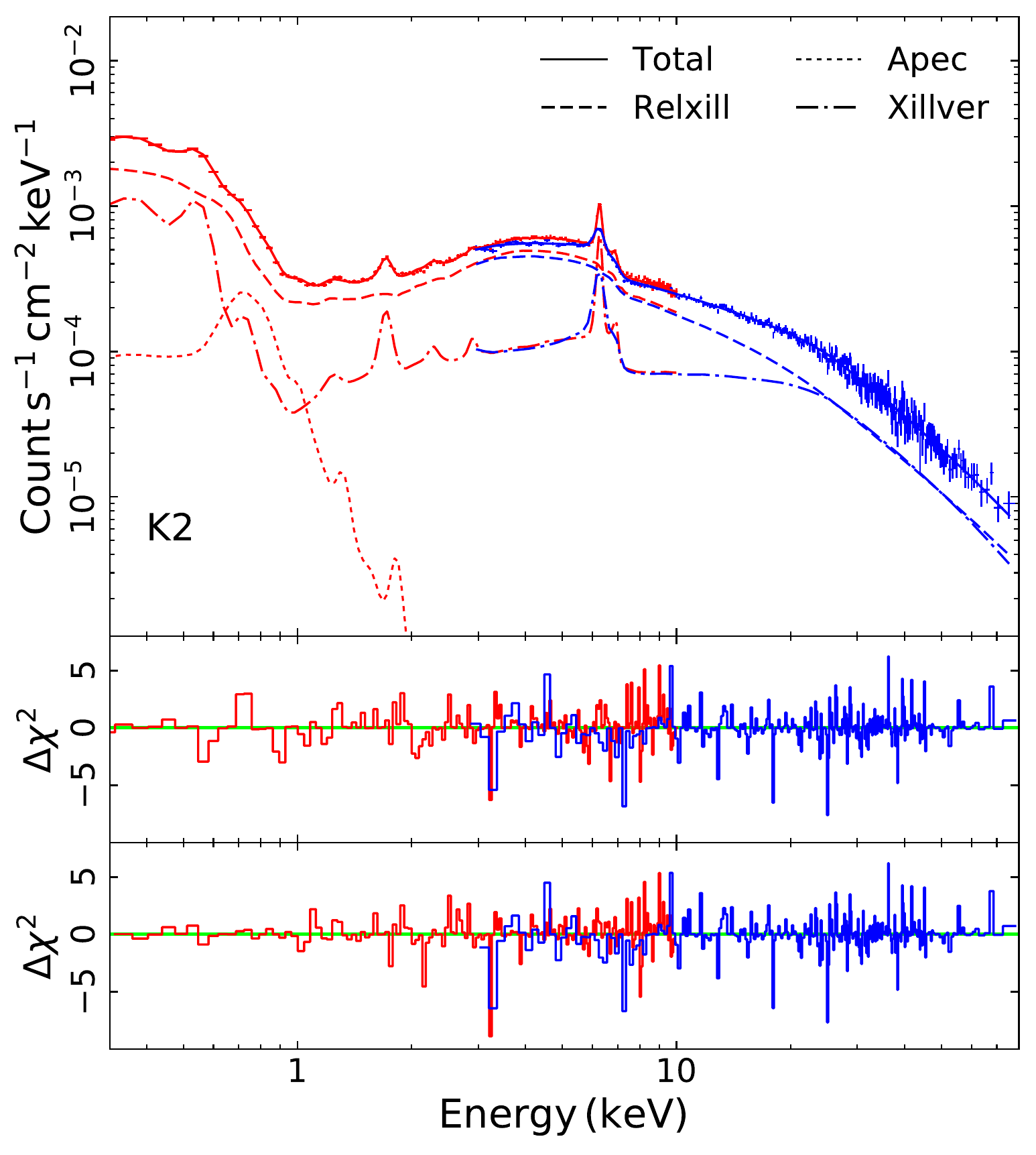}
\includegraphics[width = 0.29\textwidth]{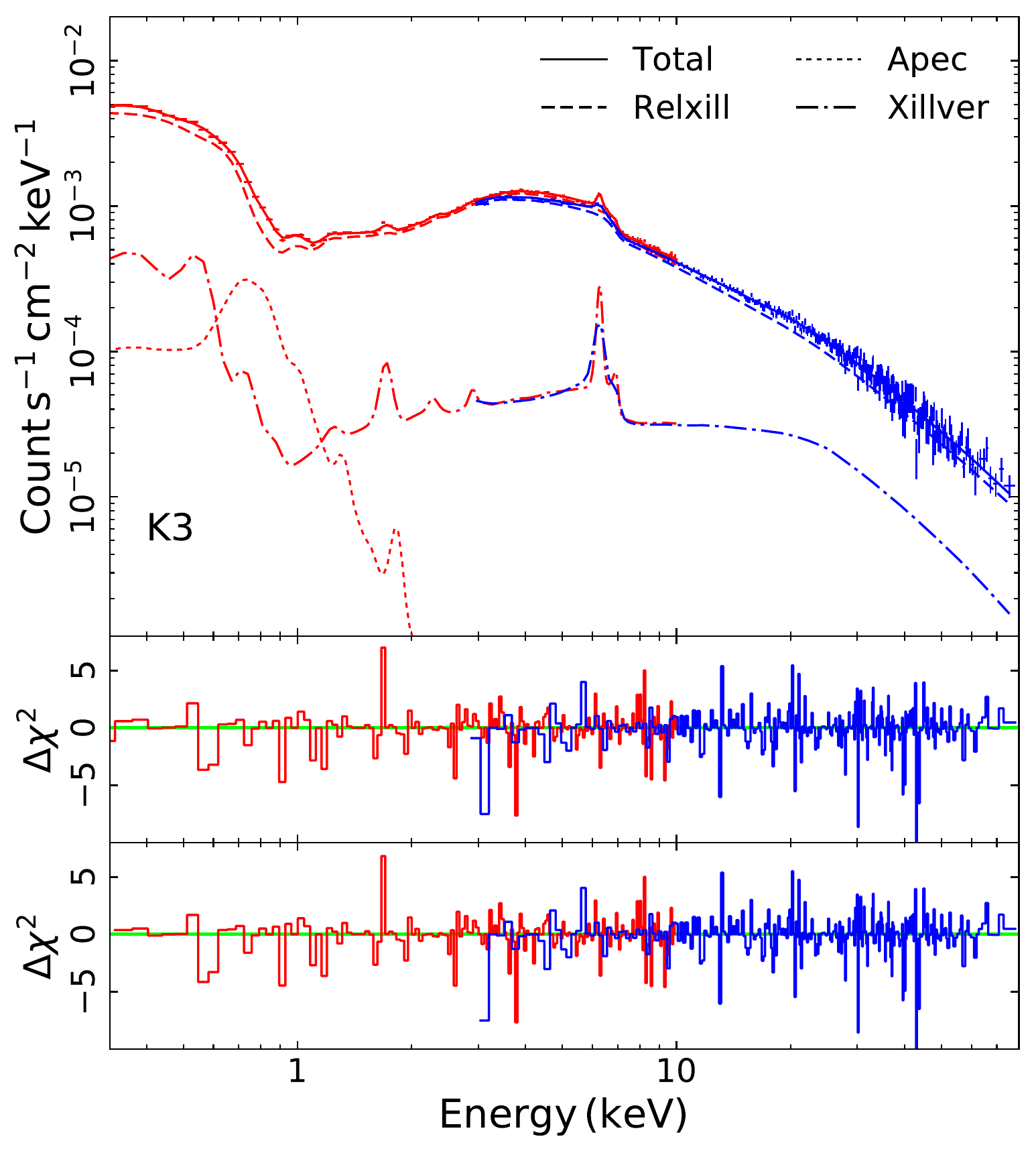}
\includegraphics[width = 0.29\textwidth]{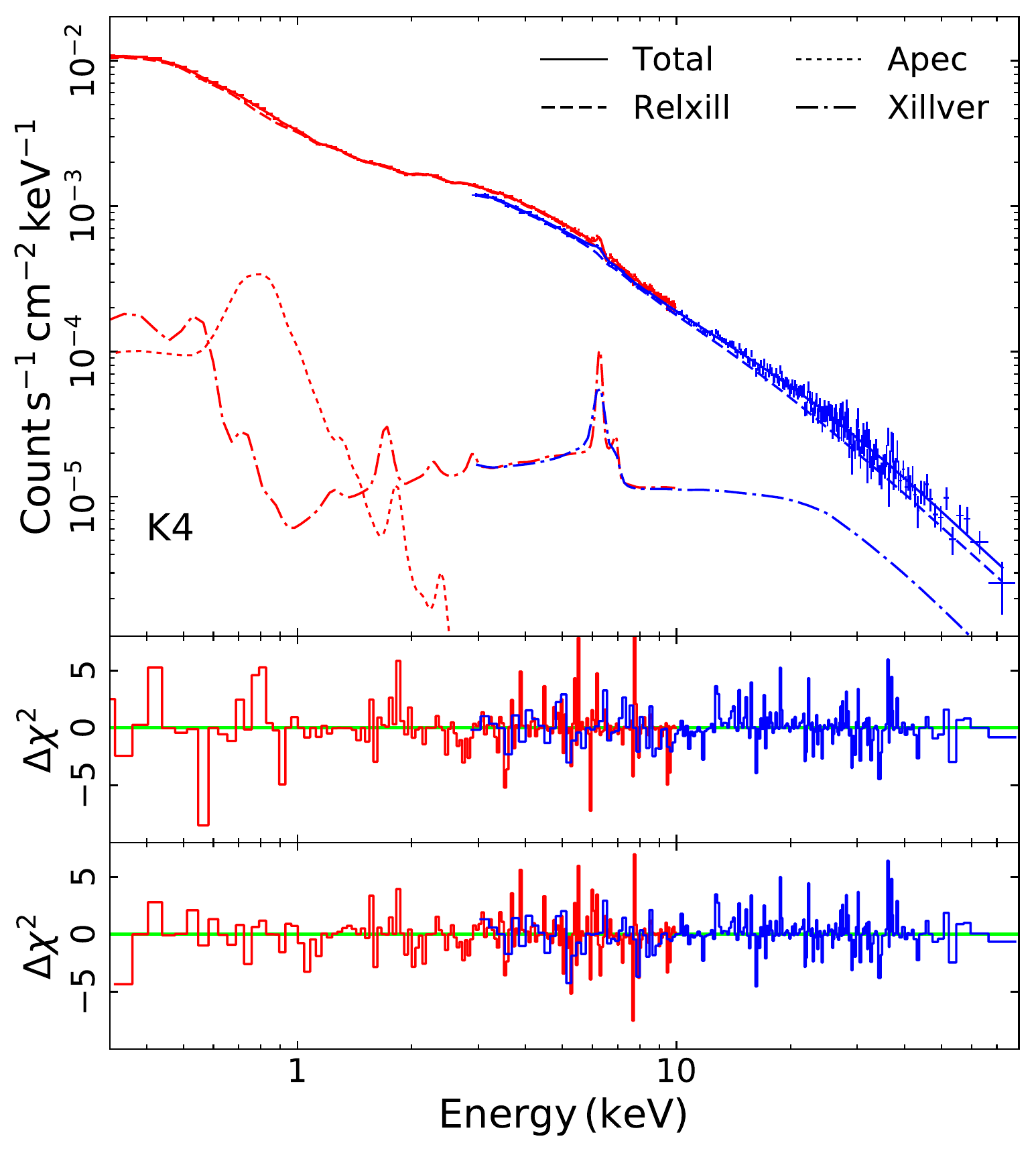}
\includegraphics[width = 0.29\textwidth]{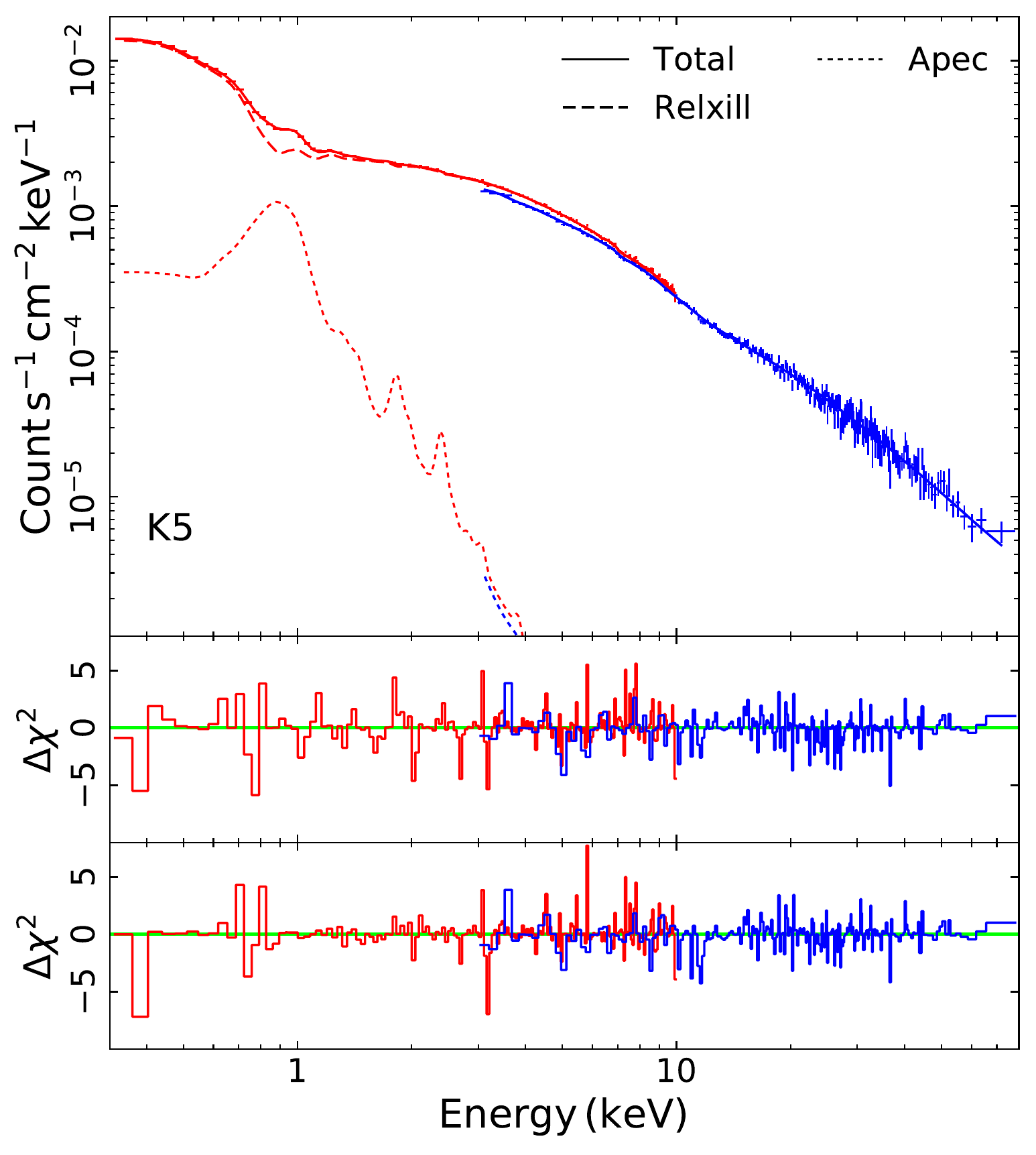}
\includegraphics[width = 0.29\textwidth]{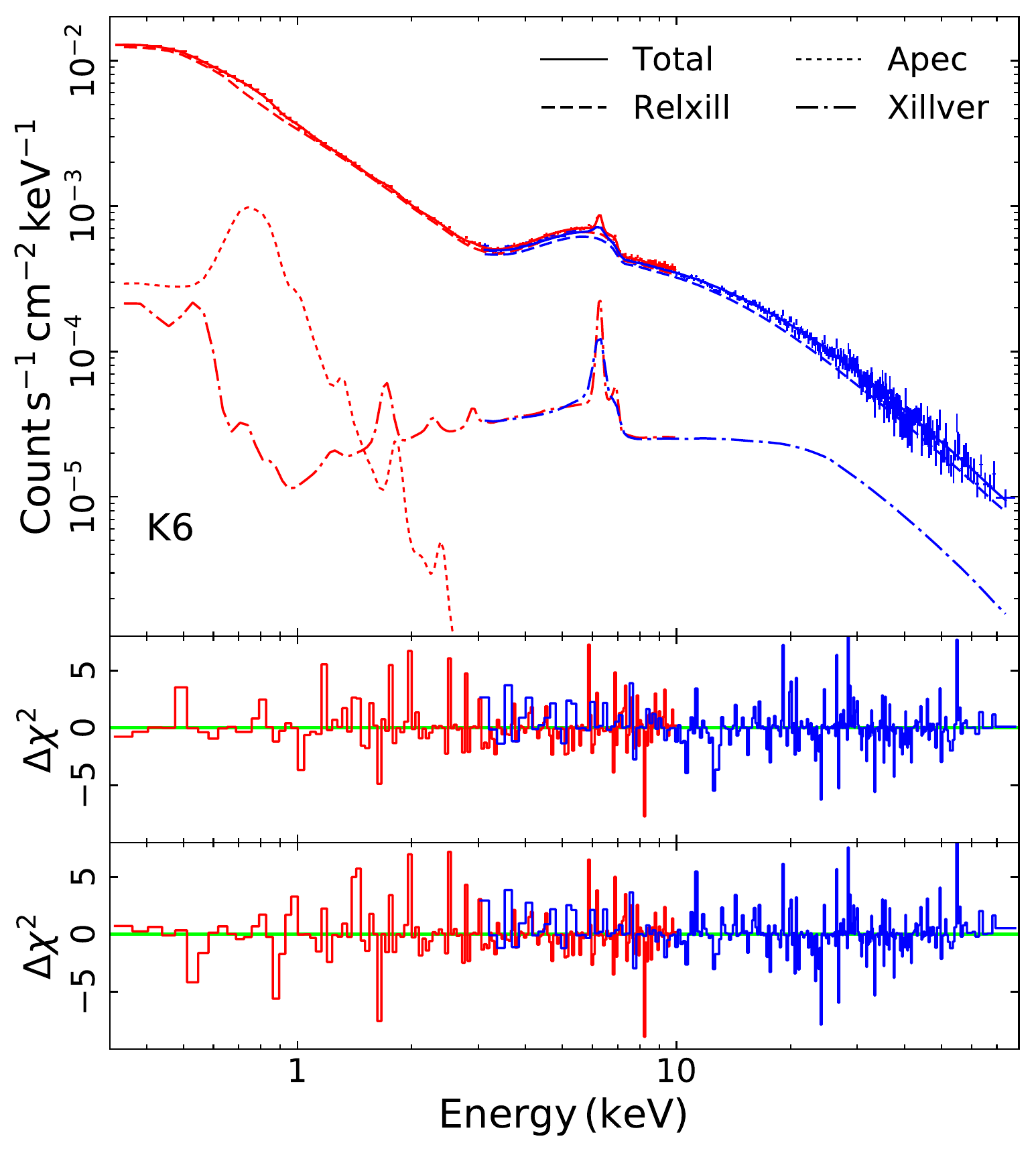}
\includegraphics[width = 0.29\textwidth]{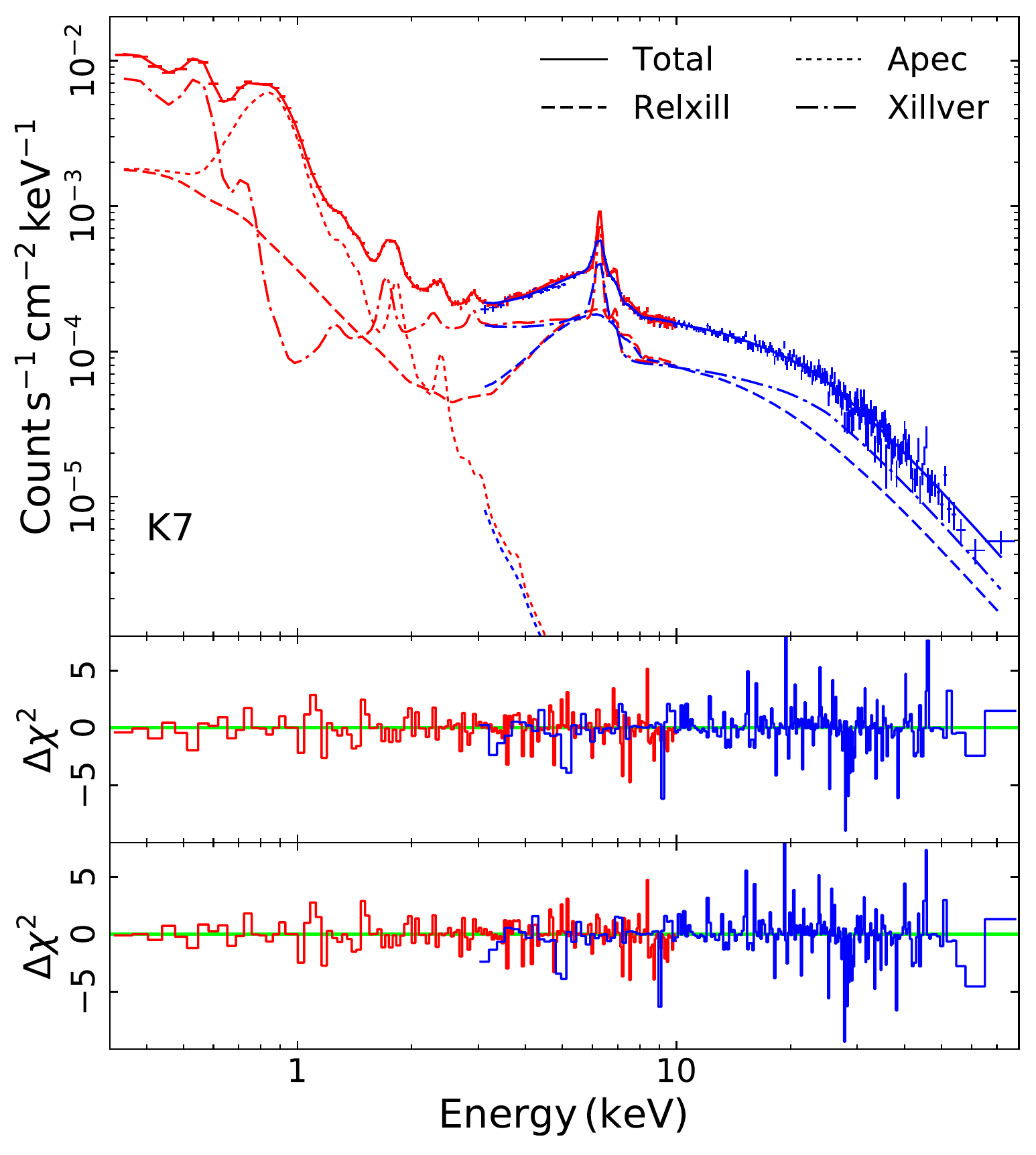}
\includegraphics[width = 0.29\textwidth]{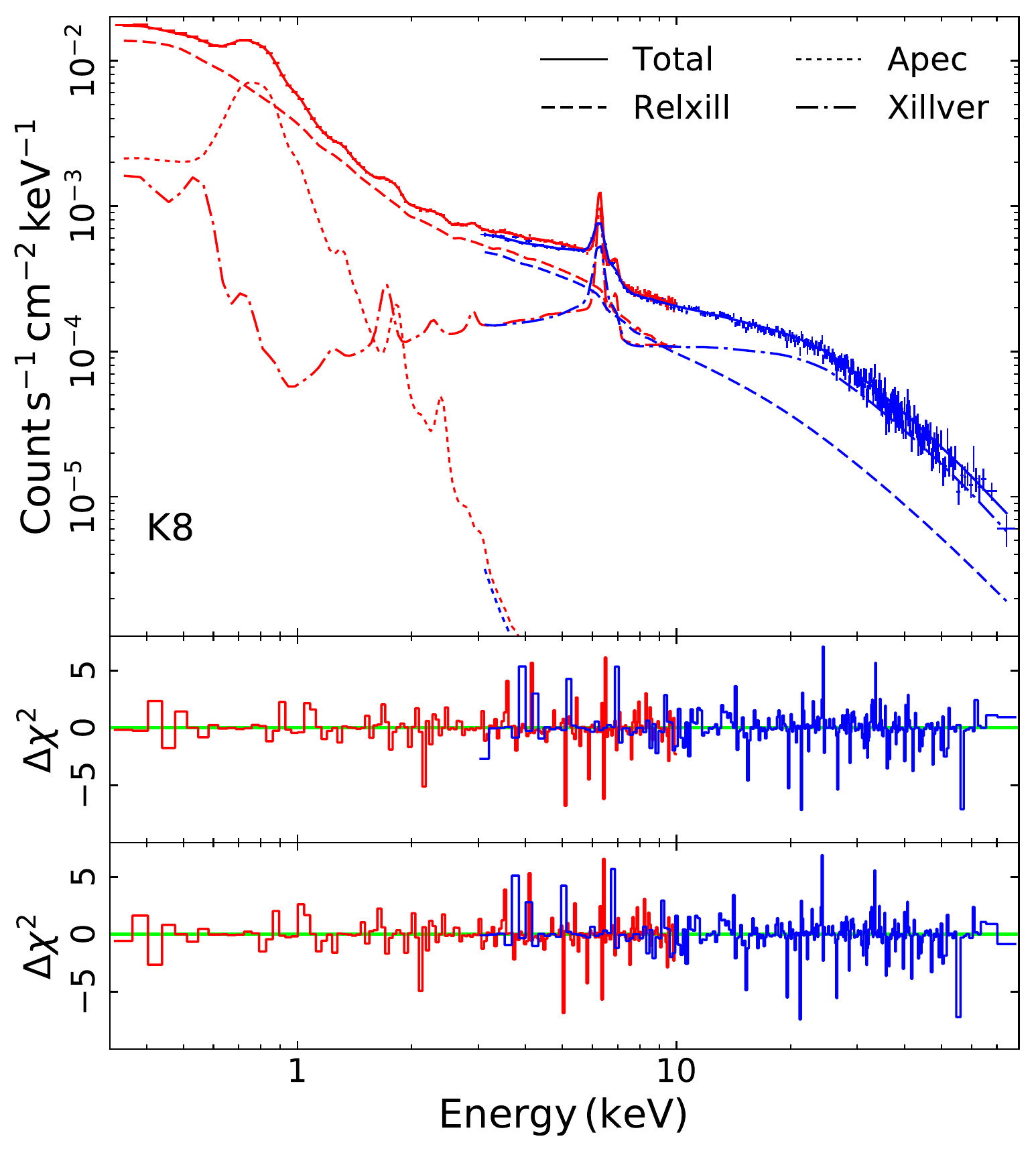}
\includegraphics[width = 0.29\textwidth]{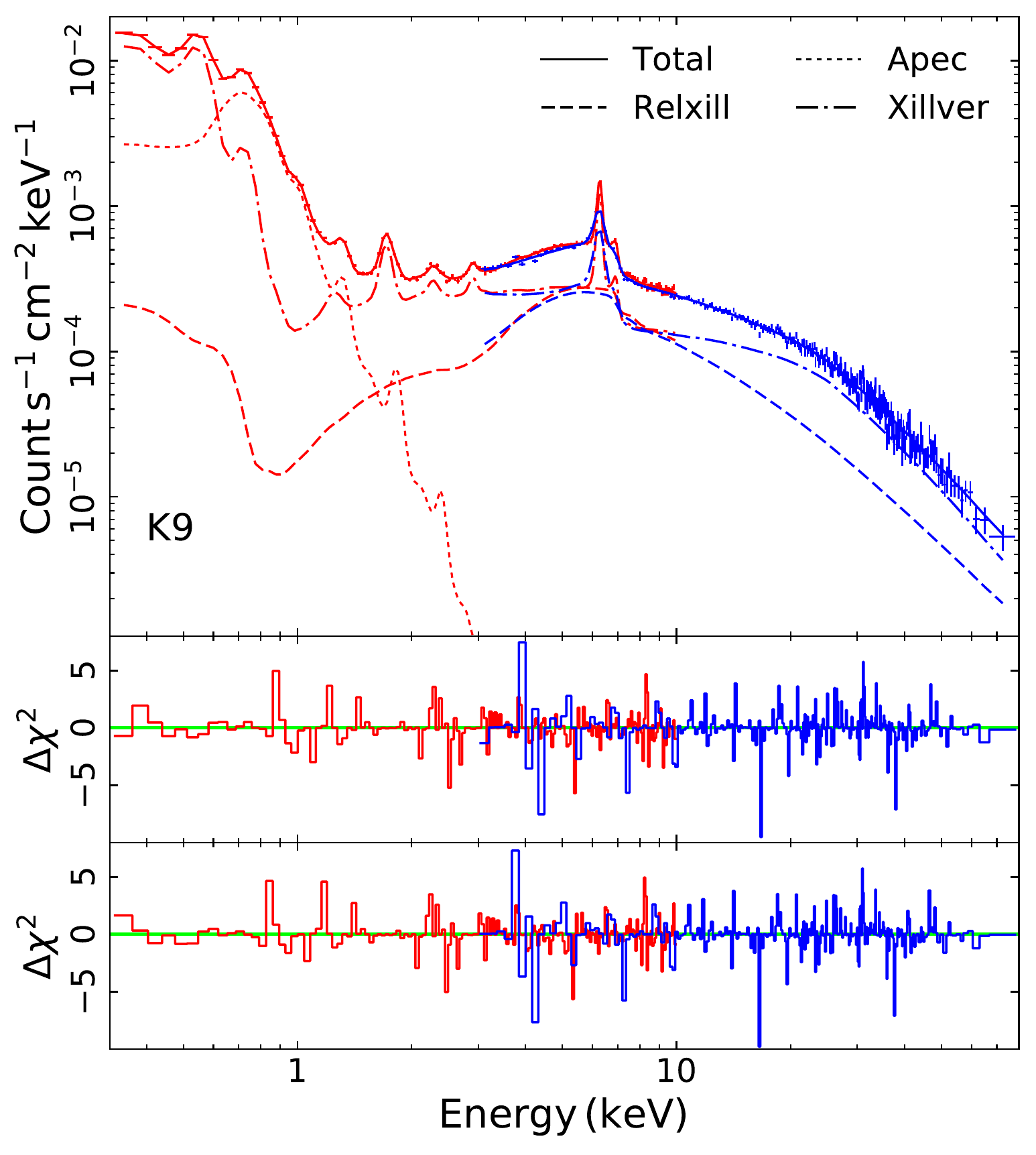}
\caption{Continued}
\end{figure*}

\begin{figure*}
\ContinuedFloat 
\centering
\includegraphics[width = 0.29\textwidth]{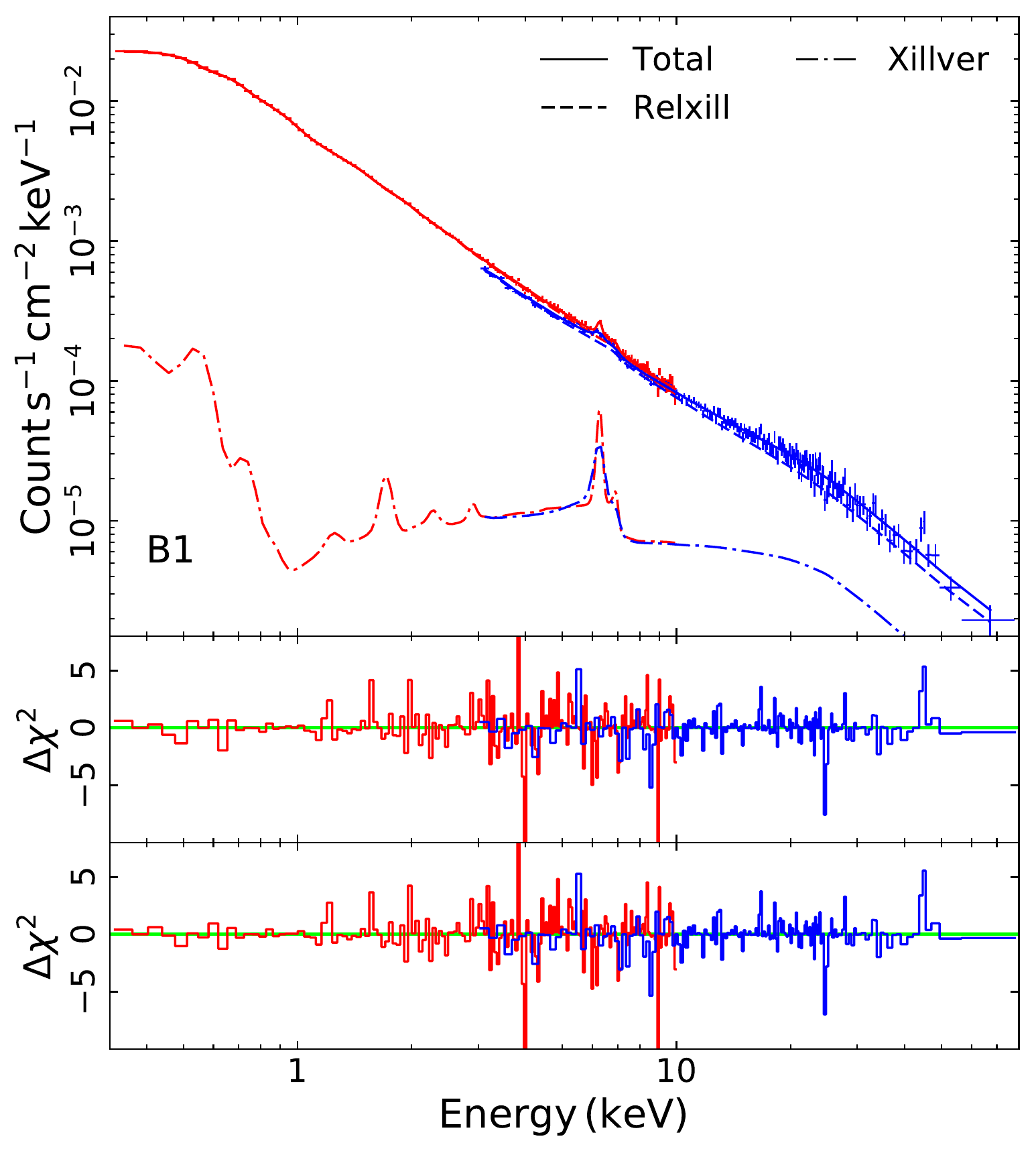}
\includegraphics[width = 0.29\textwidth]{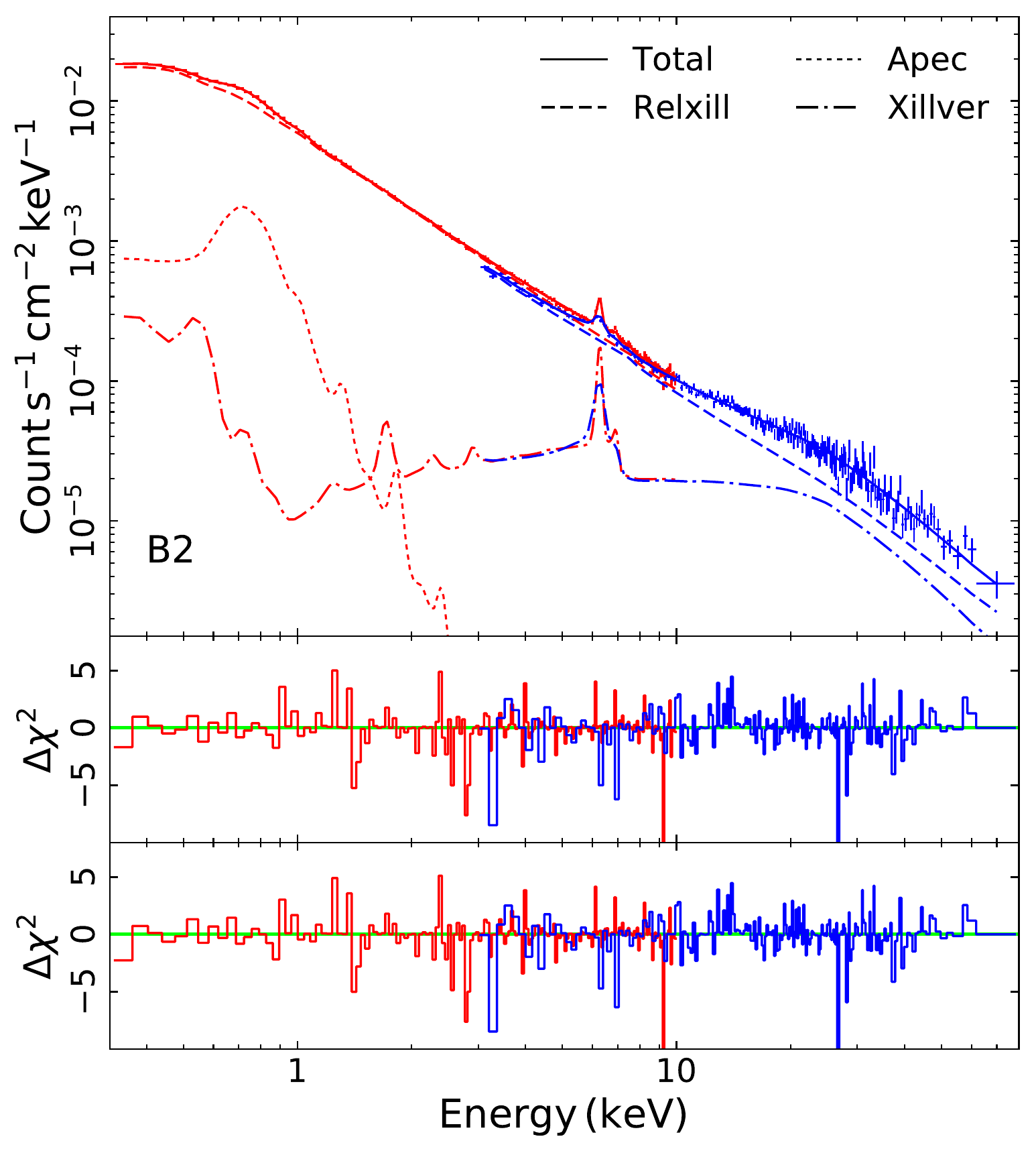}
\includegraphics[width = 0.29\textwidth]{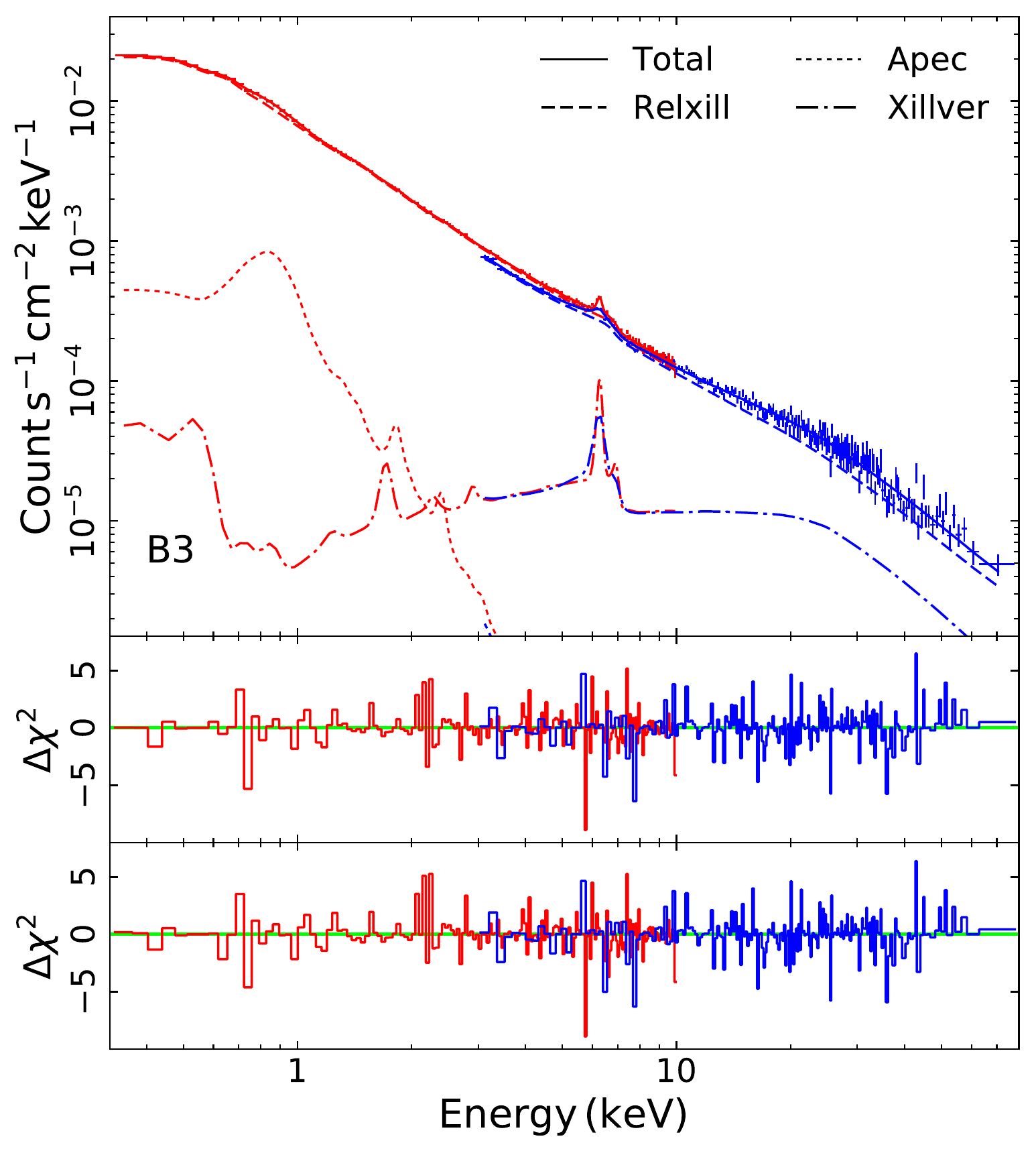}
\includegraphics[width = 0.29\textwidth]{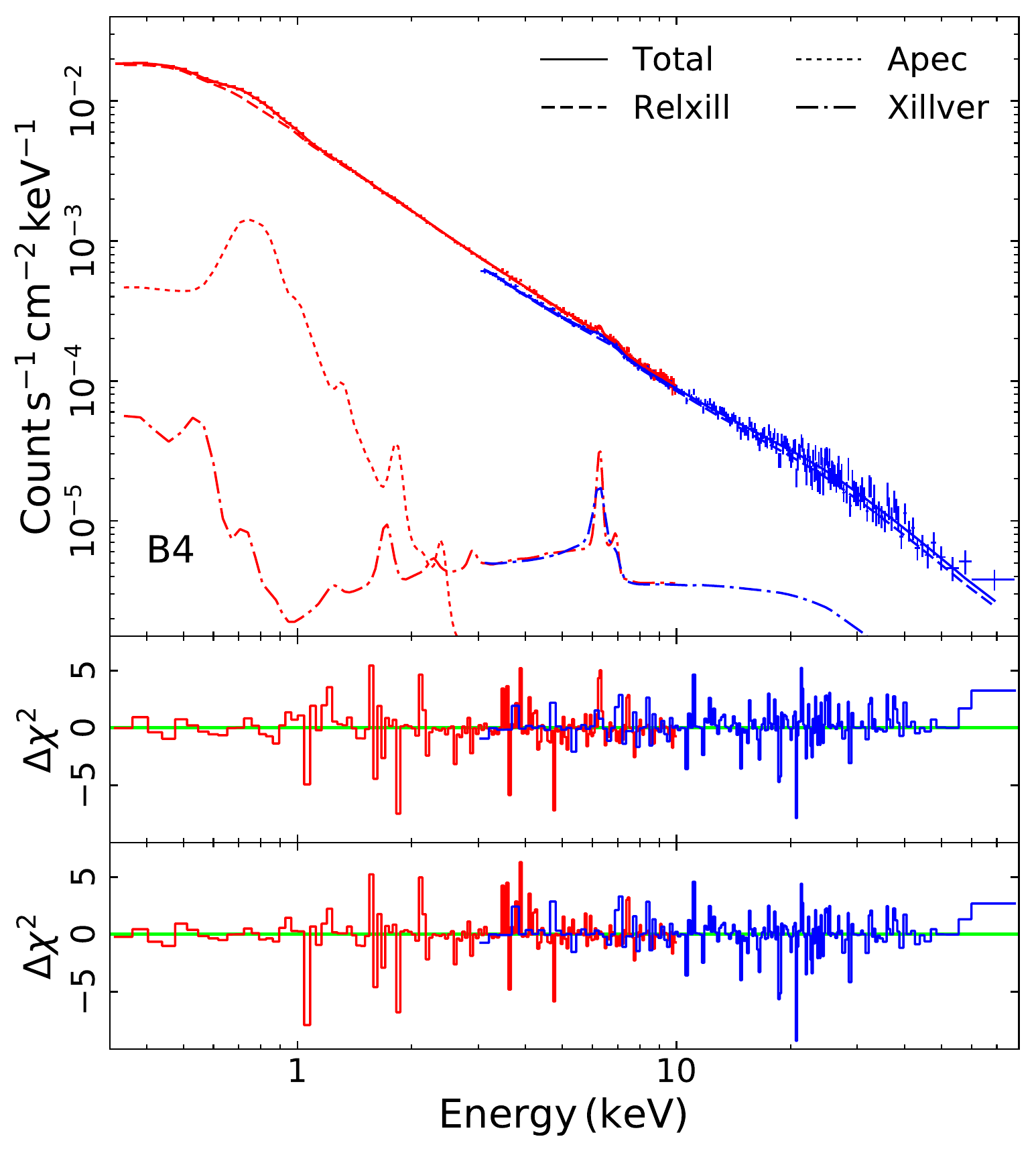}
\includegraphics[width = 0.29\textwidth]{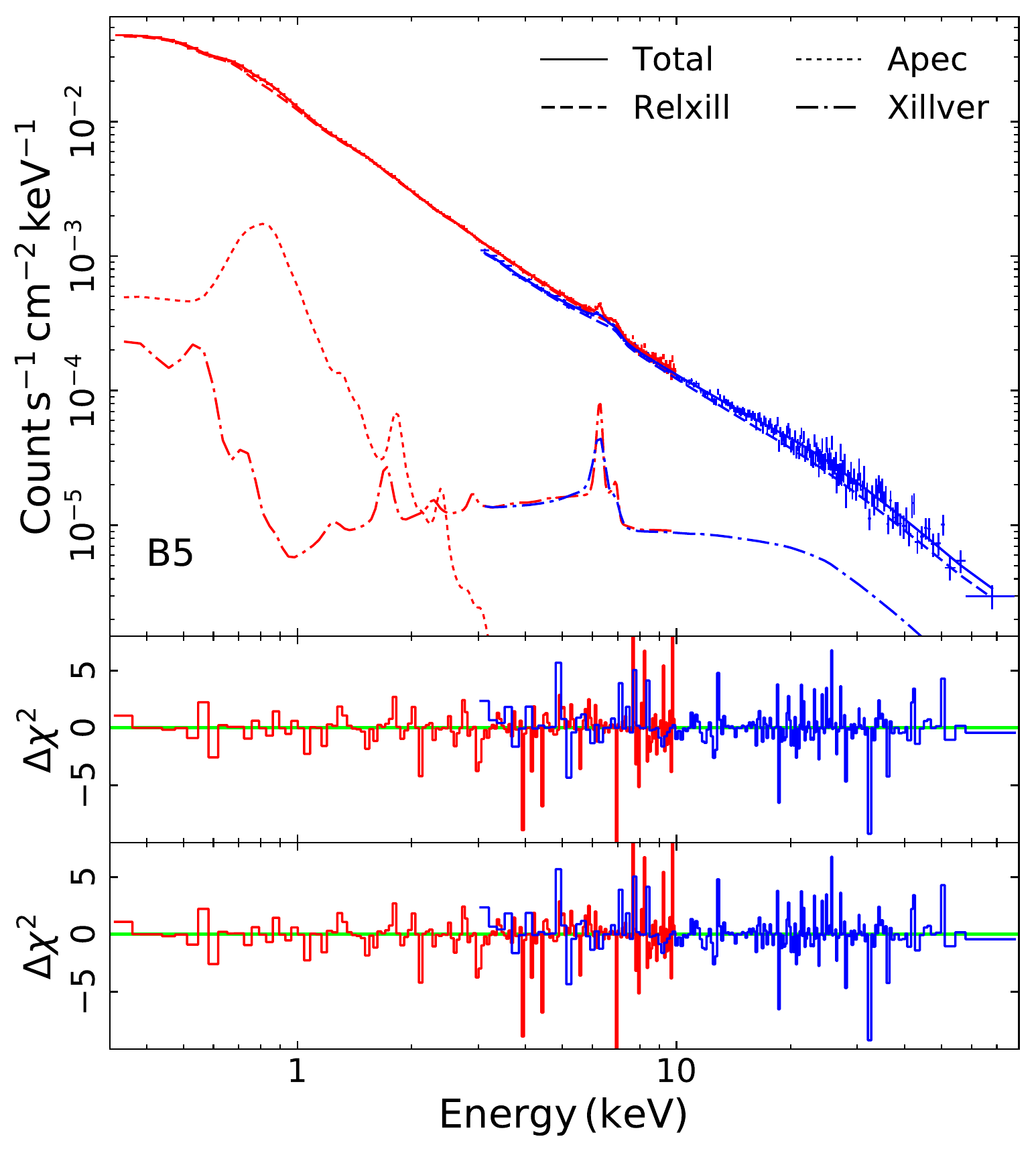}
\includegraphics[width = 0.29\textwidth]{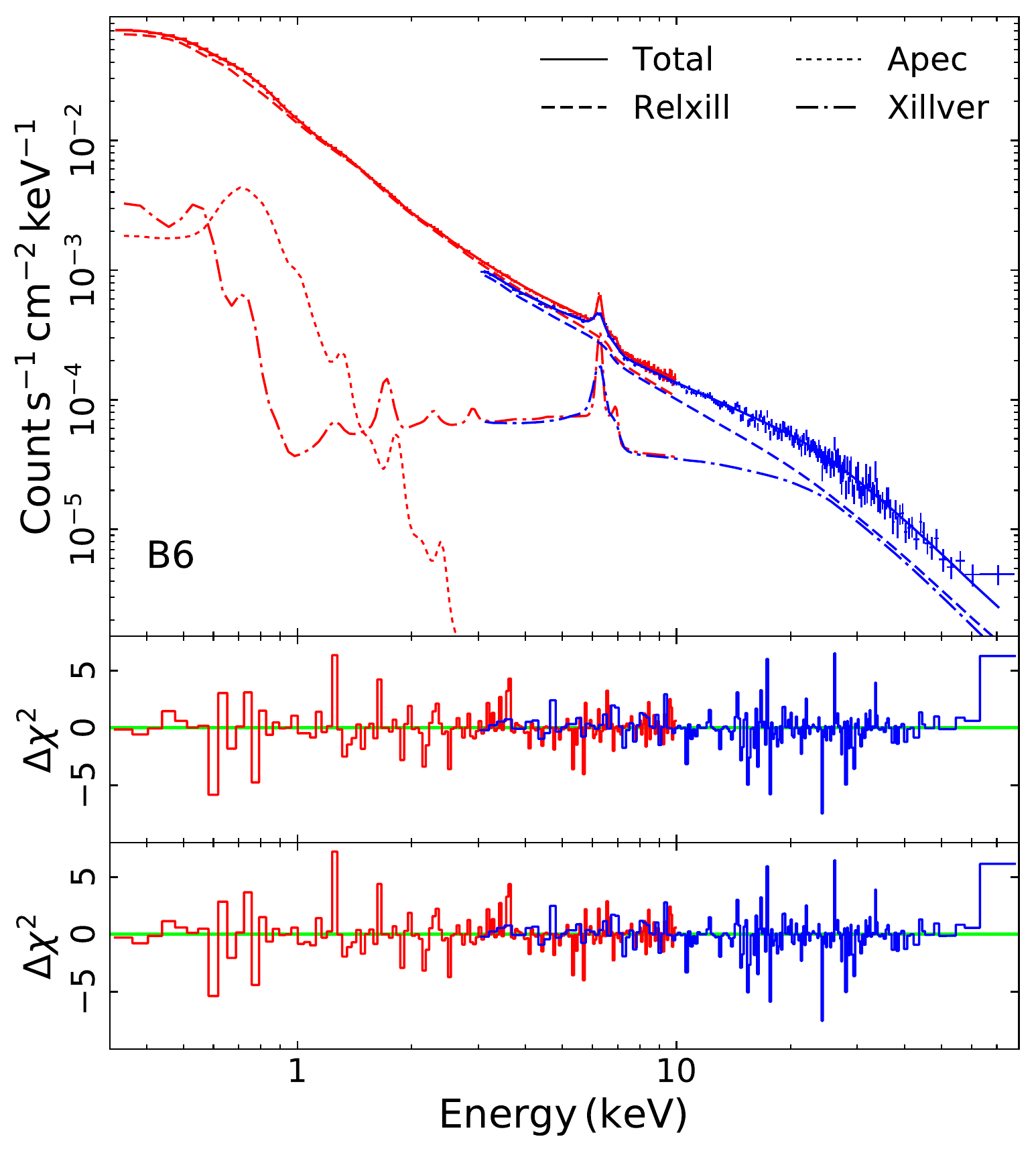}
\caption{Continued}
\end{figure*}

\end{document}